\documentclass[aps,superscriptaddress, showpacs,preprintnumbers, superscriptaddress, nofootinbibt, twocolumn]{revtex4-2}
\usepackage{eurosym}
\usepackage{amsfonts}
\usepackage{amssymb,amsmath,enumitem}
\usepackage{graphicx,color,epstopdf}
\usepackage{subfigure}
\usepackage{ulem}
\usepackage{tabularx}
\usepackage{booktabs}

\def\be{\begin{equation}}
\def\ee{\end{equation}}
\def\beg{\begin{align}}
\def\eeg{\end{align}}
\def\bea{\begin{eqnarray}}
\def\eea{\end{eqnarray}}

\def\nn{\nonumber \\}

\begin{document}

\title{Black hole solutions in the quadratic Weyl conformal geometric theory of gravity}
\author{Jin-Zhao Yang}
\email{yangjch6@mail3.sysu.edu.cn}
\affiliation{School of Physics, Sun Yat-Sen University, Xingang Road,
Guangzhou 510275, P. R. China,}
\author{Shahab Shahidi}
\email{s.shahidi@du.ac.ir}
\affiliation{School of Physics, Damghan University, Damghan, Iran,}
\author{Tiberiu Harko}
\email{tiberiu.harko@aira.astro.ro}
\affiliation{Department of Physics, Babes-Bolyai University, Kogalniceanu Street,
	Cluj-Napoca 400084, Romania,}
\affiliation{Department of Theoretical Physics, National Institute of Physics
and Nuclear Engineering (IFIN-HH), Bucharest, 077125 Romania,}
\affiliation{Astronomical Observatory, 19 Ciresilor Street,
	Cluj-Napoca 400487, Romania}

\date{\today }

\begin{abstract}
We consider numerical black hole solutions in the Weyl conformal geometry, and its associated conformally invariant Weyl quadratic gravity. In this model Einstein gravity (with a positive cosmological constant) is recovered in the spontaneously broken phase of Weyl gravity, after the Weyl gauge field ($\omega _{\mu}$)  becomes massive through a Stueckelberg mechanism,  and it decouples. As a first step in our investigations we write down the conformally invariant gravitational action, containing a scalar degree of freedom, and the Weyl vector. The field equations are derived from the variational principle in the absence of matter. By adopting a static spherically symmetric geometry, the vacuum field equations for the gravitational, scalar, and Weyl fields are obtained. After reformulating the field equations in a dimensionless form, and by introducing a suitable independent radial coordinate, we obtain their solutions numerically. We detect the formation of a black hole from the presence of a Killing horizon for the timelike Killing vector in the metric tensor components, indicating the existence of the singularity in the metric. Several models, corresponding to different functional forms of the Weyl vector, are considered. An exact black hole model, corresponding to a Weyl vector having only a radial spacelike component, is also obtained. The thermodynamic properties of the Weyl geometric type black holes (horizon temperature, specific heat, entropy and evaporation time due to Hawking luminosity) are also analyzed in detail.
\end{abstract}

\maketitle
\tableofcontents


\section{Introduction}

One of the fundamental results in theoretical relativistic astrophysics indicates that compact objects having mass functions greater than 3-4 solar masses must be black holes \citep{Mmass}. Black hole solutions are among the first that have been considered in the framework of general relativity, with the first vacuum solution of the Einstein gravitational field equations obtained more than one hundred years ago \cite{Sch}. However, the first firm observational evidence for the existence of black holes was found relatively recently, beginning with the astronomical discoveries of the 1970's. One of the first black hole candidates was observed in a binary system, consisting of the supergiant star HD 226868, associated with the High Mass X-ray Binary object Cyg X-1. The determination of the mass function $f\left(M_x\right)$, giving the mass $M_x$ of the compact object Cyg X-1 in terms of the companion star $M_c$, and of the inclination angle $i$, already showed that $M_x>4M_{\odot}$ \citep{Bo75}, indicating that Cyg X-1 may be a stellar mass black hole. In the Milky Way alone the total number of stellar mass black holes (in binaries or isolated) is estimated to be of the order of 100 million \cite{Ro98}.
On the other hand,  compact objects consisting of exotic matter (axion stars, Bose-Einstein Condensate stars, quark stars etc.) have many properties very similar to those of the stellar type black holes \citep{To02,YuNaRe04}. Hence, finding effective methods  for distinguishing between different types of compact object is a fundamental problem in present day astronomy and astrophysics.

An important, and intriguing class of black holes, is represented by the  supermassive black holes, which are located at the center of every massive galaxy \cite{Mag98}. A supermassive black hole accretes during its lifetime a significant amount of matter, which leads to the formation of an accretion disk around the compact object. For a review of the supermassive black hole properties see \cite{Mer2016}. If the supermassive black hole is in the accreting state, it is called an active galactic nucleus. 
The most important observational evidence for the existence of supermassive black holes is provided by the Very-Long Baseline Interferometry (VLBI) imaging of molecular ${\rm H_2O}$ masers. The VLBI imaging, using Doppler shift measurements that assume a Keplerian type motion of the masering source, led to the very precise estimation of the mass of the black hole at the center of the active galaxy NGC4258, which was obtained as $3.6 \times 10^7M_{\odot}$ \cite{Mi95}. The closest to Earth supermassive black hole is the variable infrared, radio and X-ray source Sgr A*, located at the center of the Milky Way \cite{Sgr1,Sgr2,Sgr3}.
The investigation of the orbital motion of the stars around Sgr A* lead to further confirmations of the predictions of the general theory of relativity. An important recent advance in black hole physics was the observation by the Event Horizon Telescope collaboration of the first image of the supermassive black hole M87*  \cite{EHT1,EHT2,EHT3}. These observations are consistent with the Kerr-like nature of the Sgr A* black hole, but still {\it deviations from the predictions of general relativity cannot be excluded a priori}.

Hence, finding black hole solutions is essential for the theoretical understanding, and the observational testing of gravitational theories. The first vacuum solution of the general relativistic field equations \cite{Sch}, found by Karl Schwarzschild in 1916, was obtained by considering a static, spherically symmetric central compact object. There is a large number of exact vacuum and black hole solutions in general relativity, and in its extensions. For a review of the exact solutions of the Einstein gravitational field equations see \cite{Rev}. One effective method to test black hole properties, as well as the nature of the gravitational force, is by using the electromagnetic emissivity properties of thin accretion disks that form around compact astrophysical type objects \cite{Accr1, Accr2, Accr3,Accr4, Accr5,Accr6,Accr7,Accr8, Accr9, Accr10, Accr11, Accr12,Accr13,Accr14, Accr15, Accr16, Accr17, Accr18, Accr19, Accr20}. For a review of the observational possibilities of testing black hole geometries by using electromagnetic radiation see \cite{Bam}.

Black hole solutions have been also investigated in modified theories of gravity. Some  investigations in this field can be found in the works \cite{Ste1,B1, B2, B3, Ste2,B4, B5, B6, B7, B8, B9, B10, B11, B12, B13, B14, B15, B16, B17, B18, B19, B20, B21, B22, B23, B24, B25, B26, B27, B28, B29, B30, B31, B32, B33, B34, B35, B36}. Static spherically symmetric solutions of the gravitational field equations obtained by including four-derivative terms of the form $\int{R_{\mu\nu}R^{\mu \nu}\sqrt{-g}d^4x}$ and $\int{R^2\sqrt{-g}d^4x}$, respectively, was considered in \cite{Ste1}.  The static, linearized solutions of the field equations are combinations of the Newtonian and Yukawa potentials, while the
acceptable  static metric solutions in the full nonlinear theory are regular at the origin. Static black hole solutions in Einstein's gravity with additional quadratic curvature terms were obtained in \cite{Ste2}, with the use of a Lichnerowicz type theorem that simplifies the analysis through the condition of a vanishing Ricci scalar curvature.
The existence of black hole solutions was proven by using numerical methods, and  their thermodynamic properties were also discussed.

 A family of exact black hole solutions on a static spherically symmetric background in second-order generalized Proca theories with derivative vector-field interactions coupled to gravity was obtained in \cite{B11}. The intrinsic vector-field derivative interactions  give rise to a secondary hair, induced by nontrivial field profiles. The deviation from General Relativity is most significant around the horizon. The properties of black holes on a static and spherically symmetric background in U (1) gauge-invariant scalar-vector-tensor theories with second-order equations of motion were studied in \cite{B23}. It was shown that in shift-symmetric theories invariant under the shift of scalar $\phi\rightarrow \phi +c$,  new hairy black hole solutions do exist.
Vacuum static spherically symmetric solutions in the hybrid metric-Palatini gravity theory, which is a combination of the metric and Palatini $f (R )$ formalisms unifying local constraints at the Solar System level and the late-time cosmic acceleration, were considered in \cite{B28}, by adopting the scalar-tensor representation of the theory, in which the scalar-tensor definition of the potential can be represented as the solution of a Clairaut differential equation.
To study the behavior of the metric functions and of the scalar field a numerical approach was adopted.
The static and spherically symmetric solutions in a gravity theory that extends the standard Hilbert-Einstein action with a Lagrangian constructed from a three-form field $A_{\alpha \beta \gamma}$, which is related to the field strength and a potential term, were investigated in \cite{B29}. The field equations were obtained explicitly for a static and spherically symmetric geometry in vacuum.
Several models, corresponding to different functional forms of the three-field potential, including the Higgs and exponential types, were considered.

 The effects of the quantum fluctuations on the spherically symmetric static black hole solutions were investigated in \cite{B30}, by considering a set of modified field equations,  and by assuming that the quantum correction tensor is given by the coupling of a scalar field to the metric tensor.  The formation of a black hole was detected from the presence of a singularity in the metric tensor components.
   Several models, corresponding to different functional forms of the scalar field potential, were considered. The thermodynamic properties of the black hole solutions were also investigated. Several classes of exact and perturbative spherically symmetric solutions in $f(T,B)$ gravity theories were considered in \cite{B33}. General methods and strategies that allow to find spherically symmetric static vacuum solutions in modified teleparallel theories of gravity, like, for example, generalized Bianchi identities,  were also presented.  Charged black hole and wormhole solutions in the Einstein-Maxwell system in the presence of a massless, real scalar field were classified, and studied in \cite{B34}.  Using a conformal transformation, the static, spherically symmetric possible structures in the minimally coupled system were analyzed. Besides wormholes and naked singularities, only a restricted class of black holes does exist, exhibiting a horizon with an infinite surface, and a timelike central singularity.

 The birth of general relativity did have a deep effect not only on theoretical and observational physics, but also on mathematics. Soon after Einstein proposed his gravitational field equations, Weyl \cite{Weyl1,Weyl2} proposed a geometric extension of the Riemannian geometry (on which Einstein's theory is based), in which the covariant divergence of the metric tensor does not vanish identically, but satisfies a relation of the form $\nabla _{\lambda}g_{\alpha \beta}=Q_{\lambda \alpha \beta}$, where $Q_{\lambda  \alpha \beta}$ is the nonmetricity tensor. In the initial formulation of its geometry, Weyl assumed that $Q_{\lambda \alpha \beta}=\omega _{\lambda}g_{\alpha \beta}$, where $\omega _{\lambda}$ was assumed to be the electromagnetic potential. By using this geometric interpretation, Weyl tried to construct a unified theory of gravitation and of the electromagnetic field, which was severely criticized by Einstein. Due to this criticism, later this unified theory was completely abandoned. However, Weyl's geometry has many attractive features, and it did open some new perspectives in physics, mainly due to the concept of conformal invariance, on which Weyl geometry is based. For a review of Weyl geometry and its applications see \cite{Weyl3}.

 The present day physical applications of Weyl geometry are obtained by adopting the essential assumption that {\it all physical laws must satisfy the principle of conformal invariance}. A local conformal transformation is defined as $d\tilde{s}^2=\Sigma ^n(x)g_{\alpha \beta}dx^{\alpha}dx^{\beta}=\tilde{g}_{\alpha \beta}dx^{\alpha}dx^{\beta}$, where $\Sigma (x)$ is the conformal factor, and $n$ is the Weyl charge. A generalization of Weyl's gravity was proposed by Dirac \cite{Di1,Di2}, and this approach was further investigated in \cite{Di3,Di4}. In Weyl-Dirac type theories ordinary baryonic matter is generated at the very beginning of the cosmological evolution as a direct consequence of the presence of the Dirac gauge function. On the other hand, in the late Universe, Dirac’s gauge function creates the dark energy that triggers the transition to a de Sitter type accelerating phase. In \cite{Ste3} it was suggested that the conformal invariance of the string theory has its origin in the worldvolume diffeomorphism invariance of the membrane theory, a result obtained by postulating an ansatz for the dimensional reduction that leaves the conformal factor as an “integration constant”. For the role of conformal symmetry in field theory and quantum gravity see \cite{C1} and \cite{C2}, respectively.

 Conformally invariant gravitational theories can be constructed by using actions constructed with the help of the Weyl tensor $C_{\alpha \beta \gamma \delta}$, and which can be obtained from the action $S=-(1/4)\int{C_{\alpha \beta \gamma \delta}C^{\alpha \beta \gamma \delta}\sqrt{-g}d^4x}$ \cite{M1,M2,M3,M4, M5, M6}. Gravitational theories constructed with the help of such actions are called {\it conformally invariant, or Weyl gravity type theories}. Another interesting application of Weyl geometry is related to the $f(Q)$ type gravity theories, in which it is assumed that {\it the basic quantity describing the gravitational field is the nonmetricity} \cite{Q1,Q2,Q3,Q4,Q5, Q6}. $f(Q)$ or $f(Q,T)$ type gravity theories may represent an attractive alternative to standard general relativity, and to the standard $\Lambda$CDM cosmological model.

 The idea of conformal invariance also plays a central role in the Conformal Cyclic Cosmology (CCC) theoretical model \cite{P1}, in which the Universe exists as o collection of eons, which are geometric structures corresponding to time oriented spacetimes. The conformal compactification of eons have spacelike null infinities. Different aspects of the CCC model were investigated in \cite{P2,P3,P4}. In \cite{H1} it was proposed that the conformal symmetry is an exact symmetry of nature that is spontaneously broken,  and that it could be as important as the Lorentz invariance of physical laws. The breaking of the conformal invariance could provide a mechanism that would allow the physical understanding of the small-scale structure of the gravitational interaction, and also of the Planck scale physics. A gravitational theory, assuming that local conformal symmetry is an exact, but spontaneously broken symmetry, was introduced in \cite{H2}.

  A conformally invariant Weyl type gravity, developed by fully using the geometrical structures of Weyl geometry, was investigated, in both metric and Palatini formulations, in \cite{Gh1, Gh2, Gh3,Gh4,Gh5,Gh6,Gh7, Gh8, Gh9, Gh10}. The physical implications of the theory for the understanding of the elementary particle physics, as well as its cosmological aspects related to the very early evolution of the Universe were studied in detail. The Weyl geometry inspired quadratic action has a spontaneous symmetry breaking through the Stueckelberg mechanism, which leads to the important result that {\it the Weyl gauge field becomes massive}. Hence, in this approach, one reobtains the standard Hilbert-Einstein action of general relativity in the presence of a positive cosmological constant. Moreover, a Proca type term  for the massive Weyl gauge field also appears in the action \cite{Gh1}. A Weyl-geometric invariant Lagrangian without ghosts, given by
\bea
L&=& \sqrt{-g}\,\Big\{ - \frac{\xi_j}{2} \, \Big[\frac{1}{6}\, \, \phi^2_j\,R+ g^{\mu\nu} \,
\partial_\mu\phi_j\, \partial_\nu\phi_j\Big]\nonumber\\
&&+(1+\xi_j)\, \frac{1}{2} g^{\mu\nu} \tilde D_\mu\phi_j\,\tilde D_\nu\phi_j
-V(\phi_j)\Big\}.
\eea
was investigated in \cite{Gh2}, where a potential term $V(\phi_j)$  for the scalars $\phi_j$ was also considered, with $V$ given by a homogeneous function of the form
$V(\phi_j)=\phi_k^4\, V(\phi_j/\phi_k), k={\rm fixed}$.

It turns out that in the Weyl geometric approach a successful description of the early Universe inflation can be achieved if one of the scalar fields is considered as the inflaton field \cite{Gh1,Gh2, Gh3}. Inflation in the Weyl-geometry inspired gravity in the presence of a scalar field gives similar results as in the Starobinsky model \cite{Star},  which is recovered in the limit of the vanishing of the non-minimal coupling term \cite{Gh3}. Hence, Weyl geometry could play a fundamental role in the early Universe, once one assumes that the effective theory at short distances is conformally invariant. Moreover, in  \cite{Gh4} it  was shown that after a particular Weyl gauge transformation (or gauge fixing),  Weyl conformal geometry acquires a geometric Stueckelberg mechanism, which is naturally built in the theory. This mechanism is broken spontaneously, and leads to the appearance of the Riemannian geometry. On the other hand,  the Stueckelberg mechanism conserves the total number of the degrees of freedom, only rearranging them.

In the Palatini formalism, quadratic Weyl gravity with Lagrangian $R^2+R_{\mu\nu}^2$,  was studied in \cite{Gh5}. In this approach, the connection and the metric are considered as independent variables, with the action having a gauged scale symmetry. For non-minimally coupled Higgs-like fields, the theory can describe well inflation. In \cite{Gh6} a comparative study of inflation in the original Weyl quadratic gravity, and in a theory obtained by replacing the Weyl connection by its Palatini counterpart, was considered.
After the Weyl vector becomes massive via the Stueckelberg mechanism, the  Planck scale, the metricity condition, and the Einstein-Proca action of the Weyl field arise in the broken phase. For large Higgs fields, inflation is also possible. In \cite{Gh8} the cosmological dynamics of the Weyl geometry inspired quadratic gravitational model was investigated in detail.

The coupling of matter to geometry in Weyl geometry inspired conformal quadratic gravity, by assuming a coupling term of the form $L_m\tilde{R}^ 2$, where $L_m$ is the ordinary matter Lagrangian, and $\tilde{R}$ is the Weyl scalar, was investigated in \cite{Gh10}. The coupling explicitly satisfies the conformal invariance of the theory. By expressing $\tilde{R}^2$ with the help of an auxiliary scalar field and of the Weyl scalar, the gravitational action can be linearized, leading in the Riemannian space to a conformally invariant $f\left(R,L_m\right)$ type theory, with the matter Lagrangian nonminimally coupled to the Ricci scalar. The cosmological implications of the theory were also considered for the case of a flat, homogeneous and isotropic geometry. The model can give a good description of the observational data for the Hubble function up to a redshift of the order of $z \approx 3$.

It is the goal of the present work to investigate black hole solutions in the Weyl type geometric gravity theories, by following the approach introduced and developed in \cite{Gh1, Gh2, Gh3, Gh4, Gh5,Gh6,Gh7,Gh8,Gh9,Gh10}. We assume, {\it as a starting point of our investigations, that the background geometry of the spacetime is of Weyl type}. Moreover, we also {\it require the conformal invariance of any gravitational theory}. To implement these  requirements, we adopt {\it the simplest possible model of a conformally invariant Weyl geometric gravity model}, by assuming that the Lagrangian density is constructed from the square of the Weyl scalar, and of the electromagnetic type tensor $F_{\mu \nu}$ only. This Lagrangian density can be linearized with the help of an auxiliary scalar field, and finally it can be written as an Einstein-Proca type Lagrangian in the ordinary Riemannian space, in the presence of a nonminimal coupling between the scalar field and the Ricci scalar, and the Weyl vector, respectively.

As the next step in our study, after obtaining, from the considered variational principle, the gravitational field equations, we consider a spherically symmetric static metric, and write down  the corresponding vacuum field equations. Due to their mathematical complexity, and of their strong nonlinear nature,  the differential equations describing the vacuum solutions of the theory  can be generally investigated only numerically. We consider three classes of models, corresponding to the three possible choices of the Weyl vector field, assumed to have only one timelike component, only one radial spacelike component, and both components. The field equations are rewritten in a dimensionless form, and, after introducing a new independent radial coordinate, defined as the inverse of $r$, we investigate their solutions numerically. Thus, three distinct classes of black hole models are obtained. We detect the presence of an event horizon from the appearance of the singularities in the metric tensor components. {\it An exact solution of the gravitational field equations, obtained in the presence of the radial component of the Weyl vector, is also obtained}.  The thermodynamic properties of the Weyl geometric type black holes,  including their horizon temperature, specific heat, entropy, as well as the evaporation time due to Hawking luminosity, are also considered in detail by using numerical methods for all considered cases.

The present paper is organized as follows. We review the foundations of the Weyl geometry, we introduce the gravitational action, and we obtain the general field equations of the Weyl geometric gravity in Section~\ref{sect1}. The spherically symmetric static field equations in the Weyl geometric model of gravity are obtained in Section~\ref{sect2} for three distinct choices of the functional form of the Weyl vector $\omega ^{\mu}$. Numerical black hole solutions in Weyl geometric gravity, corresponding to the different choices of the Weyl vector,  are considered in Section~\ref{sect3}. An exact solution of the static spherically symmetric field equations in the presence of a Weyl vector having only a radial spacelike component is also obtained. The thermodynamic properties of the obtained black hole solutions are considered in Section~\ref{sect4}.  We discuss and conclude our results in Section~\ref{sect5}. In the present paper we use a system of units with $c = 1$.

\section{Gravitational field equations and equations of motion in Weyl conformal gravity}\label{sect1}

We begin the present Section with a brief discussion of the basic elements of Weyl geometry. Then we will proceed to introduce the simplest Weyl-geometry inspired conformally invariant gravitational action. After linearizing the quadratic Weyl action,  we obtain the gravitational field equations describing the dynamical evolution of the metric, of the auxiliary scalar field, and of the Weyl vector, respectively, in their general form.

\subsection{A quick introduction to Weyl geometry}

A basic property of a  manifold is the existence of an intrinsic metric that gives the distance between two infinitesimally closed points.  In Riemannian geometry, the distance is preserved under the parallel transport with respect to the metric-compatible connection, called the Levi-Civita connection, or the Christoffel symbols, if we refer to its  components. Shortly after Einstein and Hilbert obtained the correct form of the general relativistic field equations, with Hilbert introducing the variational principle for their derivations, H. Weyl \cite{Weyl1, Weyl2} proposed the existence of a conformal transformation in every point of the spacetime manifold. In the case of the metric, such a conformal transformation takes the form $\tilde{g}_{\mu \nu } =\Sigma ^{n}(x)g_{\mu \nu }$, where $n$ is the Weyl charge. {\it In the following we will consider only the case} $n=1$.

In Weyl geometry, under parallel transport, the length $\ell$ of a vector varies when it is parallelly transported by an infinitesimal displacement between the points $x^\mu$ and $x^\mu + \delta
x^\mu$. The change in the length of a vector is given by
\begin{eqnarray}  \label{a1}
\delta \ell = \ell \omega_\mu \delta x^\mu,
\end{eqnarray}
where $\omega_\mu$ is the Weyl vector field. In the general case one can introduce the nonmetricity $Q_{\lambda \mu \nu}$ of the Weyl geometry, via the covariant derivative of the metric tensor, according to the definition,
\begin{align}\label{nmetr}
& \tilde{\nabla}_\lambda g_{\mu\nu} = - \alpha \omega_\lambda g_{\mu\nu}\equiv Q_{\lambda \mu \nu},
\end{align}
where $\alpha$ is the Weyl gauge coupling constant. For the connection of the Weyl geometry, from the nonmetricity condition (\ref{nmetr}) we obtain the following expression,
\begin{align}
& \tilde{\Gamma}^\lambda_{\mu\nu} = \Gamma^\lambda_{\mu\nu} + \frac{1}{2}
\alpha \Big[ \delta^\lambda_\mu \omega_\nu + \delta^\lambda_\nu \omega_\mu -
g_{\mu\nu} \omega^\lambda \Big],
\end{align}
where $ \Gamma^\lambda_{\mu\nu}$ is the standard Levi-Civita connection associated to the metric $g$. In the following, {\it all geometrical and physical quantities defined in the Weyl geometry will be denoted by a tilde}.

By using the Weyl connection one can easily construct the curvature tensor $\tilde{R}_{\mu \nu \sigma }^{\lambda }$, defined as,
\begin{equation}
\tilde{R}_{\mu \nu \sigma }^{\lambda }=\partial _{\nu }\tilde{\Gamma}_{\mu
\sigma }^{\lambda }-\partial _{\sigma }\tilde{\Gamma}_{\mu \nu }^{\lambda }+%
\tilde{\Gamma}_{\rho \nu }^{\lambda }\tilde{\Gamma}_{\mu \sigma }^{\rho }-%
\tilde{\Gamma}_{\rho \sigma }^{\lambda }\tilde{\Gamma}_{\mu \nu }^{\rho },
\end{equation}%
its first contraction,
\begin{equation}
\tilde{R}_{\mu \nu }=\tilde{R}_{\mu \lambda \nu }^{\lambda },\tilde{R}%
=g^{\mu \sigma }\tilde{R}_{\mu \sigma },
\end{equation}%
and its second contraction (the Weyl scalar), respectively, given by
\begin{eqnarray}
\tilde R = g^{\mu\nu} \Big( \partial_\rho \tilde{\Gamma}^\rho_{\mu\nu} -
\partial_\nu\tilde{\Gamma}^\rho_{\mu\rho} + \tilde{\Gamma}^\rho_{\mu\nu}
\tilde{\Gamma}^\sigma_{\rho\sigma} - \tilde{\Gamma}^\sigma_{\mu\rho}\tilde{%
\Gamma}^\rho_{\nu\sigma} \Big),
\end{eqnarray}
or, equivalently, by
\begin{equation}\label{R}
\tilde{R}=R-3n\alpha \nabla _{\mu }\omega ^{\mu }-\frac{3}{2}\left( n\alpha
\right) ^{2}\omega _{\mu }\omega ^{\mu },
\end{equation}
where $R$ is the Ricci scalar defined in the Riemannian geometry.

Another important geometrical and physical quantity,  the Weyl tensor, is defined according to
\begin{equation}\label{9}
\tilde{C}_{\mu \nu \rho \sigma }^{2}=C_{\mu \nu \rho \sigma }^{2}+\frac{3}{2}%
\left( \alpha n\right) ^{2}\tilde{F}_{\mu \nu }^{2},
\end{equation}
where $C_{\mu \nu \rho \sigma }$ is the Weyl tensor as introduced in the standard Riemannian geometry \cite{LaLi}.
$C_{\mu \nu \rho \sigma }^{2}$ can be obtained in terms of the Riemann and Ricci tensors, and of the Ricci scalar, as
\begin{equation}\label{10}
C_{\mu \nu \rho \sigma }^{2}=R_{\mu \nu \rho \sigma }R^{\mu \nu \rho \sigma
}-2R_{\mu \nu }R^{\mu \nu }+\frac{1}{3}R^{2}.
\end{equation}

If we apply a conformal transformation with a conformal factor $\Sigma$ in one point, the variances of the
metric tensor, of the Weyl field, and of a scalar field $\phi $ are given by,
\begin{eqnarray}  \label{a2}
\hat g_{\mu\nu} = \Sigma g_{\mu\nu}, \hat \omega _\mu = \omega_\mu -
\frac{1}{\alpha} \partial_\mu \ln\Sigma, \hat \phi = \Sigma^{-\frac{1}{2%
}} \phi.
\end{eqnarray}

With the help of the Weyl vector we can construct the Weyl field strength $F_{\mu\nu}$ of $\omega_\mu$, defined as,
\be
\tilde{F}_{\mu\nu} =  \tilde{\nabla}_{[\mu} \omega_{\nu]} =  \nabla_{[\mu}
\omega_{\nu]} =  \partial_{[\mu} \omega_{\nu]}=\partial _{\mu}\omega _\nu-\partial _\nu \omega _\mu.
\ee
\\
\subsection{Action and field equations}

In the following we  consider the simplest conformally invariant Lagrangian density,  that can be constructed in Weyl geometry, and which is given by \cite{Gh3,Gh4,Gh5,Gh6,Gh7}
\bea\label{inA}
L_0=\Big[\, \frac{1}{4!}\,\frac{1}{\xi^2}\,\tilde R^2  - \frac14\, F_{\mu\nu}^{\,2} \Big]\sqrt{-g},
\eea
with perturbative coupling  $\xi < 1$. In order to extract the scalar degree of freedom of the above Lagrangian, and to linearize it,
in $L_0$ we replace $\tilde{R}^2$ by $\tilde{R}^2\rightarrow 2 \phi_0^2\,\tilde R-\phi_0^4$
where $\phi_0$ is an auxiliary  scalar field. The new Lagrangian density is equivalent with the initial one, since
by using the solution $\phi_0^2=\tilde R$ of the equation of motion of $\phi_0$ in the new
$L_0$, we recover Eq.~(\ref{inA}). Hence, we obtain a {\it mathematically  equivalent Weyl geometric Lagrangian},
\bea\label{alt3}
L_0=\sqrt{-g} \Big[\frac{1}{12}\frac{1}{\xi^2}\,\phi_0^2\,\tilde R
-\frac14 \,F_{\mu\nu}^2-\frac{\phi_0^4}{4!\,\xi^2}
\Big].
\eea

This is {\it the simplest possible gravitational Lagrangian density with Weyl gauge symmetry, and conformal invariance},  fully implemented.  $L_0$  has a spontaneous
breaking to an Einstein-Proca Lagrangian of the Weyl gauge field. On the other hand, we would like to point out that the simplest gravitational action with conformal symmetry, realized locally, is the conformal gravity model, based on the $C_{\alpha \beta \gamma \delta}^2$ action \cite{M1,M2,M3,M4,M5,M6}, with the Weyl tensor $C_{\alpha \beta \gamma \delta}^2$  defined in the standard Riemannian geometry, and  given by Eq.~ (\ref{10}). Note that the tensor defined by Eq.~(\ref{10}) is different from the Weyl geometric tensor, as introduced in Eq.~(\ref{9}). In the simple model of conformal Weyl gravity no Weyl gauge field $\omega_\mu$, nor the scalar $\phi$ are present. However, in four dimensions the theory  is still invariant under local Weyl gauge transformations. Generally, a physical system is called conformally invariant if it satisfies the condition that the variation of the action $S\left[g_{\mu \nu},\phi\right]$ with respect to the group of the conformal transformations is zero, $\delta _cS\left[g_{\mu \nu},\phi\right]=\int{d^n x\left(\delta S/\delta \phi\right)\delta _c\phi}=0$ \cite{Kaz}. Another type of transformation is given by the Weyl rescaling, implying the simultaneous pointwise transformations of the metric and of the fields, $\tilde{g}_{\mu \nu}(x)=e^{2\sigma (x)}g_{\mu \nu}$, and $\tilde{\phi}=e^{-\Delta \sigma (x)}\phi(x)$, under which the action transforms as $\delta _\sigma S\left[g_{\mu \nu},\phi\right]=\int{d^n x\sigma \left[2\left(\delta S/\delta g_{\mu \nu}\right)g_{\mu \nu}-\Delta _n\left(\delta S/\delta \phi\right)\phi\right]}$ \cite{Kaz}. For an in depth discussion of the conformally and Weyl invariance see \cite{Kaz}.

By replacing in Eq.~(\ref{alt3})  $\tilde{R}$ as given by Eq.~(\ref{R}), and by performing a gauge transformation that allows the redefinition of the variables, we obtain an action, defined in the Riemannian space,  invariant under conformal
transformation, given by \cite{Gh3,Gh4,Gh5},
\begin{eqnarray}\label{a3}
\mathcal{S }&=& \int d^4x \sqrt{-g} \Bigg[ \frac{1}{12} \frac{\phi^2}{\xi^2}
\Big( R - 3\alpha\nabla_\mu \omega^\mu - \frac{3}{2} \alpha^2 \omega_\mu
\omega^\mu \Big) \nonumber\\
&&- \frac{1}{4!}\frac{\phi^4}{\xi^2} - \frac{1}{4} F_{\mu\nu}
F^{\mu\nu} \Bigg],
\end{eqnarray}

The variation of the action Eq.~(\ref{a3}) with respect to the metric tensor gives the field
equation,
\begin{widetext}
\begin{eqnarray}\label{b2a}
&&\frac{\phi ^{2}}{\xi ^{2}}\Big(R_{\mu \nu }-\frac{1}{2}Rg_{\mu \nu }%
\Big)-\frac{3\alpha }{2\xi ^{2}}\Big(\omega ^{\rho }\nabla _{\rho }\phi
^{2}g_{\mu \nu }-\omega _{\nu }\nabla _{\mu }\phi ^{2}-\omega _{\mu }\nabla
_{\nu }\phi ^{2}\Big)+\frac{3\alpha ^{2}}{4\xi ^{2}}\phi ^{2}\Big(\omega
_{\rho }\omega ^{\rho }g_{\mu \nu }-2\omega _{\mu }\omega _{\nu }\Big)
\notag  \\
&&-6F_{\rho \mu }F_{\sigma \nu }g^{\rho \sigma }+\frac{3}{2}F_{\rho \sigma
}^{2}g_{\mu \nu }+\frac{1}{4\xi ^{2}}\phi ^{4}g_{\mu \nu }+\frac{1}{\xi ^{2}}%
\Big(g_{\mu \nu }\Box -\nabla _{\mu }\nabla _{\nu }\Big)\phi ^{2}=0.
\end{eqnarray}
\end{widetext}

The trace of Eq.~(\ref{b2a}) gives,
\begin{equation}\label{b3n}
\Phi R+3\alpha \omega ^{\rho }\nabla _{\rho }\Phi -\Phi ^{2}-\frac{3}{2}%
\alpha ^{2}\Phi \omega _{\rho }\omega ^{\rho }-3\Box \Phi =0,
\end{equation}%
where we have denoted $\Phi \equiv \phi ^{2}$. Variation of the action given by Eq.~(\ref{a3})
with respect to the scalar field $\phi $ gives the equation of motion of $%
\phi $,
\begin{equation}\label{b4}
R-3\alpha \nabla _{\rho }\omega ^{\rho }-\frac{3}{2}\alpha ^{2}\omega _{\rho
}\omega ^{\rho }-\Phi =0.
\end{equation}%

From  Eqs.~(\ref{b3n}) and (\ref{b4}) it immediately follows that
\begin{equation}\label{b5}
\Box \Phi-\alpha \nabla _{\rho }(\Phi\omega ^{\rho })=0.
\end{equation}%

The variation of Eq.(\ref{a3}) with respect to $\omega _{\mu }$
gives the equation of motion of the Weyl vector as,
\begin{equation}\label{Fmunu}
4\xi ^{2}\nabla _{\nu }F^{\mu \nu }+\alpha ^{2}\Phi\omega ^{\mu }-\alpha \nabla
^{\mu }\Phi=0.
\end{equation}%

Applying to both sides of the above equation the operator  $\nabla _{\mu }$ we obtain equation \eqref{b5}, a result that can be seen as a consistency check of the theory.

\section{The spherically symmetric vacuum field equations}\label{sect2}

 In the following we adopt a static and spherically symmetric geometry, in which all quantities depend only on the radial coordinate $r$, in a system of coordinates defined according to $\left(t,r,\theta, \varphi\right)$. For mathematical convenience we will use geometric units in all equations. In the following we denote by a prime the derivative with respect to the radial coordinate $r$.
Hence, in the adopted geometry the spacetime interval becomes,
\begin{equation}
ds^{2}=e^{\nu (r)}dt^{2}-e^{\lambda (r)}dr^{2}-r^{2}d\Omega ^{2},
\end{equation}%
where we have denoted $d\Omega ^{2}=d\theta ^{2}+\sin ^{2}\theta d\varphi ^{2}$. Thus,  the
metric tensor components are given by,
\begin{equation}
g_{\mu \nu }=\mathrm{diag}(e^{\nu (r)},-e^{\lambda (r)},-r^{2},-r^{2}\sin
^{2}\theta ).
\end{equation}

If we consider a spherically symmetric configuration, the third and the
fourth components of the Weyl vector  vanishes identically. Hence, the Weyl
vector can be represented in a general form as $\omega _{\mu }=\left( \omega _{0},\omega
_{1},0,0\right) $.

 In the following we will write down the field equations resulting from the three
possibilities allowed by the various choices of the Weyl vector, and we also present some of their consequences.

\subsection{The case $\omega _{\mu }=\left( 0,\omega %
_{1},0,0\right) $}

In the first case we are considering, the temporal component, $\omega _{0}$,
of the Weyl vector can be taken as zero, and thus $\omega _{\mu}$ is given by $%
\omega _{\mu }=(0,\omega _{1},0,0)$. For this choice of $\omega _{\mu}$  we have $F_{\mu \nu}\equiv 0$. Hence, Eq.~(\ref{Fmunu}) immediately gives
\be\label{15}
\Phi '=\alpha \Phi \omega _1.
\ee

By taking into account that
\be
\Box \Phi=%
\frac{1}{\sqrt{-g}}\frac{\partial }{\partial x^{\alpha }}\left( \sqrt{-g}%
g^{\alpha \beta }\frac{\partial \Phi}{\partial x^{\beta }}\right) ,
\ee
and
\be
\nabla _{\alpha }\omega ^{\alpha }=\frac{1}{\sqrt{-g}}\frac{\partial }{\partial
x^{\alpha }}\left( \sqrt{-g}\omega ^{\alpha }\right),
\ee
respectively, Eq.~(\ref{b5}) becomes
\bea\label{16}
&&\frac{1}{\sqrt{-g}}\frac{d}{dr}\left( \sqrt{-g}g^{11}\frac{d\Phi}{dr}\right)
-\alpha \omega ^{1}\frac{d\Phi}{dr}
-\frac{\alpha \Phi}{\sqrt{-g}}\frac{d}{dr}%
\left( \sqrt{-g}\omega ^{1}\right) \nonumber\\
&&=0.
\eea

In order to make Eqs.~(\ref{15}) and (\ref{16}) consistent, {\it we need to impose the gauge condition on the Weyl vector}, which
generally can be formulated as
\begin{equation}
\nabla _{\alpha }\omega ^{\alpha }=\frac{1}{\sqrt{-g}}\frac{\partial }{%
\partial x^{\alpha }}\left( \sqrt{-g}\omega ^{\alpha }\right) =0,
\end{equation}%
giving for the present choice of $\omega _{\mu}$, the relation
\begin{equation}
\frac{1}{\sqrt{-g}}\frac{d}{dr}\left( \sqrt{-g}\omega ^{1}%
\right) =0,
\end{equation}
or,
\bea\label{omeg1}
\omega ^{1}=\frac{C_{1}}{\sqrt{-g_{rad}}}=C_{1}\frac{e^{-{(\nu +\lambda )}/2}%
}{r^{2}},
\eea
and
\bea
\omega _{1}=g_{11}\omega ^{1}=-C_{1}\frac{e^{-\left( \nu -\lambda
\right) /2}}{r^{2}},
\eea%
respectively, where $C_{1}$ is an arbitrary integration constant, and we have denoted $%
\sqrt{-g_{rad}}=e^{\left( \nu +\lambda \right) /2}r^{2}$. By taking into
account the gauge condition, and the explicit form of $\omega ^{1}$, Eq.~(%
\ref{16}) becomes
\be\label{eqev}
\frac{1}{\sqrt{-g}}\frac{d}{dr}\left( \sqrt{-g}g^{11}\frac{d\Phi}{dr}\right)
-\alpha \omega ^{1}\frac{d\Phi}{dr}=0,
\ee
or, equivalently,
\begin{equation}
\frac{1}{\sqrt{-g_{rad}}}\frac{d}{dr}\left( \sqrt{-g_{rad}}g^{11}\frac{d\Phi}{dr%
}\right) -\alpha \frac{C_{1}}{\sqrt{-g_{rad}}}\frac{d\Phi}{dr}=0,
\end{equation}%
giving
\begin{equation}
\sqrt{-g_{rad}}g^{11}\frac{d\Phi}{dr}-\alpha C_{1}\Phi=C,
\end{equation}%
where $C$ is an arbitrary integration constant, which for consistency with Eq.~(\ref{15}) must be taken as zero. Hence we reobtain again Eq.~(\ref{15}),
$\Phi ^{\prime }=\alpha \omega _{1}\Phi$
respectively.

By taking into account the explicit expression of $\omega ^{1}
$ as given by Eq.~(\ref{omeg1}), we obtain for $\Phi$ the differential equation
\begin{equation}\label{34}
\frac{d\Phi}{dr}=-\alpha C_{1}\frac{e^{-\left( \nu -\lambda \right) /2}}{r^{2}}\Phi.
\end{equation}

Alternatively, from Eqs.~(\ref{eqev}) and (\ref{15}) we obtain a non-trivial dynamical equation for $\Phi(r)$,
\begin{eqnarray}\label{c19}
\frac{d}{dr}\left(\sqrt{-g}g^{11}\frac{d\Phi}{dr} \right) - \sqrt{-g} g^{11}
\frac{1}{\Phi} \left(\frac{d\Phi}{dr}\right)^2 = 0.
\end{eqnarray}

Next, the gravitational field equations give the following relations
\bea\label{c15}
&& -1+e^{\lambda }-\frac{1}{4}e^{\lambda }r^{2}\Phi-\frac{2r\Phi^{\prime }}{\Phi}+%
\frac{3r^{2}}{4}\frac{\Phi^{\prime 2}}{\Phi^{2}}+r\lambda ^{\prime }\nonumber\\
&&+\frac{%
r^{2}\lambda ^{\prime }}{2}\frac{\Phi^{\prime }}{\Phi }-\frac{r^{2}\Phi^{\prime \prime
}}{\Phi}=0,
\eea
\bea \label{c16}
&& 1-e^{\lambda }+\frac{1}{4}e^{\lambda }r^{2}\Phi+\frac{2r\Phi^{\prime }}{\Phi}+\frac{%
3r^{2}}{4}\frac{\Phi^{\prime 2}}{\Phi^{2}}+r\nu ^{\prime }\left(1
+\frac{r}{2}\frac{\Phi^{\prime }}{\Phi}\right)\nonumber\\
&&=0,
\end{eqnarray}
and
\bea\label{e22}
\hspace{-0.5cm}&&2(\nu ^{\prime
}-\lambda ^{\prime })+(4-2r\lambda ^{\prime }+2r\nu ^{\prime })\frac{\Phi ^{\prime }}{\Phi }\nonumber\\
\hspace{-0.5cm}&&+r\left(e^{\lambda }\Phi +4\frac{\Phi ^{\prime \prime }}{\Phi }-3\frac{\Phi
^{\prime 2}}{\Phi ^{2}}-\lambda ^{\prime }\nu ^{\prime }+\nu ^{\prime
2}+2\nu ^{\prime \prime }\right)=0,
\eea
respectively.

It can easily be proved that Eq.~\eqref{e22} is a linear combination of the other two field equations. By adding Eqs.~(\ref{c15}) and (\ref{c16}) we obtain,
\be\label{sum}
\frac{\Phi ''}{\Phi}-\frac{3}{2}\frac{\Phi ^{\prime 2}}{\Phi ^2}-\frac{\nu '+\lambda '}{2}\frac{\Phi '}{\Phi}-\frac{\nu '+\lambda '}{r}=0.
\ee

\subsection{The case $\omega _{\mu }=\left(\omega_0,0,0,0\right) $}

We consider now the case in which it is possible to neglect all the effects of the spatial components $\omega_1$ in the Weyl vector. Thus, only the $\omega_0$ component of the Weyl vector is nonzero. Thus, Eq.~(\ref{Fmunu}) immediately gives,
\begin{align}\label{c23}
	&-\frac{\alpha^2}{4\xi^2} e^\lambda \Phi \omega_0 + \frac{-4+r \lambda^\prime + r \nu^\prime}{2r}\omega_0^\prime - \omega^{\prime\prime}_0 = 0,
\end{align}
and
\begin{eqnarray}\label{c24}
	\Phi^\prime = 0,
\end{eqnarray}
respectively. Hence, the gauge condition imposed on the Weyl vector field in the former Section is now fulfilled trivially. In this case the gravitational field equations give the following relations,
\begin{eqnarray}\label{feqomega0}
	&& -1 + e^\lambda + \frac{1}{4} e^\lambda r^2 \Phi - \frac{3}{4} \alpha^2 e^{\lambda-\nu} r^2 \omega_0^2 +
	 r \lambda^\prime + \frac{3\xi^2 e^{-\nu} r^2 \omega_0^{\prime 2}}{\Phi}\nonumber\\
&& = 0,\label{c25}\\
	&& 1 - e^\lambda - \frac{1}{4} e^\lambda r^2 \Phi - \frac{3}{4} \alpha^2 e^{\lambda-\nu} r^2 \omega_0^2 +
	 r \nu^\prime - \frac{3\xi^2 e^{-\nu} r^2 \omega_0^{\prime 2}}{\Phi}\nonumber\\
&&=0.\label{c26}
\end{eqnarray}
It is also easy to prove that Eq.~(\ref{b4}) gives a trivial constraint for $\Phi$, if Eq.~(\ref{c24}) is used to express $\Phi$.
	
\subsection{The case $\omega _{\mu }=\left(\omega_0,\omega_1,0,0\right) $}

In this case, Eq.~\eqref{Fmunu} gives
\begin{align}\label{sec31a}
\Phi '=\alpha \Phi \omega _1,
\end{align}
and
\begin{align}\label{sec31u}
&- \frac{\alpha^2}{4\xi^2} e^\lambda \Phi \omega_0 + \frac{-4+r \lambda^\prime + r \nu^\prime}{2r}\omega_0^\prime - \omega^{\prime\prime}_0 = 0,
\end{align}
respectively.

By substituting $\omega_1$ from Eq.~\eqref{sec31a}, one can simplify the field equations Eq.~\eqref{b2a} as
\begin{align}\label{sec315}
& -1+e^{\lambda }+\frac{1}{4}e^{\lambda }r^{2}\Phi-\frac{2r\Phi^{\prime }}{\Phi}+%
\frac{3r^{2}}{4}\frac{\Phi^{\prime 2}}{\Phi^{2}}+r\lambda ^{\prime }\nonumber\\
&+\frac{%
	r^{2}\lambda ^{\prime }}{2}\frac{\Phi^{\prime }}{\Phi }-\frac{r^{2}\Phi^{\prime \prime
}}{\Phi}-\frac34\alpha^2 e^{\lambda-\nu}r^2\omega_0^2+\frac{3\xi^2}{\Phi}e^{-\nu}r^2\omega^{\prime 2}_0=0,
\end{align}
and
\begin{align} \label{sec316}
& 1-e^{\lambda }-\frac{1}{4}e^{\lambda }r^{2}\Phi+\frac{2r\Phi^{\prime }}{\Phi}+\frac{%
	3r^{2}}{4}\frac{\Phi^{\prime 2}}{\Phi^{2}}+r\nu ^{\prime }\nonumber\\
&+\frac{r^{2}\nu ^{\prime
}}{2}\frac{\Phi^{\prime }}{\Phi}-\frac34\alpha^2 e^{\lambda-\nu}r^2\omega_0^2-\frac{3\xi^2}{\Phi}e^{-\nu}r^2\omega^{\prime 2}_0=0,
\end{align}
respectively.
By adding  Eqs.~\eqref{sec315} and \eqref{sec316} one obtains
\begin{align}\label{scal1}
\frac{\Phi ''}{\Phi}-\frac{3}{2}\frac{\Phi ^{\prime 2}}{\Phi ^2}-\frac{\nu '+\lambda '}{2}\frac{\Phi '}{\Phi}-\frac{\nu '+\lambda '}{r}+\frac32\alpha^2 e^{\lambda-\nu}\omega_0^2=0.
\end{align}

\subsection{Asymptotic and near horizon behavior}

We consider now the asymptotic and near horizon behavior of the field equations of the geometric Weyl gravity. In particular, the asymptotic values of the metric tensor components, and of the scalar and Weyl vector fields allow us to fix the conditions at infinity for the numerical integration and analysis of the field equations. We will consider independently the three cases corresponding to the three different choices of the Weyl vector considered in the previous Subsection. We assume that at infinity the metric can either be asymptotically flat, satisfying the conditions
\be
\lim_{r\rightarrow \infty}\nu(r)=0, \lim_{r\rightarrow \infty}\lambda(r)=0,
\ee
and
\be
\lim_{r\rightarrow \infty}\nu '(r)=0, \lim_{r\rightarrow \infty}\lambda '(r)=0,
\ee
respectively, or of the de Sitter type, so that
\be
\lim_{r\rightarrow \infty}e^{\nu (r)}=\lim_{r\rightarrow \infty}e^{-\lambda}=1-\frac{r^2}{\mu ^2},
\ee
and
\be
\lim_{r\rightarrow \infty}\nu '(r)=-\lim_{r\rightarrow \infty}\lambda '(r)=-\frac{2r/\mu ^2}{1-r^2/\mu ^2},
\ee
respectively, where $\mu $ is a constant. In both cases at infinity the metric tensor components satisfy the conditions $\lim_{r\rightarrow \infty}\left( \nu (r)+\lambda (r)\right) =0$, and $\lim_{r\rightarrow \infty}\left( \nu '(r)+\lambda '(r)\right) =0$, respectively.  For a correct and complete definition of the asymptotic limits of the components of the metric tensor one should also consider some bounds of these functions, like, for example, that for large $r$ they fall faster than $1/r$.

In the general case the static spherically symmetric field equations of the Weyl geometric gravity are too complicated to be solved analytically. Therefore, to obtain their solutions, we must resort to numerical methods. In our investigations {\it we assume that the field equations have a black hole solution}, whose presence is indicated by the presence of a horizon at a radius $r = r_0 > 0$, where the metric functions $e^{\nu}$ and $e^{\lambda}$  become singular. Then, near the horizon, the metric functions can be approximated by their Taylor expansions \cite{Ste1,Ste2},
\be
e^{\nu}=A_1\left(r-r_0\right)+A_2\left(r-r_0\right)^2+A_3\left(r-r_0\right)^3+....,
\ee
and
\be
e^{-\lambda}=B_1\left(r-r_0\right)+B_2\left(r-r_0\right)^2+B_3\left(r-r_0\right)^3+...,
\ee
respectively, where $A_i$ and $B_i$, $i=1,2,3,...$ are constants that can be determined recursively after substitution into the field equations.

\subsubsection{Models with radial component of the Weyl vector}

\paragraph{Asymptotic limits.} In the presence of a Weyl vector having only a radial component, in the limit of large $r$, for both the asymptotically flat and de Sitter cases Eq.~(\ref{sum}) takes the form
\be
\Phi \Phi ''=\frac{3}{2} \Phi ^{\prime 2},
\ee
with the solution
\be
\Phi (r)= \frac{K_1}{\left(r+K_2\right)^2},
\ee
where $K_1$ and $K_2$ are arbitrary integration constants. Hence, for the scalar and the vector fields we obtain the general asymptotic conditions
\bea
\lim_{r\rightarrow \infty}\Phi (r)&=&0, \lim_{r\rightarrow \infty}\Phi ' (r)=0, \nonumber\\
\lim_{r\rightarrow \infty}\omega _1 (r)&=&-\frac{2}{\alpha}\lim_{r\rightarrow \infty}\frac{1}{r+K_2}=0.
\eea
In the asymptotically flat case Eq.~(\ref{c15}) gives $-16K_2r+\left(K_1-4\right)r^2\approx 0$, which is identically satisfied for $K_2=0$ and $K_1=4$, respectively. For the de Sitter type asymptotic behavior we find $-16K_2r+r^2\left[-12K_2^2+\left(K_1-4\right)\alpha ^2\right]/\alpha ^2\approx 0$, a condition that is again satisfied for $K_2=0$, and $K_1=4$, respectively.

However, in the asymptotic limit, Eq.~(\ref{sum}) also has the solution $\Phi={\rm constant}$, implying that {\it at infinity the scalar field can take arbitrary values}.  By taking into account the relation between the scalar field and the radial component of the Weyl vector, as given by Eq.~(\ref{15}), it follows that $\lim_{r\rightarrow \infty}\omega _1(r)=0$, that is, the Weyl vector vanishes at infinity, independently of the asymptotic values of $\Phi$.

\paragraph{Near horizon behavior.} In order to consider the near horizon behavior of the model we also assume that near $r_0$ the metric tensor components and the scalar field admit Taylor expansions of the form \cite{Ste1,Ste2}
\be
f_i(r)=K_{i1}\left(r-r_0\right)+K_{i2}\left(r-r_0\right)^2+K_{i3}\left(r-r_0\right)^3+...,
\ee
where $f_i=\left\{e^{\nu}, e^{\lambda}, \Phi\right\}$, respectively, while $K_{ij}$, $i,j=1,2,3,...$ are constants.

As a simple example of the near horizon behavior of Weyl geometric black holes in the presence of the radial component of the Weyl vector, we assume that the metric tensor components can be represented near the singularity as
\be
e^{\nu (r)}=\left(r-r_0\right)^2\left[K_{11}+K_{12}\left(r-r_0\right)\right]^2,
\ee
and
\be
e^{\lambda (r)}=\left(r-r_0\right)^2\left[K_{21}+K_{22}\left(r-r_0\right)\right]^2,
\ee
respectively, where $K_{ij}$, $i,j=1,2$ are constants. For the near horizon behavior of the scalar field we adopt the functional form
\be
\Phi(r)=\Phi_0e^{-A/r}r^B\left[C+D\left(r-r_0\right)\right]^{-B},
\ee
where $\Phi_0,A,B,C,D$ are constants. By substituting the above expressions of metric tensor components and of the scalar field into Eq.~(\ref{34}), it follows that the equation is identically satisfied if the unknown coefficients satisfy the conditions
\be
C=K_{11},D=K_{12},
\ee
\be
A\left(C-r_0D\right)+\alpha C_1\left(K_{21}-r_0K_{22}\right)=0,
\ee
and
\be
AD+B\left(C-r_0D\right)+C_1K_{22}\alpha=0,
\ee
respectively, giving the relations
\be
K_{21}=\frac{B r_0 \left(K_{12} r_0-K_{11}\right)-A K_{11}}{\alpha C_1},
\ee
\be
K_{22}=-\frac{A K_{12}+B\left( K_{11}- K_{12} r_0\right)}{\alpha  C_1}.
\ee
By substituting the above representations of the metric tensor and of the scalar field into the field equations Eq.~(\ref{c15}) and (\ref{c16}), one obtains a system of ordinary nonlinear algebraic equations for the determination of the parameters of the solution. The resulting equations can be generally solved only approximately, due to their extremely complicated character. The series solutions can be extended to any order of approximation. Hence, at least in principle, one can obtain a complete power series solution of the Weyl geometric field equations near the black hole horizon.

An alternative approximate solution can be obtained by neglecting the nonlinear term $\Phi^{\prime 2}/\Phi ^2$ in the scalar field equation (\ref{scal1}), and in the gravitational field equations (\ref{c15}) and (\ref{c16}). In this case we obtain a system of algebraic nonlinear equations that can be solved to recursively obtain the values of the coefficients in the Taylor series expansions of the physical and geometrical quantities. Thus, a series solution of the static spherically symmetric field equations of Weyl geometric gravity in the presence of the radial component of the Weyl vector can be constructed in both its linearized and nonlinear versions.

\subsubsection{Models with temporal component of the Weyl vector}

\paragraph{Asymptotic limits.} In the case in which the Weyl vector has only a nonzero temporal component $\omega _0$, the scalar field is a constant, $\Phi=\Phi_0={\rm constant}$. In the asymptotic limit of the flat spacetimes, with $\lambda =0$, and $\nu'+\lambda'=0$,  Eq.~(\ref{c23}) becomes
\be
\omega _0^{\prime \prime}+\frac{2}{r}\omega _0^{\prime}+\frac{\alpha ^2\Phi_0}{4\xi ^2}\omega _0=0,
\ee
with the general solution given by
\be
\omega _0=\frac{c_1\cos( \sqrt{A} r)}{r}+\frac{c_2 \sin(\sqrt{A} r)}{r},
\ee
where we have denoted $A=\alpha ^2\Phi_0/4\xi^2$, and $c_1$ and $c_2$ are constants of integration.
Thus, we obtain
\be\label{condom0}
\lim_{r\rightarrow \infty}\omega_0=0, \lim_{r\rightarrow \infty}\omega_0^{\prime}=0.
\ee

By assuming that asymptotically the metric is de Sitter, with $e^{\lambda}=\left(1-r^2/\mu ^2\right)^{-1}$, Eq.~(\ref{c23}) takes the form
\be
\omega _0^{\prime \prime}+\frac{2}{r}\omega _0^{\prime}+\frac{A}{1-r^2/\mu ^2}\;\omega _0=0,
\ee
with the general solution given by
\begin{eqnarray}
\omega _{0}(r)& =&c_{2}\,_{2}F_{1}\left( \delta _{-},\delta _{+};\frac{3}{2};%
\frac{r^{2}}{\mu ^{2}}\right)  \nonumber\\
&&-\frac{ic_{1}\mu \,_{2}F_{1}\left( -\delta _{+},-\delta _{-};\frac{1}{2};%
\frac{r^{2}}{\mu ^{2}}\right) }{r},
\end{eqnarray}%
where $c_{1}$ and $c_{2}$ are arbitrary integration constants, $%
_{2}F_{1}\left( a,b;c;z\right) $ is the hypergeometric function, and we have
denoted

\begin{equation}
\delta _{-}=\frac{1}{4}\left( 1-\sqrt{1+4A\mu ^{2}}\right) ,\delta _{+}=%
\frac{1}{4}\left( 1+\sqrt{1+4A\mu ^{2}}\right) .
\end{equation}

Since $\omega _{0}$ is a real quantity, we need to take $c_{1}=0$ in the
solution, thus obtaining
\begin{equation}\label{om0a}
\omega _{0}(r)=c_{2}\,_{2}F_{1}\left( \delta _{-},\delta _{+};\frac{3}{2};%
\frac{r^{2}}{\mu ^{2}}\right) .
\end{equation}

 Here, the hypergeometric function $_{2}F_{1}(a,b;c;z)$, representing the regular solution of the hypergeometric differential equation,  is defined for $|z|<1$ by a power series of the form $_{2}F_{1}(a,b;c;z)=\sum_{k=0}^{\infty}{\left[(a)_k(b)_k/(c)_k\right]z^k/k!}$, where $(q)_k$ is the (rising) Pochhammer symbol \cite{AS}. For $c$ not a negative integer, the function $_{2}F_{1}(a,b;c;z)$ converges for all of $|z|<1$, and, if ${\rm Re}(c-a-b)>0$ also on the unit circle $|z|=1$.  Different solutions of the hypergeometric differential equations can also be derived for other values of $z$, not restricted to the range $|z|<1$, and these solutions are valid in different regions of the complex plane.

Thus, by taking into account that the radius of convergence of the hypergeometric function in the expression (\ref{om0a}) of $\omega _0(r)$ is $r^{2}<\mu
^{2}$, we find
\begin{equation}
\lim_{r\rightarrow \mu }\omega _{0}(r)=-\frac{2\sqrt{\pi }c_{2}}{A\mu
^{2}\Gamma \left( \delta _{-}\right) \Gamma \left( \delta _{+}\right) },
\end{equation}%
where $\Gamma (z)=\int_{0}^{\infty}{t^{z-1}e^{-t}dt}$ is the Euler gamma function. The value $r=\mu$ of the radial coordinate defines a cosmological horizon for the present model.

For the limit at infinity of the  derivative of $\omega _0$ we obtain
\begin{equation}
\lim_{r\rightarrow \mu }\omega _{0}^{\prime }(r)=-\frac{1}{3}Ac_{2}\mu
\,_{2}F_{1}\left( 1+\delta _{-},1+\delta _{+};\frac{5}{2};1\right) .
\end{equation}

In the case of an asymptotically flat geometry, with the use of the conditions (\ref{condom0}), the field equations (\ref{feqomega0}) as evaluated at infinity give $\Phi=0$. Hence, it follows that if one assumes the presence of a non-zero scalar field, the asymptotic limit of the metric cannot be flat.

\subsubsection{Models with both temporal and radial components of the Weyl vector}

\paragraph{Asymptotic limits.} Finally, we consider the behavior at infinity of the static spherically symmetric Weyl geometric models in the presence of both temporal and radial components of the Weyl field. In the case of the asymptotically flat geometry, the coupled system of equations describing the behavior of the scalar field and of the temporal component of the Weyl vector are given by
\be\label{64}
\omega _0^{\prime \prime}+\frac{2}{r}\omega _0^{\prime}+\frac{\alpha ^2}{4\xi ^2}\Phi \omega _0=0,
\ee
and
\be\label{65}
\frac{\Phi ^{\prime \prime}}{\Phi}-\frac{3}{2}\frac{\Phi^{\prime 2}}{\Phi ^2}+\frac{3}{2}\alpha ^2\omega _0^2=0,
\ee
respectively. By neglecting the nonlinear term $\Phi \omega _0$ in Eq.~(\ref{64}), we find
\be
\omega_0^{\prime}(r)=\frac{c_4}{r^2},\omega _0(r)=c_3-\frac{c_4}{r}.
\ee
Hence,
\be
\lim_{r\rightarrow \infty}\omega _0^{\prime}=0, \lim_{r\rightarrow \infty}\omega _0=c_3.
\ee

Substitution on Eq.~(\ref{65}) gives for $\Phi (r)$ the equation
\be\label{67}
\frac{\Phi ^{\prime \prime}}{\Phi}-\frac{3}{2}\frac{\Phi^{\prime 2}}{\Phi ^2}+\frac{3}{2}\alpha ^2\left(c_3-\frac{c_4}{r}\right)^2=0.
\ee
By neglecting the nonlinear term $\Phi^{\prime 2}/\Phi ^2$, and in the limit of large $r$, Eq.~(\ref{67}) becomes
\be
\Phi ^{\prime \prime}+\frac{3}{2}\alpha ^2c_3^2\Phi=0,
\ee
with the general solution given by
\be
\Phi (r)=C_5\cos \left(\sqrt{\frac{3}{2}}\alpha c_3r+C_6\right),
\ee
where $C_5$ and $C_6$ are arbitrary integration constants. Hence, at least in the approximation considered, the scalar field has an oscillatory behavior at infinity. For the Weyl vector component we obtain
\be
\omega _1(r)\approx -\frac{3}{2}\alpha c_3\tan\left(\sqrt{\frac{3}{2}}\alpha c_3r+C_6\right).
\ee
Thus, the component of the Weyl have an oscillatory behavior at infinity.

In the case of an asymptotic de Sitter geometry, the scalar field equation (\ref{scal1}) can be approximated as
\be
\Phi ^{\prime \prime}+\frac{3}{2}\frac{\alpha ^2c_3^2}{\left(1-r^2/\mu ^2\right)^2}\Phi=0,
\ee
 and it has the general solution
 \bea
 &&\Phi (r)= c_5 (r-\mu )^{\frac{1}{4} \left(\sqrt{4-6 \alpha ^2 c_3^2
   \mu ^2}+2\right)} (r+\mu )^{\frac{1}{4} \left(2-\sqrt{4-6 \alpha ^2 c_3^2 \mu
   ^2}\right)}\nonumber\\
  && -\frac{c_6 (r-\mu )^{\frac{1}{4} \left(2-\sqrt{4-6 \alpha ^2 c_3^2
   \mu ^2}\right)} (r+\mu )^{\frac{1}{4} \left(\sqrt{4-6 \alpha ^2 c_3^2 \mu
   ^2}+2\right)}}{\mu  \sqrt{4-6 \alpha ^2 c_3^2 \mu ^2}},
 \eea
 where $c_5$ and $c_6$ are integration constants. In the limit $r\rightarrow \zeta$, the scalar field tends to zero. Similarly, in the limit of large $r$, the derivative of the scalar field tends to zero. However, we would like to point out that the above results are approximate, but, even so, they indicate that both arbitrary non-zero, as well as very small (zero) numerical values are possible for the scalar and Weyl vector fields at infinity.

\section{Black hole solutions in Weyl geometric conformal gravity}\label{sect3}

In order to simplify the mathematical formalism we introduce now the following representation for $e^{-\lambda}$,
\begin{align}
	& e^{-\lambda } = 1-\frac{2GM(r)}{c^{2}r}=1-\frac{2nGM_{\odot}m(r)}{c^{2}r},
	\end{align}
where
\be
 m(r) = \frac{M(r)}{n M_{\odot}},
\ee
and $M_{\odot}$ denotes the mass of the Sun. $M(r)$ and $m(r)$ are effective masses incorporating both the effects of the scalar field and of the Weyl vector, while $n$ is a positive integer.

We define now a group of dimensionless quantities $(\eta,m,\nu,\psi,\zeta,\Omega_0,\Theta_0)$, given by
\begin{align}\label{c22}
	& r =  \frac{2r_g}{\eta},\;\; \psi = r_g^2\Phi,
	\;\;  \zeta = r_g^3 \frac{d\Phi}{dr},\;\; \Omega_0 = \alpha r_g \omega_0, \nn
	& \hspace{-0.2cm} \Theta_0 = \xi r_g^2 \frac{d\omega_0}{dr},\;\; e^{-\lambda} = 1- m(\eta)\eta,
\end{align}
where we have denoted $r_g=nGM_{\odot}/c^{2}$. Hence, for the derivative with respect to $\eta$ we obtain
\be
\frac{d}{dr} = -\frac{\eta^2}{2r_g} \frac{d}{d\eta},
\ee
and
\begin{align}
	& \frac{d\lambda}{dr} = -\frac{\eta^2}{2r_g} \frac{d\lambda}{d\eta} = - \frac{\eta^2}{2r_g}\frac{\frac{dm(\eta)}{d\eta}\eta + m(\eta)}{1-m(\eta)\eta},
\end{align}
respectively. We also define a new constant $\gamma$ as $\gamma = \alpha/\xi$.

\subsection{Black hole solutions with radial component of the Weyl vector field}

We will consider first black hole solutions in which the Weyl vector has only a radial component $\omega _1$. In this case it is possible to find an exact solution of the field equations. Numerical black hole solutions are also obtained for a specific set of initial conditions of the Weyl vector and of the scalar field at infinity.

\subsubsection{Exact black hole solutions}

In the following we will first look for exact black hole solutions satisfying the condition
\be
\nu (r)+\lambda (r) =0, \forall r>0.
\ee

Then Eq.~(\ref{sum}) immediately gives for $\Phi$ the equation
\be
\Phi''=\frac{3}{2}\frac{\Phi ^{\prime 2}}{\Phi},
\ee
with the general solution given by
\be\label{40}
\Phi(r)=\frac{C_1}{\left(r+C_2\right)^2},
\ee
where $C_1$ and $C_2$ are arbitrary constants of integration. Eq.~(\ref{c15}) can be reformulated as
\bea\label{41}
&&-1+\frac{d}{dr}\left(re^{-\lambda}\right)-\frac{1}{4}r^2\Phi+2re^{-\lambda}\frac{\Phi'}{\Phi}-\frac{3}{4}r^2e^{-\lambda}\frac{\Phi ^{\prime 2}}{\Phi ^2}\nonumber\\
&&-\frac{r^2\lambda 'e^{-\lambda}}{2}\frac{\Phi'}{\Phi}+r^2e^{-\lambda}\frac{\Phi ''}{\Phi}=0.
\eea

By taking into account the identity
\be
-\frac{r^2\lambda 'e^{-\lambda}}{2}\frac{\Phi'}{\Phi}=-\frac{re^{-\lambda}}{2}\frac{\Phi'}{\Phi}+\frac{r}{2}\frac{d}{dr}\left(re^{-\lambda}\right)\frac{\Phi'}{\Phi},
\ee
Eq.~(\ref{41}) becomes
\bea
&&\left(1+\frac{r}{2}\frac{\Phi '}{\Phi}\right)\frac{d}{dr}\left(re^{-\lambda}\right)+ \left(\frac{3}{2}\frac{\Phi'}{\Phi}-\frac{3}{4}r\frac{\Phi^{\prime 2}}{\Phi ^2}+r\frac{\Phi ''}{\Phi}\right)\left(re^{-\lambda}\right)\nonumber\\
&&-\frac{1}{4}r^2\Phi-1=0.
\eea

With the use of the expression of $\Phi (r)$ as given by Eq.~(\ref{40}), we obtain
\bea\label{44}
&&\left(1-\frac{r}{C_2+r}\right)
   u'(r)+3\left(\frac{ r}{(C_2+r)^2}-\frac{1}{C_2+r}\right) u(r)\nonumber\\
 &&  -\frac{C_1 r^2}{4 (C_2+r)^2}-1=0,
\eea
where we have denoted $u=re^{-\lambda}$. Eq.~(\ref{44}) has the exact general solution
\bea\label{58}
re^{-\lambda}&=&\frac{r^2 \left(12 C_3 C_2^2-C_1-4\right)}{4 C_2}+r \left(3 C_3
   C_2^2-\frac{C_1}{4}-2\right)\nonumber\\
  && +C_3 C_2^3-\frac{1}{12}
   (C_1+12) C_2+C_3 r^3,
\eea
where $C_3$ is an integration constant. Depending on the values of the integration constants $C_1$, $C_2$ and $C_3$, the metric (58) can take two distinct forms. If we assume the condition
\be
3 C_3C_2^2-\frac{C_1}{4}-2=1,
\ee
or, equivalently, $C_3C_2^3-C_1C_2/12=C_2$, the metric (\ref{58}) becomes
\be
e^{-\lambda}=e^{\nu}=1+\frac{2}{C_2}r+C_3r^2,
\ee
and it represents a generalization of the static cosmological de Sitter solution. The solution is not asymptotically flat. If one imposes the condition
\be
3 C_3C_2^2-\frac{C_1}{4}-2=2\beta
\ee
where $\beta $ is a constant, the metric tensor components (\ref{58}) become
\be
e^{-\lambda}=e^{\nu}=2\beta +\frac{1+2\beta}{C_2}r-\frac{C_2\left(1-2\beta\right)}{3}\frac{1}{r}+C_3r^2.
\ee

By assuming, in analogy with the Schwarzschild metric,  that $C_2\left(1-2\beta\right)/3=r_g$, the metric tensor components (\ref{58}) take the form
\be\label{dm1}
e^{-\lambda}=e^\nu=2\beta +\frac{1-4\beta ^2}{3}\frac{r}{r_g}-\frac{r_g}{r}+C_3r^2.
\ee

If we denote $1-2\beta=\delta$, the metric (\ref{dm1}) can be written as
\be\label{metrW}
e^{-\lambda}=e^\nu=1-\delta +\frac{\delta(2-\delta)}{3}\frac{r}{r_g}-\frac{r_g}{r}+C_3r^2.
\ee

\begin{figure*}
	\includegraphics[scale=0.9]{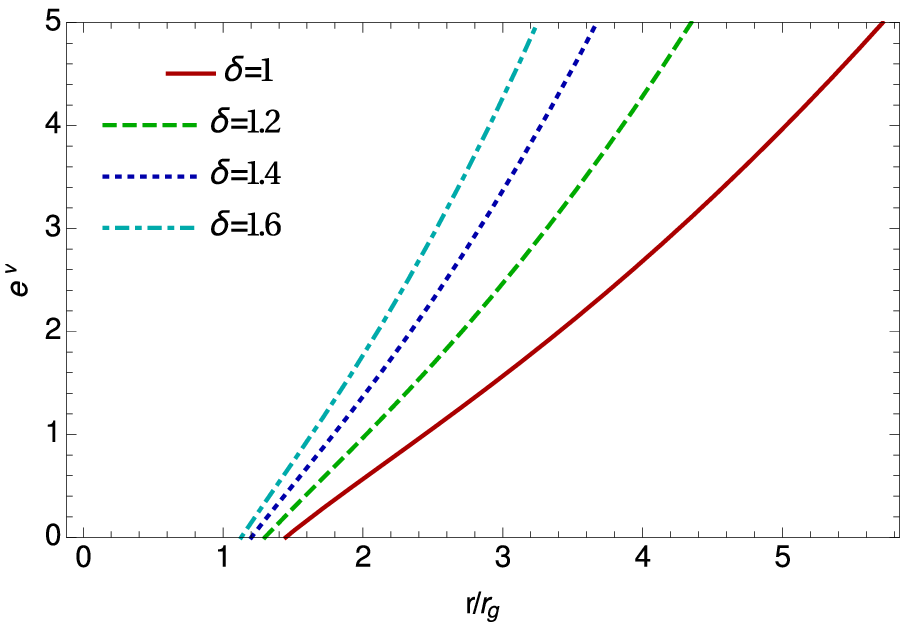}~~~\includegraphics[scale=0.9]{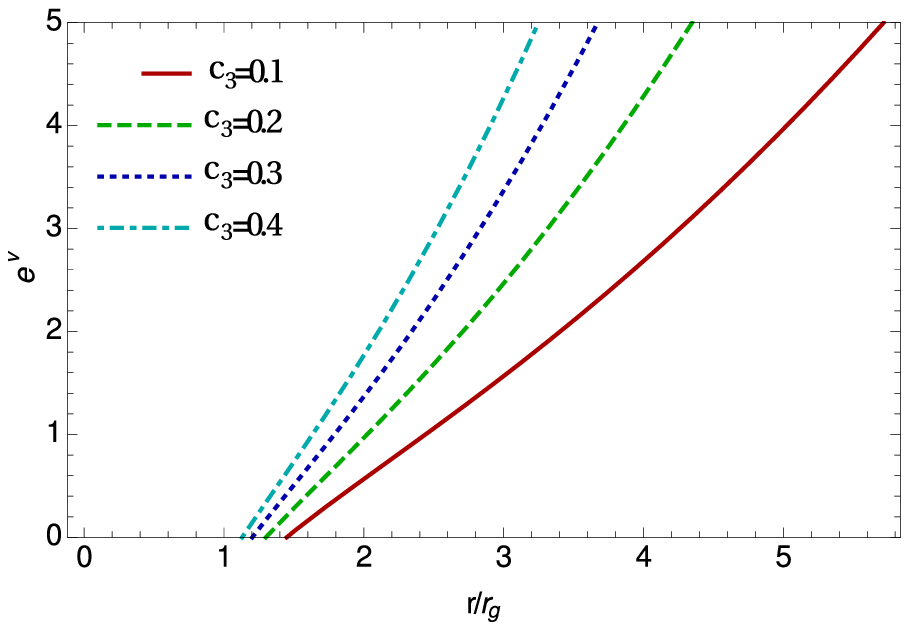}
	\caption{Variation of the metric tensor component $e^{\nu}$ for the exact solution of the Weyl geometric gravity in the presence of the radial component of the Weyl vector only, for $c_3=0.01$, and different values of $\delta$ (left panel), and for $\delta =1$, and different values of $c_3$ (right panel).} \label{figexact}
\end{figure*}

In Fig.~ (\ref{figexact}) we have plotted the behavior of the metric tensor $e^\nu$ as a function of $r/r_g$ for different values of the constants  $c_3\equiv C_3r_g^2$ and $\delta$, respectively. In each of these cases a singularity in the metric tensor component does appear, indicating the formation of an event horizon, and the presence of a black hole, respectively.

It is interesting to note that very similar solutions of the vacuum field equations have been obtained in the framework of other modified gravity theories. In particular, in conformal Weyl gravity \cite{M1,M2,M3,M4,M5,M6}, for a  static spherically symmetric metric having the standard form $ds^2=-B(r)dt^2+B^{-1}(r)dr^2+r^2d\Omega$, the theory admits vacuum solutions of the form \cite{M1}
\be\label{Mann}
B(r)=1-3\beta \gamma -\frac{\beta\left(2-3\beta \gamma\right)}{r}+\gamma r+kr^2,
\ee
where $\beta$, $\gamma$ and $k$ are arbitrary integration constants \cite{M1}. With the use of the metric (\ref{Mann}) in \cite{M1} it was suggested  that Weyl conformal gravity can explain the flat rotation curves of galaxies without invoking the presence of dark matter. Another interesting modified gravity theory,  the dRGT massive gravity theory \cite{Mgrav1, Mgrav2,Mgrav3}, with action
\begin{equation}
S = \int d^4 x \sqrt{-g} \frac{M_{Pl}^2}{2} \left[R + m_g^2 \mathcal{U}(g,f) \right] + S_m \,,
\end{equation}
where $M_{Pl}$ is the reduced Planck mass, $R$ is the Ricci scalar, $m_g$ is the graviton mass, and $\mathcal{U}$ is the self-interacting potential of the gravitons, respectively, also admits a static spherically symmetric solution of the form  (see \cite{Mgrav3}, and references therein),
\begin{equation}
B(r) = 1 - \frac{2 G m(r)}{r} - \frac{\Lambda r^2}{3} + \gamma r + \zeta,
\end{equation}
\\
where $m(r)$ is the  mass within radius $r$, and
$\Lambda \equiv - 3 m^2_g (1 + \alpha + \beta)$, $\gamma \equiv - m^2_g C (1 + 2 \alpha + 3 \beta)$, and $\zeta \equiv m_g^2 C^2 (\alpha + 3 \beta)$.

Here $C$, $\gamma$, and $\zeta$ are constants depending on the graviton mass $m_g$, $\Lambda$ corresponds to the cosmological constant, while $\alpha =3\alpha _3+1$, and $\beta =4\alpha_4-(1-\alpha)/3$ are coefficients related to the decomposition of the self-interacting potential of the graviton as $\mathcal{U}=\mathcal{U}_2+\alpha_3\mathcal{U}_3+\alpha_4\mathcal{U}_4$, with $\mathcal{U}_2 \equiv  [\mathcal{K}]^2 - [\mathcal{K}^2]$, $\mathcal{U}_3 \equiv [\mathcal{K}]^3 - 3 [\mathcal{K}][\mathcal{K}^2] + 2 [\mathcal{K}^3]$, and $\mathcal{U}_4 \equiv [\mathcal{K}]^4 - 6 [\mathcal{K}]^2 [\mathcal{K}^2] + 3[\mathcal{K}^2]^2 + 8 [\mathcal{K}][\mathcal{K}^3] - 6 [\mathcal{K}^4]$, respectively, where $\mathcal{K}^{\mu}_{\nu} \equiv \delta^{\mu}_{\nu} - \sqrt{g^{\mu\lambda} \partial_{\lambda} \varphi^a \partial_{\nu} \varphi^b f_{ab}}$, and we have denoted $[\mathcal{K}]=\mathcal{K}_{\mu}^{\mu}$. Here $g_{\mu\nu}$ is the physical metric,  $f_{\mu\nu}$ is a reference (fiducial) metric, and $\varphi^a$ are the St$\Ddot{\rm u}$ckelberg fields \cite{Mgrav3}. For $m_g = 0$, we recover  the standard Schwarzschild solution of general relativity.

The existence of mathematically identical vacuum solutions in several distinct gravity theories raises the question of the possible universality of metrics of the form (\ref{metrW}), representing the most general extension of the Schwarzschild metric.

\subsubsection{Numerical black hole solutions}

We will proceed now to obtaining numerical black hole solutions in Weyl geometric gravity in the presence of a Weyl vector field having only a radial component. In a static and spherically symmetric configuration, Eqs.~(\ref{c19})-(\ref{c16}) become,
\begin{widetext}
\begin{eqnarray}\label{e1}
	\frac{dm}{d\eta} &=& -\frac{7(1-\eta m)\zeta^3+\eta(5-4\eta m)\zeta^2\psi+\zeta(3\eta^3 m-4\eta^2-3\psi)\psi^2-\eta\psi^4}{\eta^3\psi(\zeta+\eta\psi)^2},\\
	\frac{d\nu}{d\eta} &=& \frac{\zeta(1-\eta m)(3\zeta+4\eta \psi)-(\eta^3 m+\psi)\psi^2}{\eta^2\psi(1-\eta m)(\zeta+\eta\psi)},\\
	\frac{d\psi}{d\eta} &=& -\frac{2\zeta}{\eta^2},\\
	\frac{d\zeta}{d\eta} &=& -\frac{\zeta^2(1-\eta m)(5\zeta+2\eta\psi)+\zeta(3\eta^3 m-4\eta^2-\psi)\psi^2}{\eta^2\psi(1-\eta m)(\zeta+\eta\psi)},
\end{eqnarray}
\end{widetext}

Note that the above system of ordinary, strongly nonlinear system of differential equations is independent of the coupling constants $\alpha$ and  $\xi$. We integrate the system from infinity, corresponding to $\eta =\lim_{r\rightarrow \infty}1/r=0$, up to the event horizon of the black hole.  We find a numerical black hole solution by detecting a singularity in metric tensor components  $e^{\nu}$ and $e^{\lambda}$. The singularities in the $e^{\nu}$ and $e^{\lambda}$ metric tensor components are indicated by the fulfillment of the conditions
\be
\left.e^{\nu(\eta)}\right|_{\eta=\eta_{\textit{hor}}}=0,\;\left.e^{-\lambda(\eta)}\right|_{\eta=\eta_{\textit{hor}}}=1-\left.\left[m(\eta)\eta\right]\right|_{\eta =\eta _{\textit{hor}}}=0,
\ee
where $\eta _{\textit{hor}}$ is the horizon of the black hole. For a Schwarzschild black hole the position of the event horizon corresponds to $\eta_{\textit{hor}}=1$.

\paragraph{The initial conditions.} In order to numerically integrate the gravitational field equations in the variable $\eta$, we need to fix the initial conditions at $\eta =0$, corresponding to the asymptotic values of the scalar and vector fields, and of the metric. As we have seen in the discussion of the asymptotic limits of this model, at infinity the values of the scalar field and of its derivative can be taken either as zero, so that $\psi (0)=0$, and $\zeta (0)=0$, or as having some arbitrary numerical values. As for the metric tensor components, we assume that the metric can be either asymptotically flat, corresponding to $\nu (0)=\lambda (0)=0$, or it can have a de Sitter form, $\left.e^{-\lambda}\right|_{\eta =\eta _{0}}=\left.\left[1-\left(4r_g^2/\mu ^2\right)\left(1/\eta ^2\right)\right]\right|_{\eta =\eta_{0}}$, giving $\left.m(\eta)\right|_{\eta =\eta _{0}}=\left(4r_g^2/\mu ^2\right)\left(1/\eta ^3\right)_{\eta =\eta _0}=m_0$.

Hence, in the de Sitter case we fix the initial values of $(m,\nu)$ at the {\it physical infinity} $\eta =\eta _0$ as $m\left(\eta_0\right)=m_0$, and $\nu=\left.\ln \left(1-m(\eta)\eta\right)\right|_{\eta =\eta _0}=\ln \left(1-m_0\eta _0\right)$.
\\
To obtain the black hole solutions we vary the initial values of the dimensionless scalar field $\psi (0)$, and of its derivative $\zeta (0)$, and, through the numerical integration of the field equations, we obtain the variations of the metric tensor components, of the scalar field, and of the Weyl vector field as functions of the dimensionless (inverse) radial coordinate $\eta$.

\paragraph{Numerical results.} In the following we consider numerical black hole solutions obtained by numerically integrating the gravitational field equations by assuming the presence of the spatial component of the Weyl vector only. We consider two types of models, obtained by fixing the initial value of $\zeta $ at infinity, and by varying the values of the scalar field, and by fixing the scalar field at infinity, and varying its derivative $\zeta (0)$.

\begin{figure*}[htbp]
	\centering
	\includegraphics[width=8.5cm]{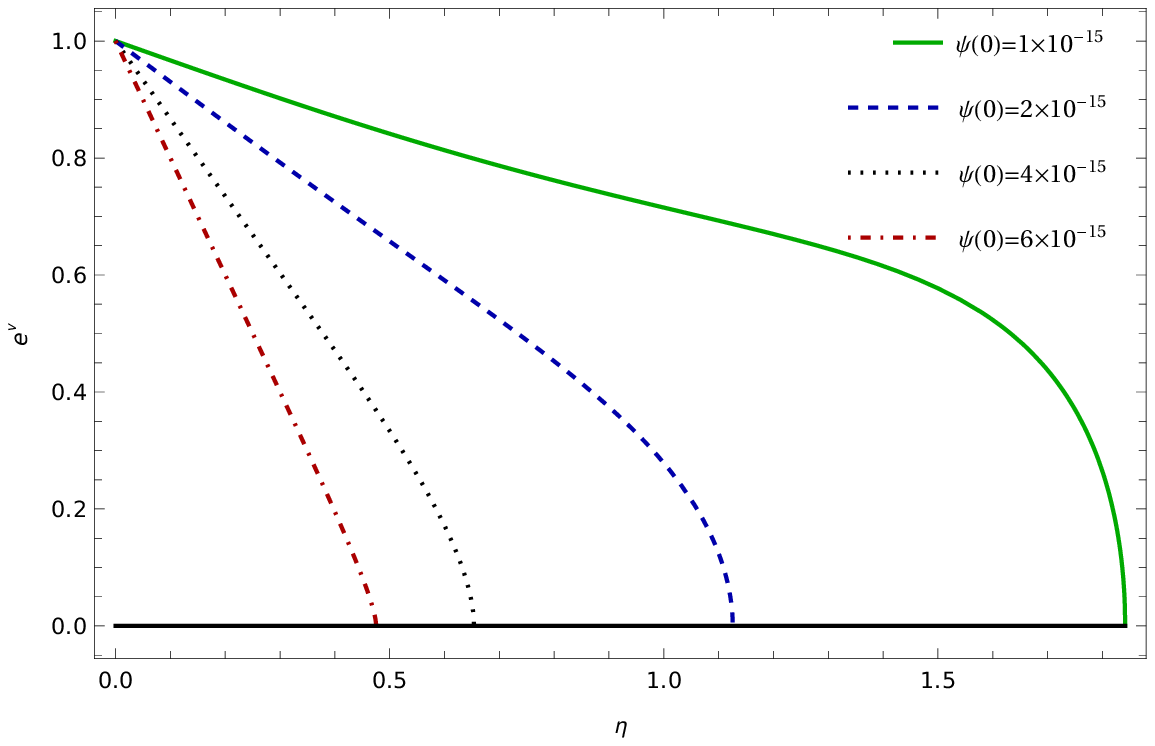}
	\includegraphics[width=8.5cm]{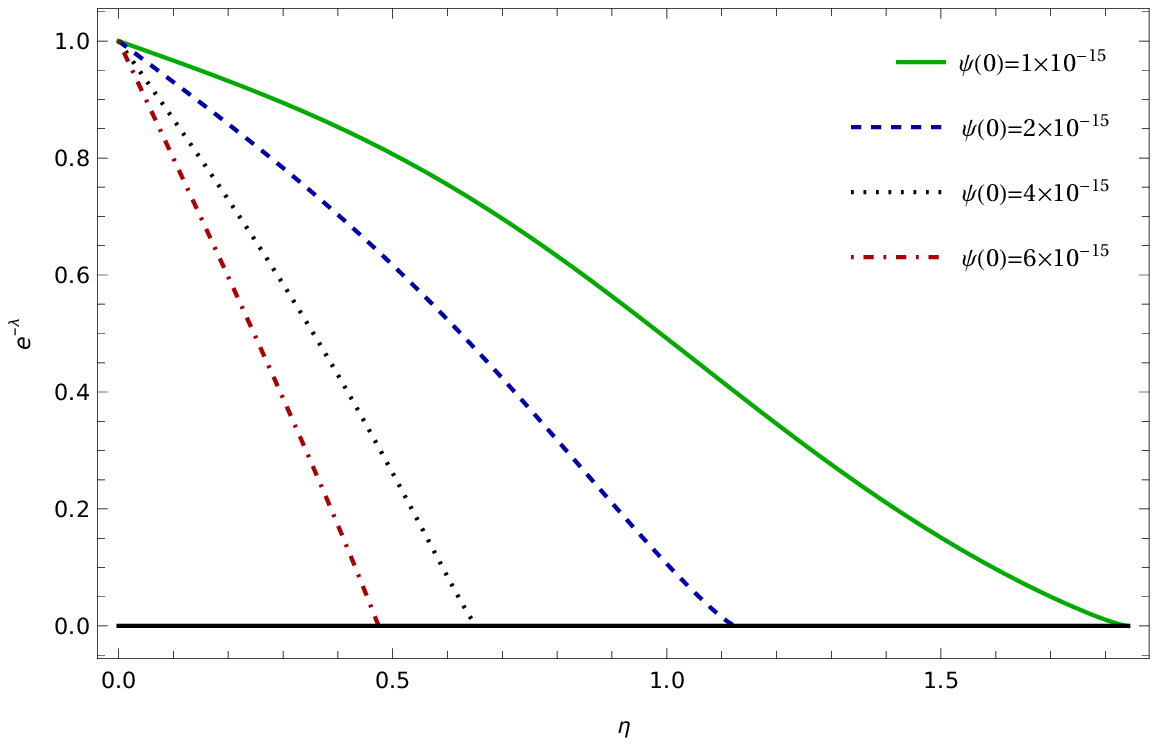}
\newline
\newline
	\includegraphics[width=8.5cm]{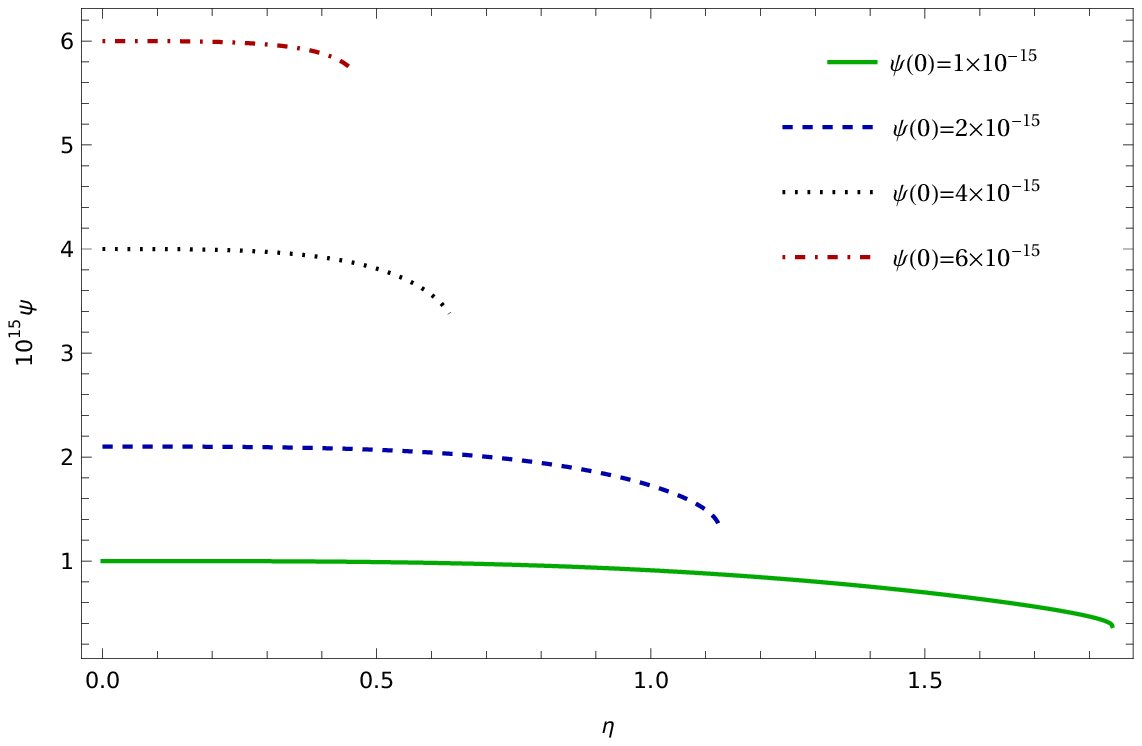}
	\includegraphics[width=8.5cm]{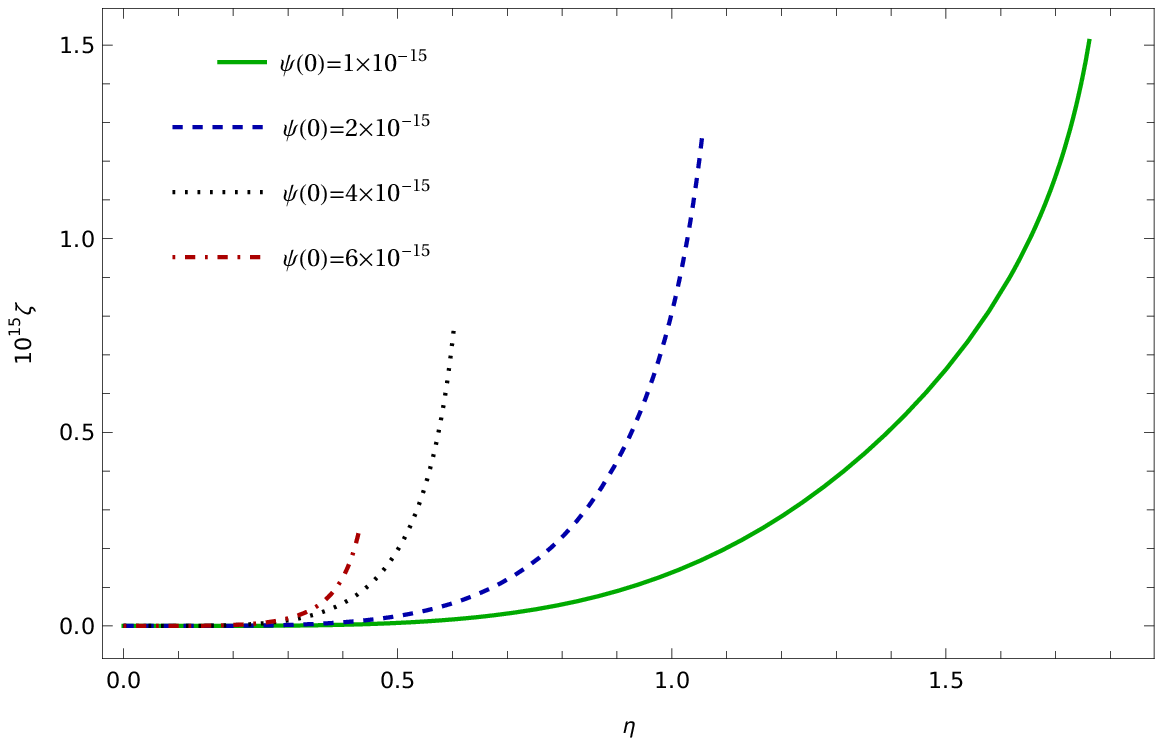}
	\caption{Variation of the metric tensor components $e^\nu$ (upper left panel) and $e^{-\lambda}$ (upper right panel), of  $\psi$ (lower left panel), and of  $\zeta$ (lower right panel) as a function of $\eta$, for a Weyl geometric black hole in the presence of the radial component of the Weyl vector only, for $\zeta(0)=1\times 10^{-35}$, and for different values of $\psi (0)$, presented in the legends of the Figures.
}\label{caseapsi}
\end{figure*}

\subparagraph{Models with fixed $\zeta (0)$ and varying $\psi (0)$.}
In Fig.~\ref{caseapsi} we present the results of the numerical integration of the field equations in the presence of a radial component of the Weyl vector only, obtained by fixing the initial value of the derivative of the scalar field $\zeta$ at infinity as $\zeta (0)=1\times 10^{-35}$, and by varying the initial values of the scalar field  $\psi(0)$. We restrict our analysis to the case of the flat asymptotic conditions.
As one can see from the upper panels Fig.~\ref{caseapsi}, the metric tensor components are decreasing functions of $\eta$, and they  become singular for a finite value of the radial coordinate, indicating the formation of a black hole. The position of the event horizon is very sensitive to the small variations of the initial conditions. The scalar field is a decreasing function of $\eta$, taking only positive values, while $\zeta$ increases with increasing $\eta$.

Selected values of $\eta_{\textit{hor}}$ corresponding to different values of  $\psi(0)$, and for a fixed $\zeta (0)$, are presented in Table~\ref{tabcaseapsi}.

\begin{table*}[tbp]
	\centering
	\begin{tabular}{|c|c|c|c|c|}
		\hline
		~~~~$\psi (0)$~~~~ & $1\times 10^{-15}$ & $2\times 10^{-15}$ & $4\times 10^{-15}$ & $6\times 10^{-15}$\\
		\hline
		$\eta_{\textit{hor}}$ & $1.841$ & $1.126$ & $0.654$ & $0.475$ \\
  		\hline
	\end{tabular}
	\caption{Variation of the position of the event horizon $\eta_{\textit{hor}}$ of the Weyl geometric black hole with radial component of the Weyl vector for $\zeta (0)=1\times 10^{-35}$, and different initial values of $\psi (0)$.}\label{tabcaseapsi}
\end{table*}
As one can see from Table~\ref{tabcaseapsi}, the position of the event horizon of the black hole is strongly dependent on the values of the scalar field at infinity, and large variations of its location are possible, ranging from half to twice of the Schwarzschild values. Thus, for example, for $\psi(0)=10^{-15}$, the physical radius of the event horizon, $r_{hor}$, is located at $r_{hor}\approx 0.54$, while $r_{hor}\approx 2.10$ for $\psi (0)=6\times 10^{-15}$.

\begin{figure*}[htbp]
	\centering
	\includegraphics[width=8.5cm]{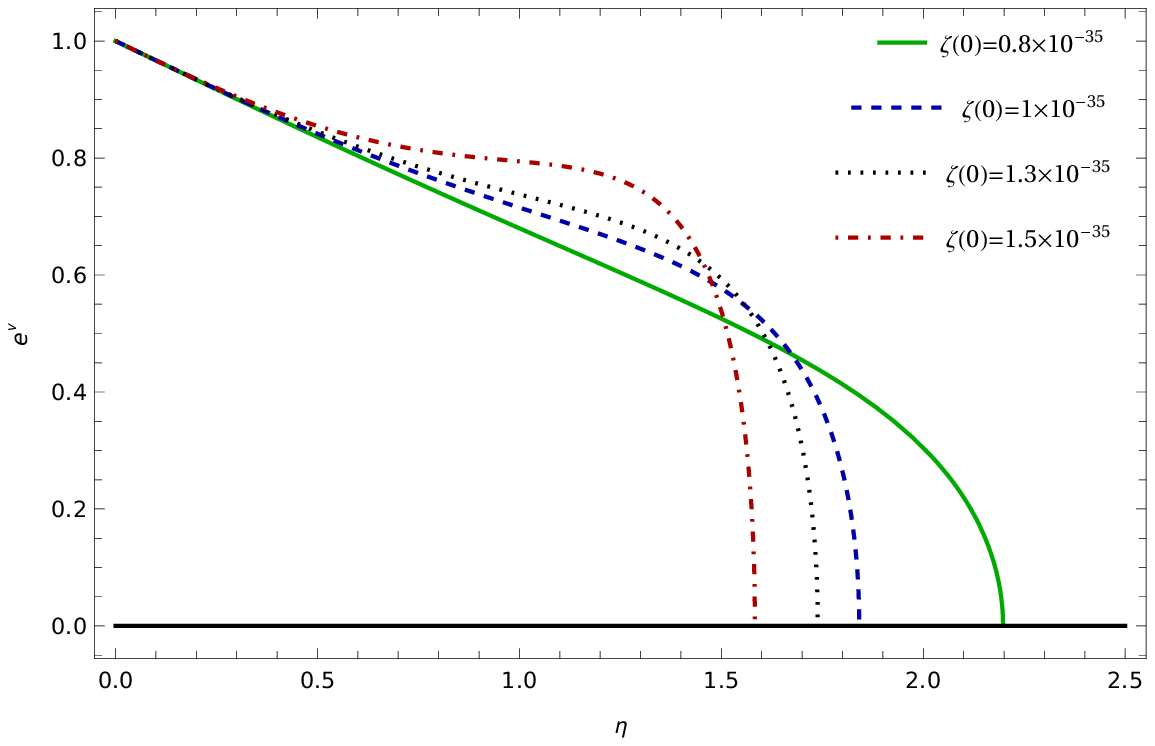}
	\includegraphics[width=8.5cm]{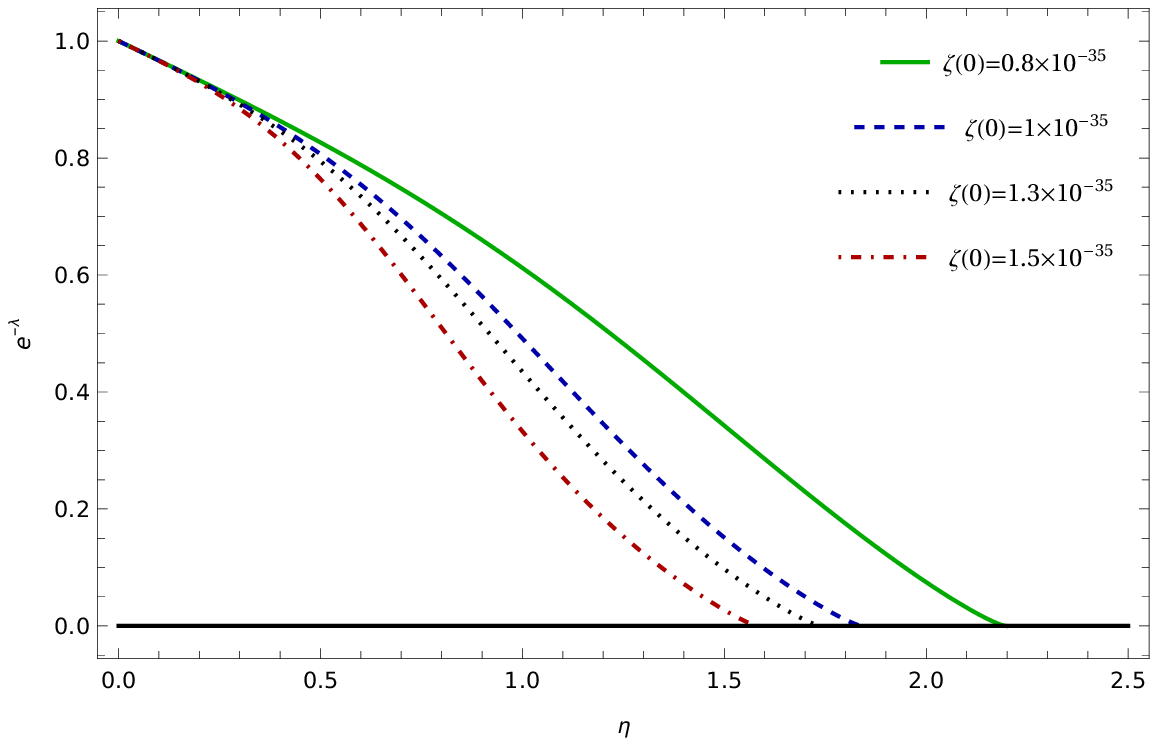}
\newline
\newline
	\includegraphics[width=8.5cm]{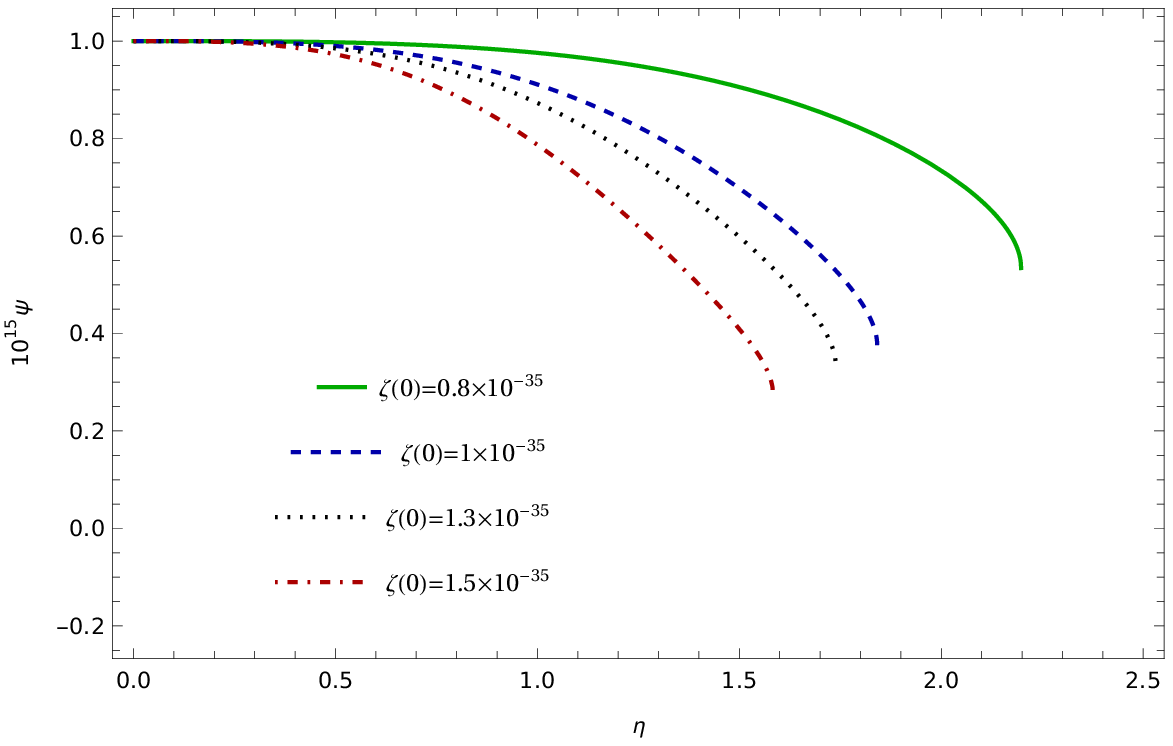}
	\includegraphics[width=8.5cm]{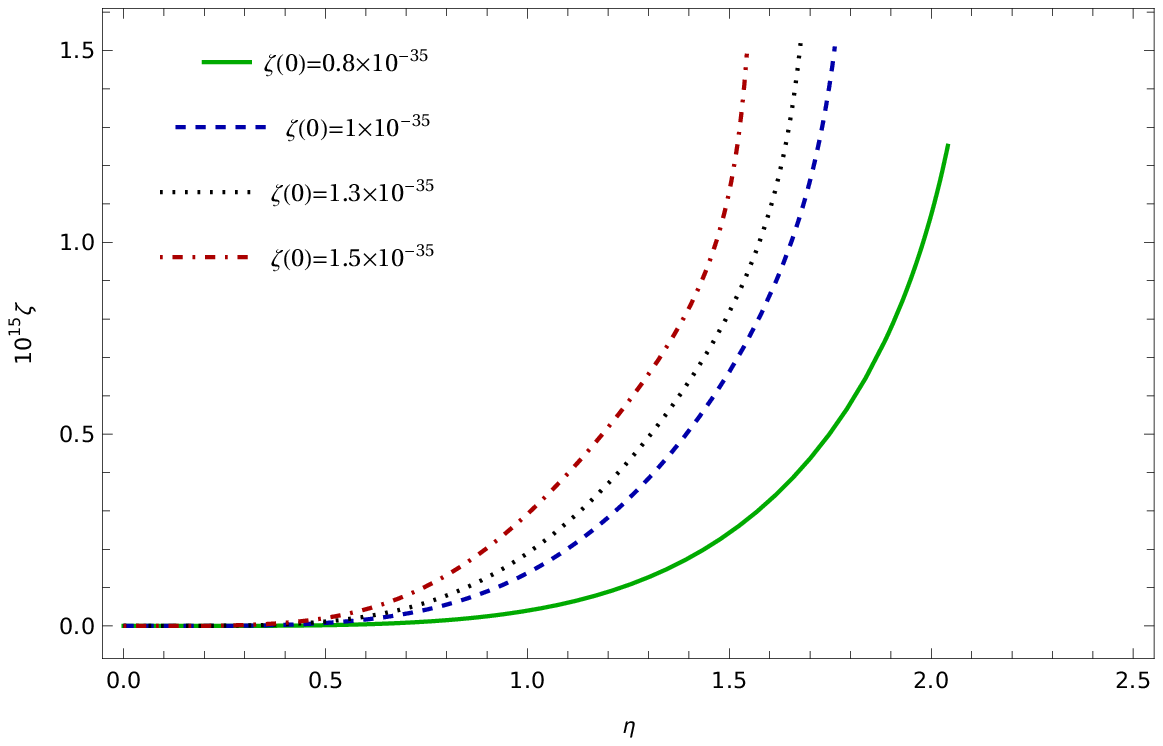}
	\caption{Variation of the metric tensor components $e^\nu$ (upper left panel) and $e^{-\lambda}$ (upper right panel), of $\psi$ (lower left panel), and $\zeta$ (lower right panel) as a function of $\eta$, for a Weyl geometric black hole in the presence of the radial component of the Weyl vector only,   for $\psi (0)=1\times 10^{-15}$, and for different values of $\zeta (0)$, presented in the legends of the Figures.
} \label{caseazeta}
\end{figure*}

\subparagraph{Models with fixed $\psi(0)$ and varying $\zeta (0)$.} In Fig.~\ref{caseazeta} we present the results of the numerical integration of the field equations in the presence of a radial component of the Weyl vector only, obtained by fixing the initial value of the scalar field $\psi (0)$ at infinity as $\psi (0)=1\times 10^{-15}$, and by varying the initial value of its derivative $\zeta (0)$.

As one can see from the upper panels of Fig~\ref{caseazeta}, a singularity does appear in the metric tensor components, corresponding to the formation of a black hole. The position of the event horizon is strongly dependent on the derivatives of the scalar field at the (physical) infinity. The scalar field is a decreasing function of $\eta$, while its derivative monotonically increases towards the event horizon.

The variation of $\eta_{\textit{hor}}$ with respect to the changes in the values of the derivative of the scalar field at infinity $\zeta (0)$, for a fixed $\psi$, are presented in Table~\ref{tabcaseazeta}.

\begin{table*}[tbp]
	\centering
	\begin{tabular}{|>{\centering}p{50pt}|>{\centering}p{50pt}|>{\centering}p{50pt}|>{\centering}p{50pt}|>{\centering\arraybackslash}p{50pt}|}
		\hline
		$\zeta (0)$ & $0.8\times 10^{-35}$ & $1\times 10^{-35}$ & $1.3\times 10^{-35}$ & $1.5\times 10^{-35}$ \\
		\hline
		$\eta_{\textit{hor}}$ & $2.198$ & $1.841$ & $1.739$ & $1.583$ \\
		\hline
	\end{tabular}
	\caption{Variation of the position of the event horizon $\eta_{\textit{hor}}$ of the Weyl geometric black hole with radial component of the Weyl vector for $\psi (0)=1\times 10^{-15}$, and different initial values $\zeta (0)$.\label{tabcaseazeta}}
\end{table*}

As one can see from the Table~\ref{tabcaseazeta}, the position of the event horizon of the black hole decreases as a function of $\eta$ (increases as a function of $r$) with increasing initial values of the derivative of the scalar field $\zeta(0)$. Thus, for $\zeta(0)=0.8\times 10^{-35}$, the physical radius of the event horizon has the value $r_{hor}\approx 0.45$, while $r_{hor}\approx 0.63$ for $\zeta (0)=1.5\times 10^{-35}$.  Hence, higher values of $\zeta (0)$ generate black holes having higher radii. However, we would like to pint out that in these considered examples the event horizons of the black holes are located at much smaller radii than the event horizons of their standard general relativistic counterparts.

\subsection{Black hole solutions with temporal component of the Weyl vector field}

We consider now a second class of Weyl type geometric black holes, in which the Weyl vector has only a temporal component, with $\omega_{\mu}$ given by  $\omega _{\mu}=\left(\omega _0, 0,0,0\right)$. In a static and spherically symmetric gravitational field configuration, and in the presence of a Weyl vector with a temporal component only, the gravitational field equations Eqs.~(\ref{c23})-(\ref{c26}) become,
\begin{widetext}
	\begin{eqnarray}\label{e2}
		\frac{dm}{d\eta} &=&\frac{\psi}{\eta^4} +\frac{12(1-\eta m)\Theta_0^2-3\psi\Omega_0^2}{e^\nu \eta^4 \psi},\\
		\frac{d\nu}{d\eta} &=&- \frac{e^\nu\psi(\psi+\eta^3m)+12(1-\eta m)\Theta_0^2-3\psi\Omega_0^2}{e^\nu \eta^3 \psi (1-\eta m)},\\
		\frac{d\psi}{d\eta} &=& 0,\\
		\frac{d\Omega_0}{d\eta} &=& - \frac{2\gamma\Theta_0}{\eta^2},\\
		\frac{d\Theta_0}{d\eta} &=& \frac{-\gamma\eta\psi\Omega_0+\Theta_0(4\eta^2(1-\eta m)-6e^{-\nu}\Omega_0^2)}{2\eta^3(1-\eta m)},\label{e2a}
	\end{eqnarray}
\end{widetext}

\subsubsection{The initial conditions}

In this case the model depends on six parameters, the initial conditions at infinity of $m$ and $\nu$, describing the metric tensor components, as well as of the scalar field $\psi$, of the Weyl vector $\Omega _0$, and of its derivative $\Theta _0$. Moreover, a coupling constant $\gamma$ is also present in the model. In the following, we investigate numerically only the asymptotically flat case, thus assuming $\nu (0)=\lambda (0)=0$. As one can see from Eqs.~(\ref{condom0}), asymptotically  the Weyl vector temporal component, and its derivative, tend to zero, and hence we have $\Omega _0(0)=0$, and $\Theta _0(0)=0$, respectively. For the initial value of $m$ we adopt the condition $m(0)=0$. The scalar field $\psi $ is a constant, and its value can be fixed arbitrarily.  From the numerical point of view we investigate the effects of the variation of the coupling constant, and of the numerical value of the scalar field, on the position of the event horizon.

\subsubsection{Numerical results}

We present now the numerical results of the numerical integration of the field equations in the presence of a temporal component of the Weyl vector only.

\begin{figure*}[htbp]
	\centering
	\includegraphics[width=8.5cm]{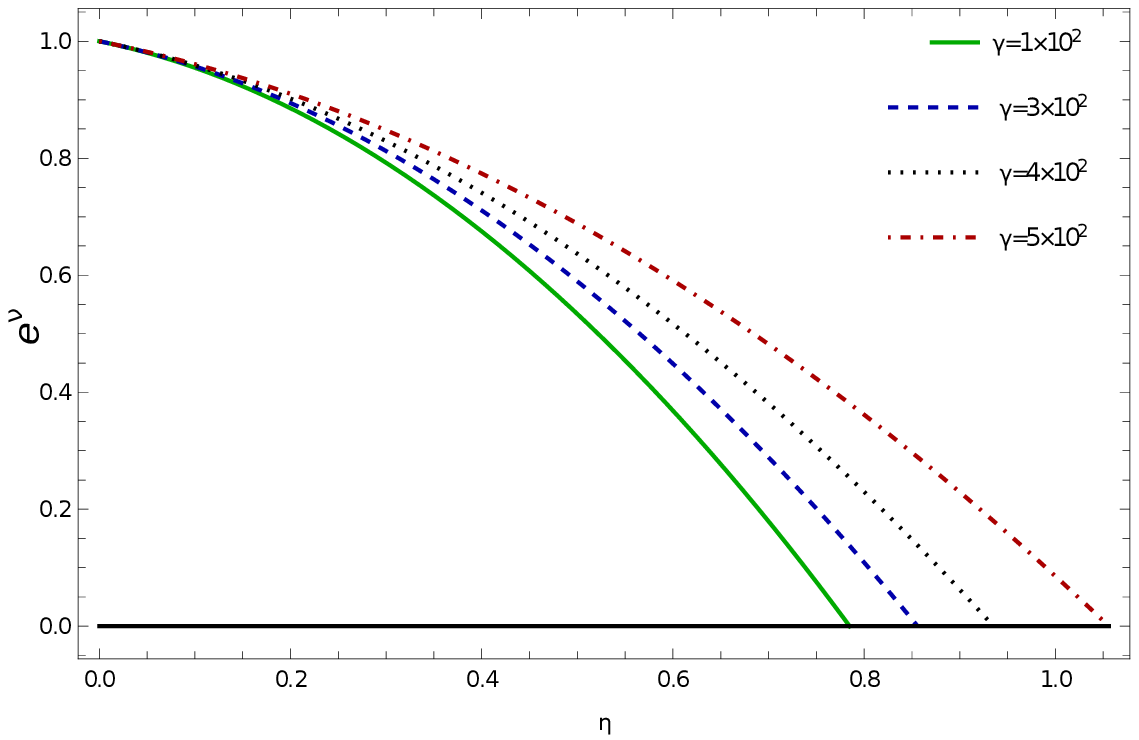}
	\includegraphics[width=8.5cm]{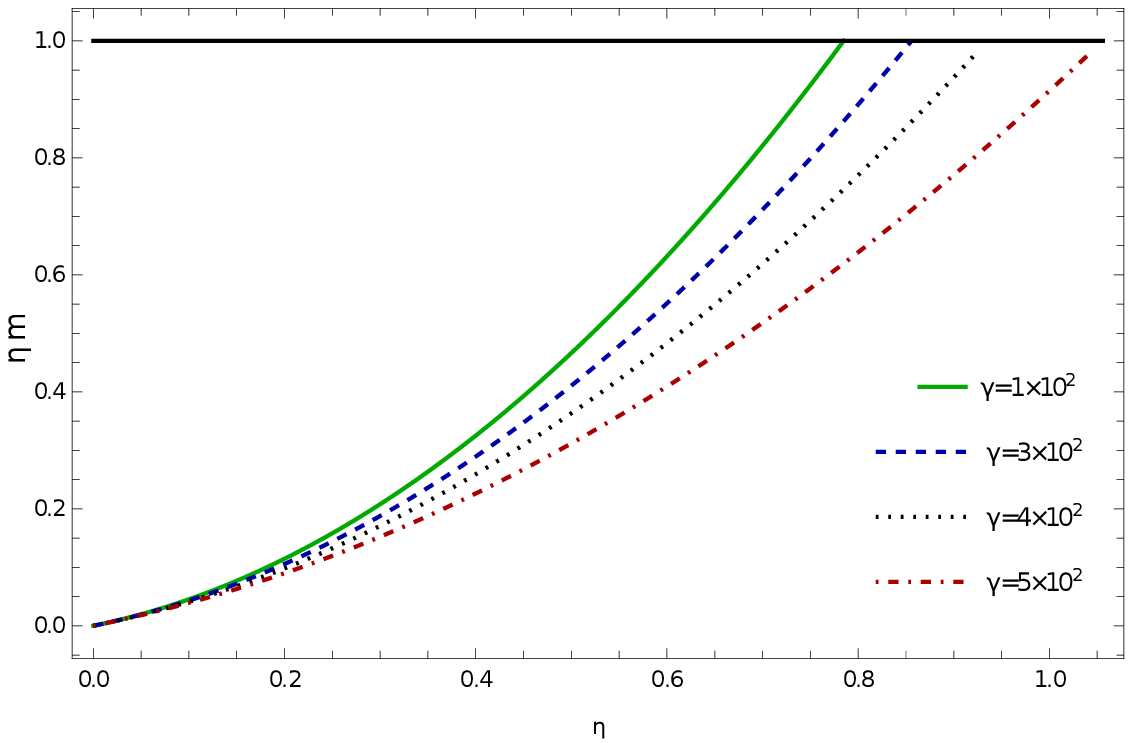}
\newline
\newline
	\includegraphics[width=8.5cm]{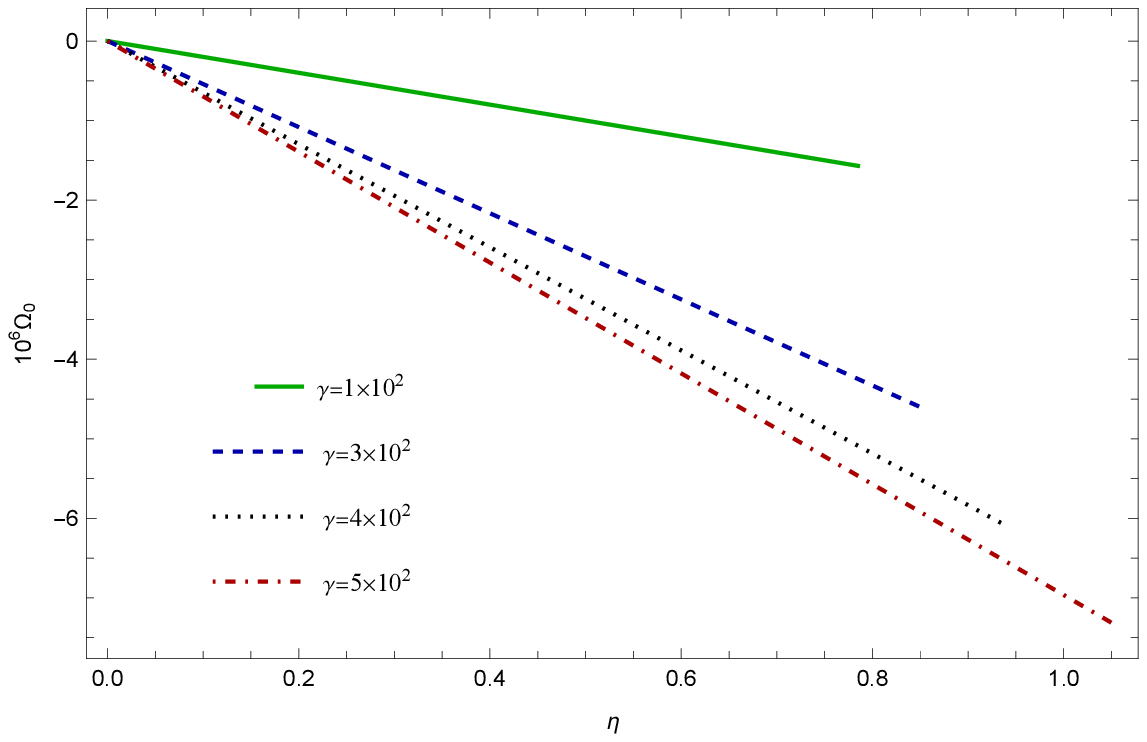}
\includegraphics[width=8.5cm]{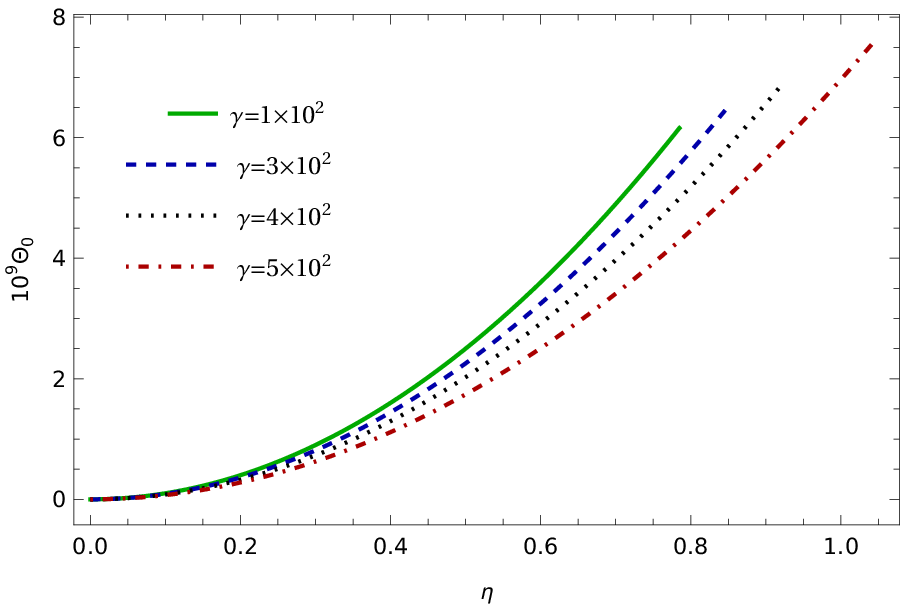}
	\caption{Variations of the metric tensor components $e^\nu$ (upper left panel), of the effective mass $\eta m$ (upper right panel), of  $\Omega_0$ (lower left panel), and of $\Theta_0$ (lower right panel) as a function of $\eta$ in the presence of the temporal component of the Weyl vector only for $ \psi (0)=1\times 10^{-15}$, $\Omega_0(0)=1\times 10^{-11}$, $\Theta_0(0)=1\times 10^{-18}$, and for different values of $\gamma$, presented in the legends of the Figures.
}\label{d1ba}
\end{figure*}

\paragraph{Varying the value of the coupling constant $\gamma$.}
 As a first example of black hole solutions with temporal Weyl vector we consider models in which only the coupling parameter $\gamma$ varies, with all other parameters fixed. The results of the numerical integration of the field equations (\ref{e2})-(\ref{e2a}) are represented in Fig.~\ref{d1ba}.

As one can see from Fig.~\ref{d1ba}, the metric tensor components do present a singular behavior, indicating the formation of the event horizon. There is a strong dependence on the numerical values of $\gamma$ of $\eta _{\textit{hor}}$. The behavior and the numerical values of the effective mass are also strongly dependent on the coupling constant $\gamma$, indicating an increase in the effective mass when approaching the event horizon of the black holes.
The variation of the Weyl vector temporal component $\Omega _0$ is represented in the lower right panel of Fig.~\ref{d1ba}. As a function of $\eta$, the Weyl vector field is a linearly decreasing function, while its derivative is an increasing function of the inverse of the radial variable. The effects of the variation of the coupling constant $\gamma$ on the behavior of the fields are significant.

The positions of the event horizon of the black holes are presented in Table~\ref{d9} for different values of $\gamma$, with all other initial conditions fixed as in Fig.~\ref{d1ba}.

\begin{table*}[htbp]
	\centering
	\begin{tabular}{|>{\centering}p{50pt}|>{\centering}p{50pt}|>{\centering}p{50pt}|>{\centering}p{50pt}|>{\centering\arraybackslash}p{50pt}|}
		\hline
		$\gamma$ & $100$ & $300$ & $400$ & $500$\\
		\hline
		$\eta_{\textit{hor}}$ & $0.785$ & $0.856$ & $0.935$ & $1.056$\\
		\hline
	\end{tabular}
	\caption{The position of the event horizon of the Weyl geometric type black holes in the presence of the temporal component of the Weyl vector only for different values of $\gamma$, and with the other parameters fixed like in the plots in Fig.~\ref{d1ba}. } \label{d9}
\end{table*}

There is a significant dependence of the position of the event horizon of the black hole on the numerical value of the coupling $\gamma$, with the position of the physical radius $r_{hor}$ of the event horizon of the black hole decreasing with increasing $\gamma$. Thus, for $\gamma =100$, $r_{hor}\approx 1.27$, while for $\gamma =500$, $r_{hor}\approx 0.95$. Hence, Weyl geometric black holes with radii larger than their Schwarzschild counterparts can be obtained for small values of the coupling constant $\gamma$.

\begin{figure*}[htbp]
	\centering
	\includegraphics[width=8.5cm]{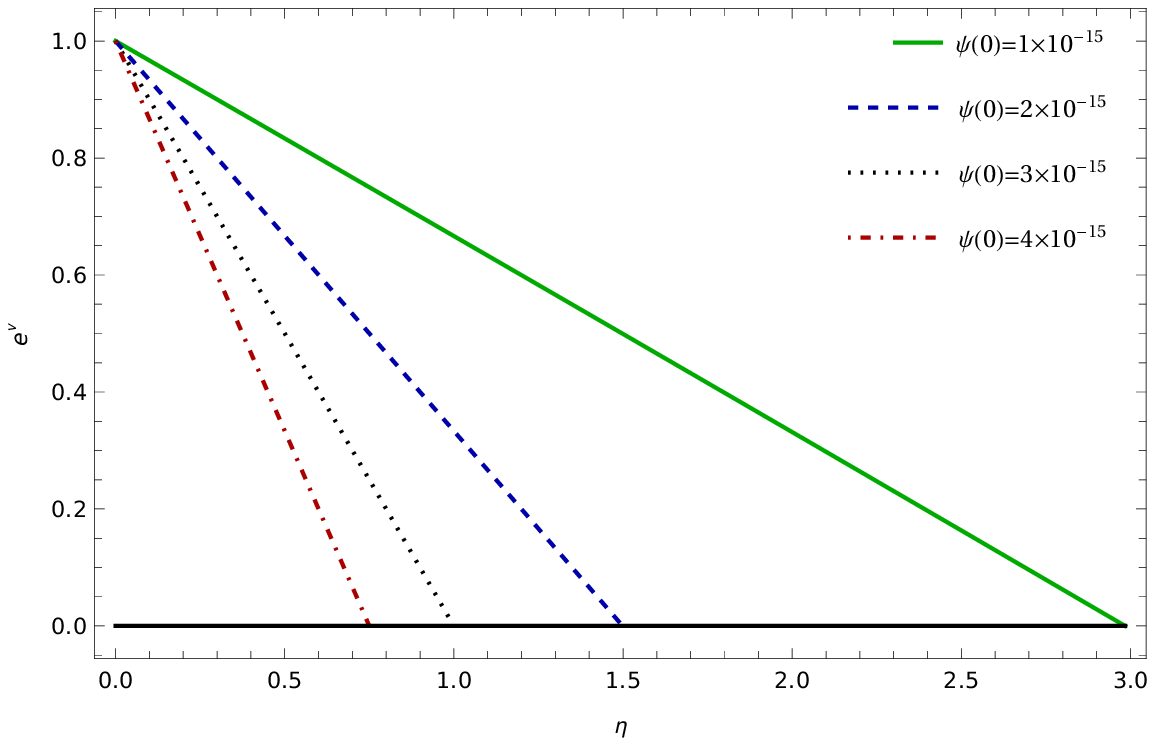}
	\includegraphics[width=8.5cm]{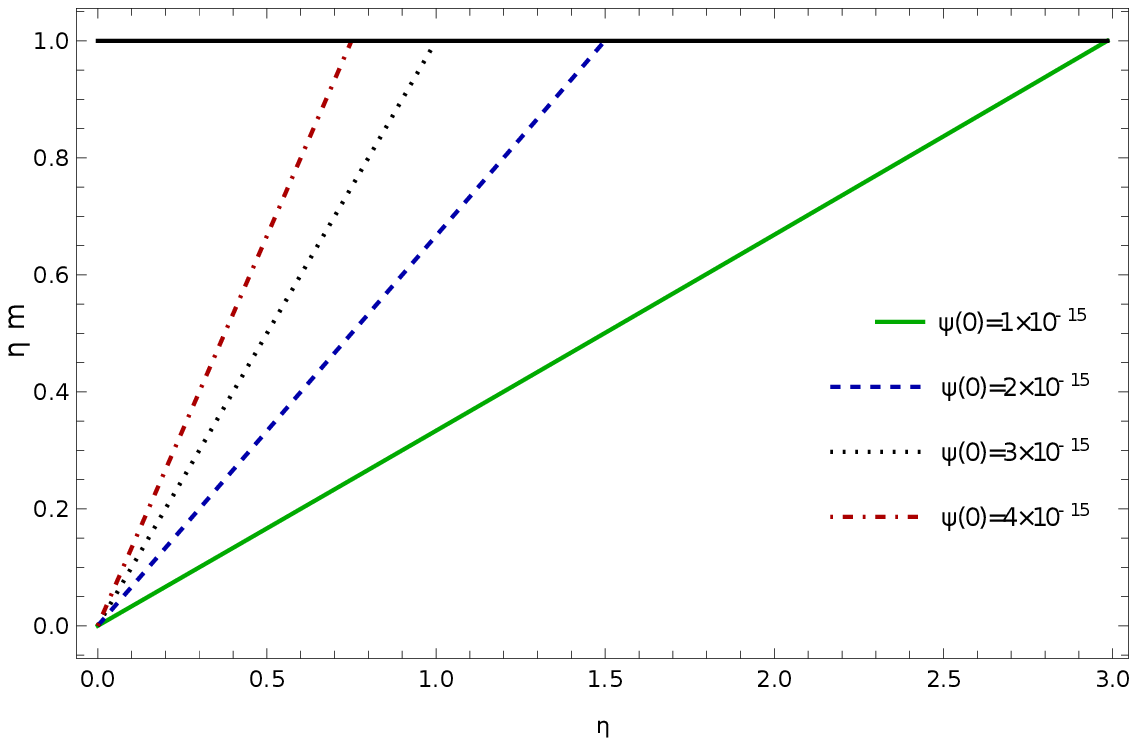}
\newline
\newline
	\includegraphics[width=8.5cm]{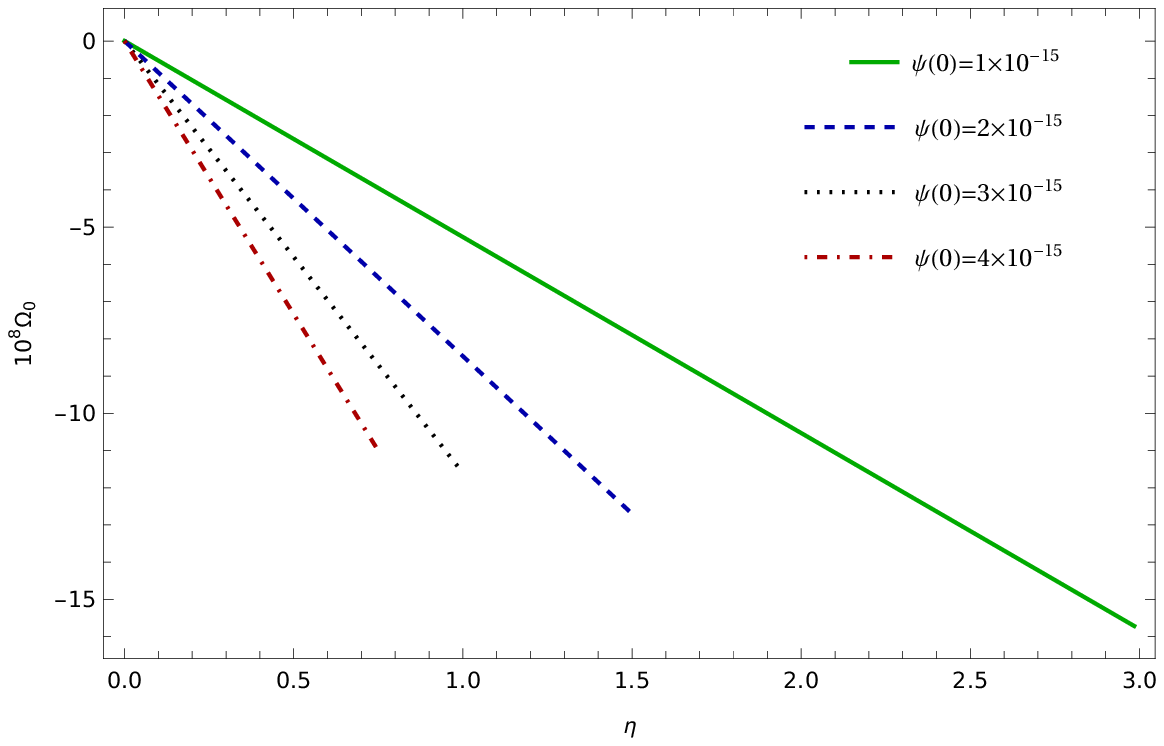}
\includegraphics[width=8.5cm]{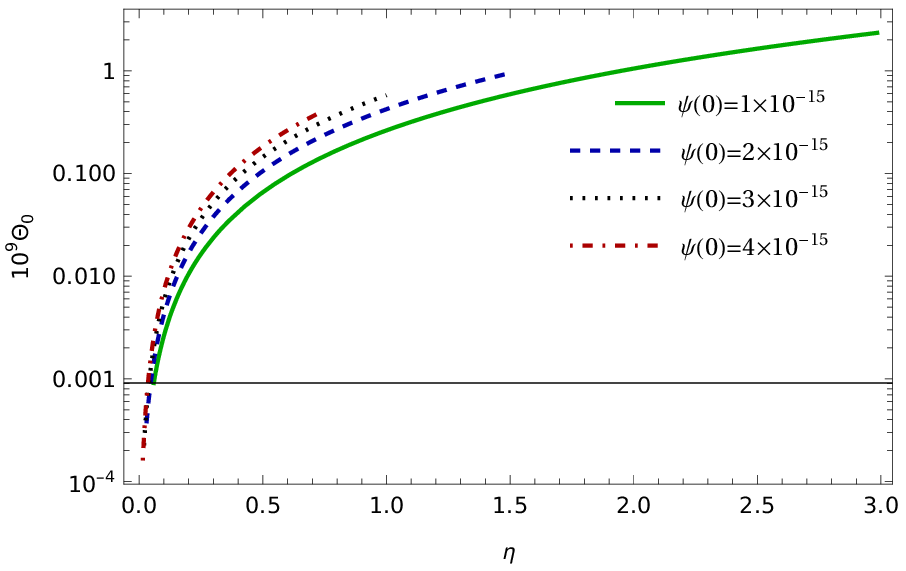}
	\caption{Variations of the metric tensor components $e^\nu$ (upper left panel), and of the effective mass $\eta m$ (upper right panel), of  $\Omega_0$ (lower left panel), and of $\Theta_0$ (lower right panel) as a function of $\eta$ of a Weyl black hole in the presence of the temporal component of the Weyl vector only,  for $\gamma=100$, $\Omega_0 (0)=1\times 10^{-11}$, $\Theta_0(0)=1\times 10^{-20}$, and for different values of $\psi (0)$, presented in the legends of the Figures.
}\label{d100}
\end{figure*}

\paragraph{Varying the value of the scalar field.}
We consider now black hole models in Weyl geometric type gravity in the presence of the temporal component of the Weyl vector obtained by varying the numerical values of the constant scalar field $\psi$, while keeping all the other numerical values at infinity of the physical and geometrical quantities fixed. The variations with respect to $\eta$ of the metric tensor components and of the effective mass are represented for this case in Fig.~\ref{d100}.
As one can see from  Fig.~\ref{d100}, for this choice of parameters the position of the event horizon is strongly dependent on the values of the scalar field, which as a constant, acts as a cosmological background. The Weyl vector $\Omega _0$ is linearly decreasing with respect to $\eta$ (linearly increasing as a function of $r$), while its derivative is a monotonically increasing function of $\eta$ (monotonically decreasing with respect to $r$).

The positions of the event horizon are presented, for different initial values of the scalar field $\psi (0)$,  in Table~\ref{d12}.

\begin{table*}[htbp]
	\centering
	\begin{tabular}{|>{\centering}p{50pt}|>{\centering}p{50pt}|>{\centering}p{50pt}|>{\centering}p{50pt}|>{\centering\arraybackslash}p{50pt}|}
		\hline
		$\psi (0)$ & $1\times 10^{-15}$ & $2\times 10^{-15}$ & $3\times 10^{-15}$ & $4\times 10^{-15}$\\
		\hline
		$\eta_{\textit{hor}}$ & $2.984$ & $1.4996$ & $0.999$ & $0.750$\\
		\hline
	\end{tabular}
	\caption{Variation of the position of the event horizon of the Weyl geometric black holes in the presence of a temporal component of the Weyl vector for different initial value of the scalar field $\psi _0(0) $. The numerical values of the other quantities are fixed like in the plots in Fig.~\ref{d100}.}
	\label{d12}
\end{table*}

There is a significant dependence on the position of the event horizon on the values of the scalar field, with $\eta_{\textit{hor}}$ decreasing with the increase of the numerical values of $\psi (0) $. The physical radius $r_{hor}$ increases with the increase of $\psi_0(0)$, from $r_{hor}\approx 0.33$ ($psi (0)=10^{-15}$), to $r_{hor}\approx 1.33$, for $\psi(0)=4\times 10^{-15}$.

\begin{figure*}[htbp]
	\centering
	\includegraphics[width=8.5cm]{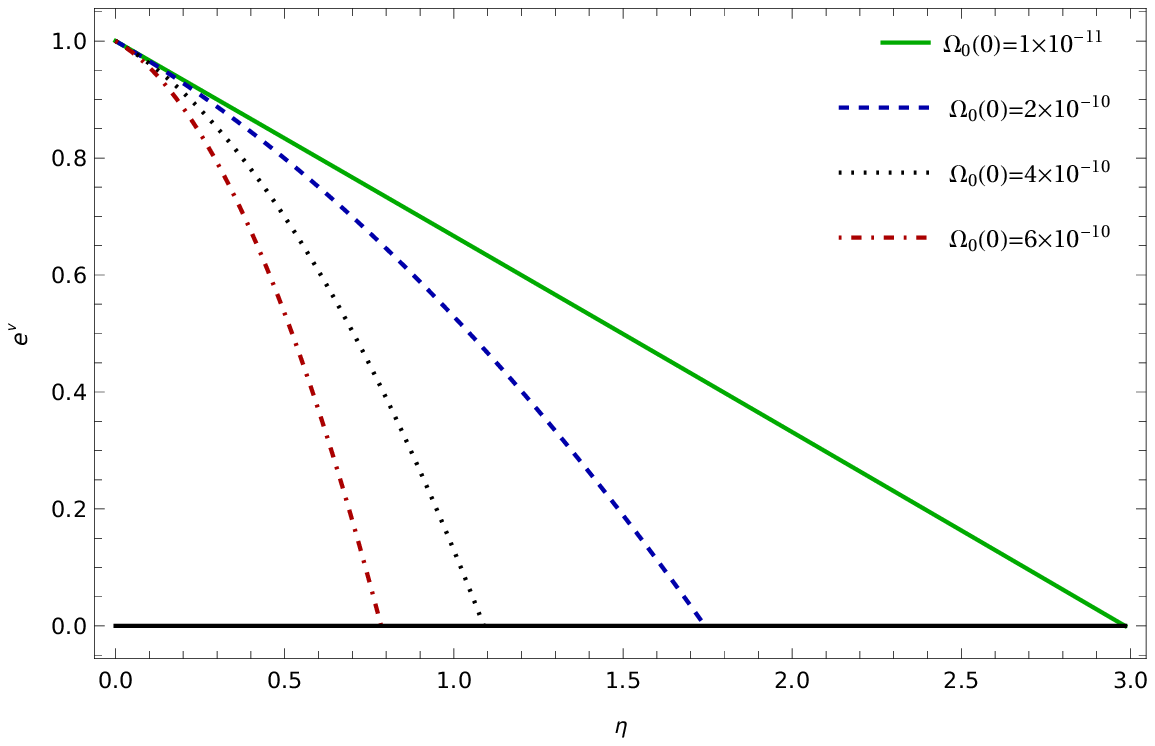}
	\includegraphics[width=8.5cm]{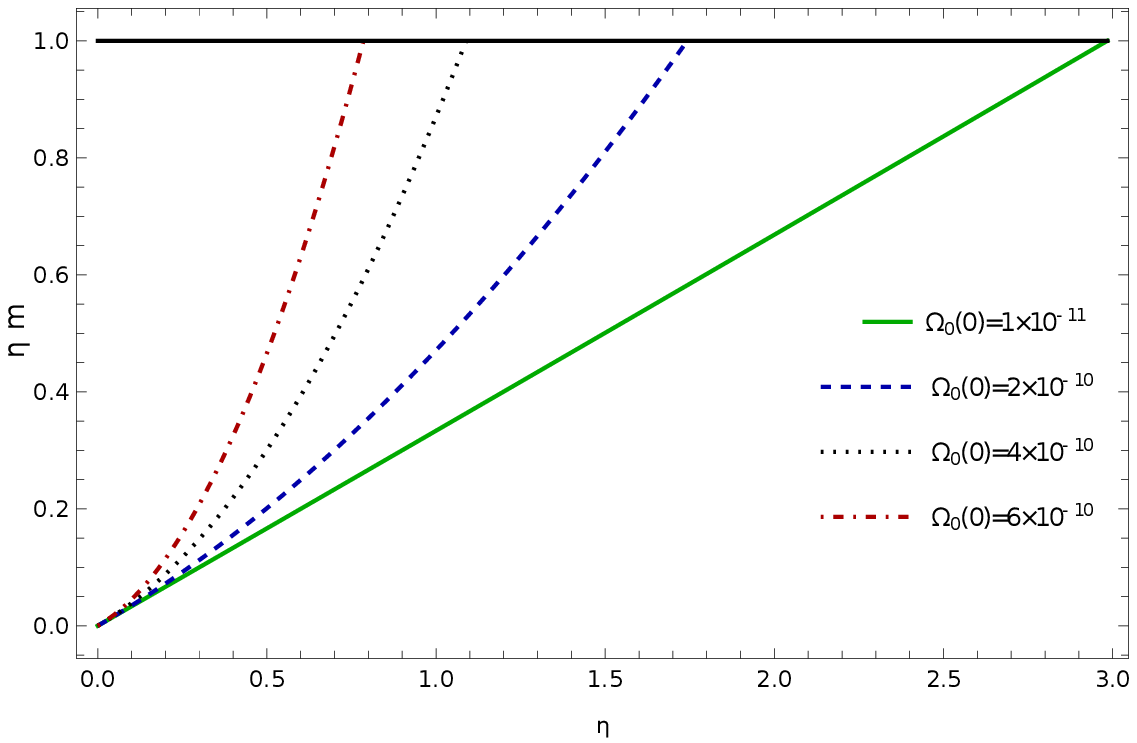}
\newline
\newline
	\includegraphics[width=8.5cm]{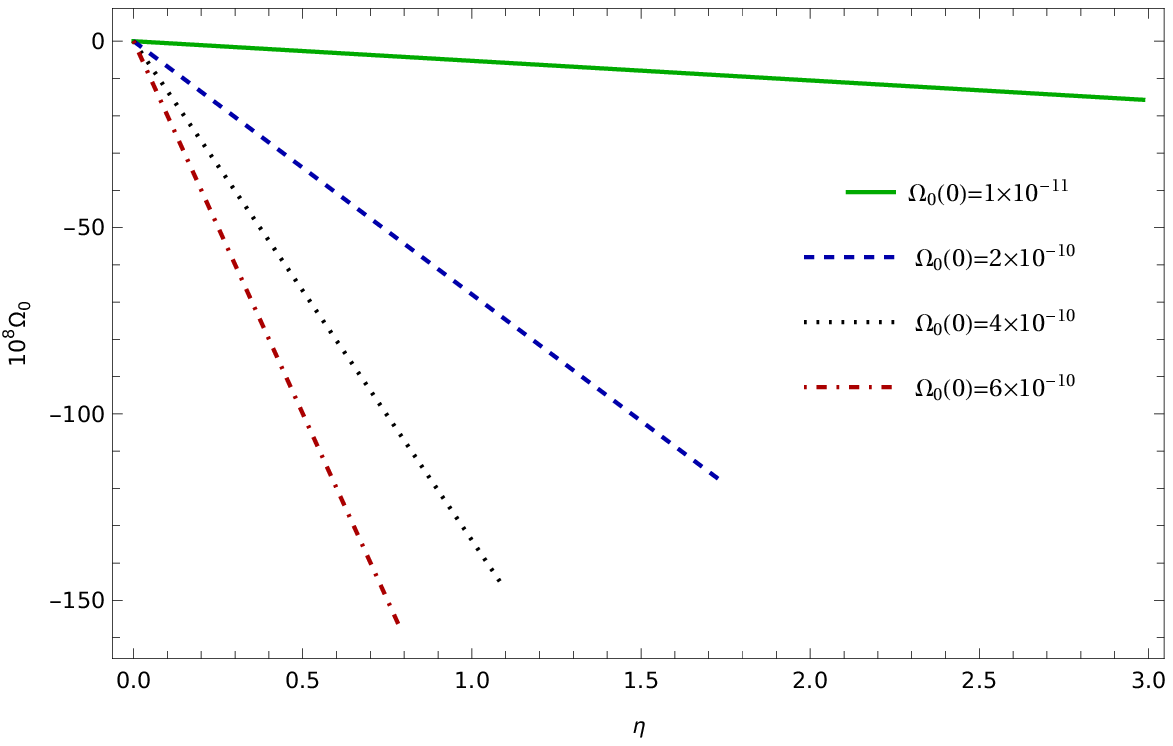}
\includegraphics[width=8.5cm]{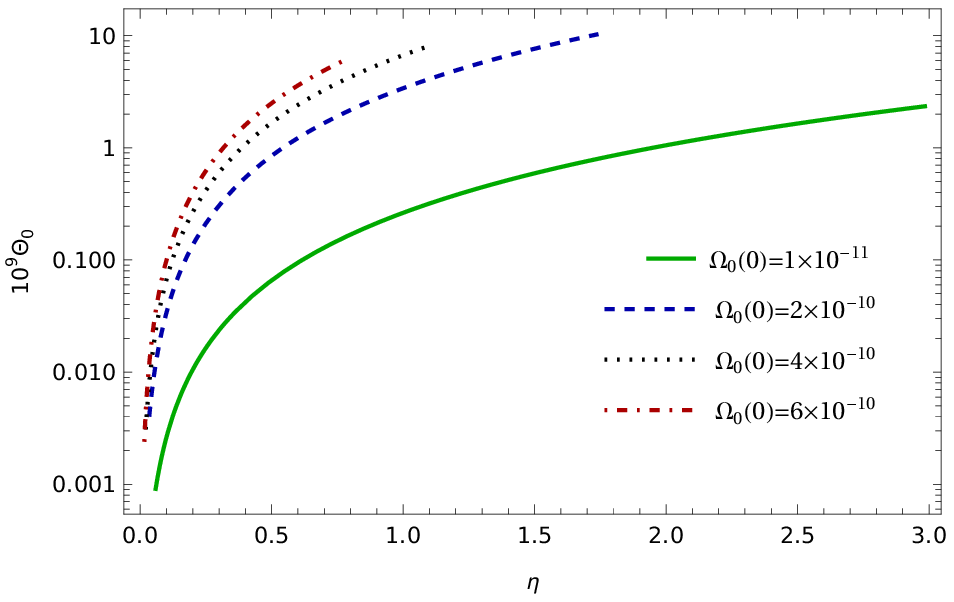}
	\caption{Variations of the metric tensor component $e^\nu$ (upper left panel), of the effective mass $\eta m$ (upper right panel), of  $\Omega_0$ (lower left panel), and of $\Theta_0$ (lower right panel) as a function of $\eta$ for a Weyl black hole in the presence of a temporal component of the Weyl vector only, for $\gamma=100$, $\psi (0)=1\times 10^{-15}$, $\Theta_0(0)=1\times 10^{-20}$, and for different values of $\Omega_0(0)$.
}\label{d13}
\end{figure*}

\paragraph{Varying the initial value $\Omega _0(0)$ of the temporal component of the Weyl vector.}
We consider now Weyl geometric type black holes obtained by varying  only the initial value of $\Omega_0$, while fixing the values at infinity of the other physical and geometrical quantities. The variations of the metric tensor components and of the effective mass are presented in Fig.~\ref{d13}. As one can see from the Figure, the presence of singularities in the metric tensor components indicate the formation of an event horizon, and therefore, the presence of a black hole. The position of the singularities is strongly dependent on the values at infinity of the Weyl vector.
 This dependence is also apparent in the behavior of the effective mass $m$, which increases with increasing $\eta$. The variations of the temporal component of the Weyl vector is represented in the lower right panel in Fig.~\ref{d13}. The Weyl field decreases linearly towards the event horizon, and its values significantly depend on the asymptotic value of the field. The derivative of the Weyl vector is an increasing function of $\eta$ (a decreasing function of $r$).

The explicit values of the positions of the event horizon are presented in Table~\ref{d15}. As one can see from Table~\ref{d15}, there is a very strong dependence of $\eta_{hor}$ on $\Omega _0(0)$.
The event horizon location $\eta _{\textit{hor}}$ decreases with increasing $\Omega_0(0)$, thus indicating an important effect of the Weyl vector on the global properties of the black holes. On the other hand, the physical radius of the event horizon $r_{hor}$ increases from $r_{hor}\approx 0.33$ ($\Omega _0(0)=10^{-11}$), to $r_{hor}\approx 1.27$, for $\Omega _0(0)=6\times 10^{-10}$, respectively.

\begin{table*}[htbp]
	\centering
	\begin{tabular}{|>{\centering}p{50pt}|>{\centering}p{50pt}|>{\centering}p{50pt}|>{\centering}p{50pt}|>{\centering\arraybackslash}p{50pt}|}
		\hline
		$\Omega_{0}(0)$ & $1\times 10^{-11}$ & $2\times 10^{-10}$ & $4\times 10^{-10}$ & $6\times 10^{-10}$\\
		\hline
		$\eta_{\textit{hor}}$ & $2.984$ & $1.742$ & $1.089$ & $0.785$\\
		\hline
	\end{tabular}
	\caption{Variation of the position of the event horizon for a Weyl geometric black hole in the presence of the temporal component of the Weyl vector $\Omega _0$ with respect to  $\Omega_0(0)$. The numerical values of the other quantities are fixed like in the plots in Fig.~\ref{d13}.}
	\label{d15}
\end{table*}

\subsection{Black hole solutions in the presence of both temporal and radial components of the Weyl vector}

We consider now the general case in which both components of the  Weyl vector are nonzero, and therefore $\omega_{\mu}=\left(\omega _0,\omega _1,0,0\right)$. Then, the field  equations (\ref{sec31a}) and (\ref{sec31u}) can be  reformulated in a dimensionless form as a first order dynamical system given by
\begin{widetext}
	\begin{eqnarray}\label{e3}
		\frac{dm}{d\eta} &=& \frac{1}{\eta^4 \psi^2 (\zeta + \eta \psi)} \Big(12e^{-\nu}\Theta_0^2\psi(1-\eta m)(2\zeta+\eta\psi)-\zeta^2(1-\eta m)(4\zeta+5\eta\psi)+\zeta\psi^2(\eta^3m+2\psi)+\eta\psi^4 \Big),\\
		\frac{d\nu}{d\eta} &=& \frac{1}{\eta^2 \psi (1-\eta m) (\zeta + \eta \psi)} \Big( \zeta(1-\eta m)(3\zeta+4\eta \psi)-(\eta^3m+\psi)\psi^2-3e^{-\nu}\psi\big(4(1-\eta m)\Theta_0^2+\psi\Omega_0^2\big)\Big),\\
		\frac{d\psi}{d\eta} &=& - \frac{ 2\zeta}{\eta^2},\\
		\frac{d\zeta}{d\eta} &=& -\frac{\zeta}{\eta^3\psi^2} \bigg[ 2\zeta(\zeta+2\eta\psi)-12e^{-\nu}\psi\Theta_0^2-\frac{\psi^2}{1-\eta m}\big(\eta^2(2-\eta m)-1\big) \bigg],\\
		\frac{d\Omega_0}{d\eta} &=& - \frac{2\gamma \Theta_0}{\eta^2},\\
		\frac{d\Theta_0}{d\eta} &=&  \frac{1}{2\eta^3\psi^2(1-\eta m)(\zeta + \eta\psi)} \Big( 2\Theta_0(1-\eta m)(12e^{-\nu}\zeta\Theta_0^2\psi-2\zeta^3-\eta\zeta^2\psi+2\eta^3\psi^3)-6e^{-\nu}\eta\Theta_0\psi^3\Omega_0^2\nn&&+2\eta^2\Theta_0\zeta\psi^2(4-3\eta m)+\gamma\psi^3\Omega_0(\zeta+\eta\psi)+2\Theta_0\zeta\psi^3 \Big),
	\end{eqnarray}
\end{widetext}

\subsubsection{The initial conditions}

For this model the general solution of the system depends on six parameters $\left(\gamma, m(0),\psi(0),\zeta(0),\Omega_{0}(0),\Theta_{0}(0)\right)$, representing the numerical values of the coupling constant $\gamma$, and of the initial conditions at infinity. As we have already discussed, the scalar field and the $\omega _1$ component have an oscillatory behavior for large $r$, and hence at infinity they do not converge to a single value. On the other hand, the $\omega _0$ component of the field becomes an arbitrary integration constant at infinity. However, we will select only the set of initial conditions that is consistent with the previously analyzed particular cases, and which lead to astrophysical black holes similar to the general relativistic Schwarzschild ones, with similar values for the positions of the event horizon. These conditions imply again very small initial values of the scalar and Weyl vector fields, and of their derivatives.

\subsubsection{Numerical results}

We consider now the results of the numerical integration of the static, spherically symmetric field equations of Weyl geometric gravity, obtained by varying the numerical values of the coupling constant, and of the initial conditions at infinity.

\begin{figure*}[htbp]
	\centering
	\includegraphics[width=8.5cm]{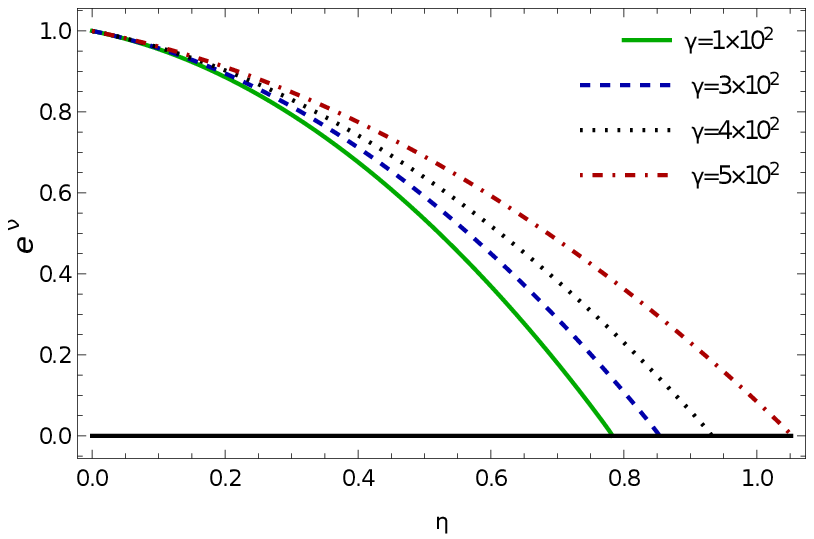}
	\includegraphics[width=8.5cm]{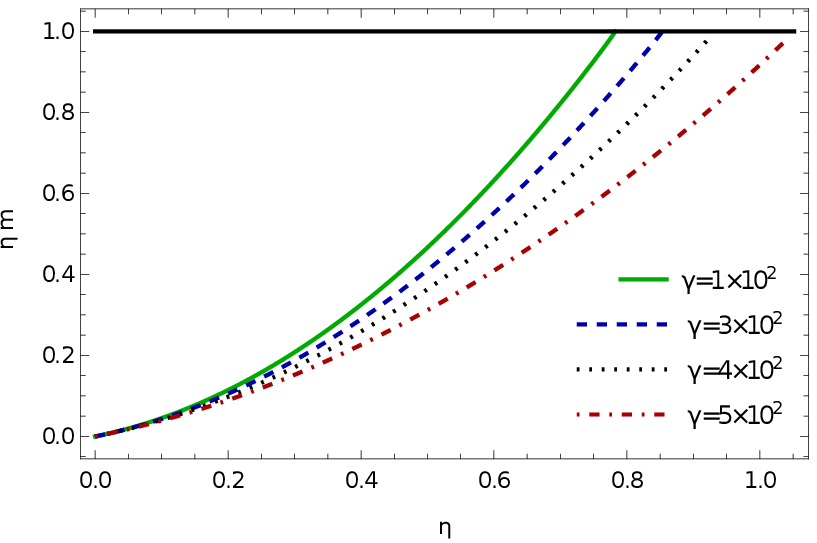}
\newline
\newline
	\includegraphics[width=8.5cm]{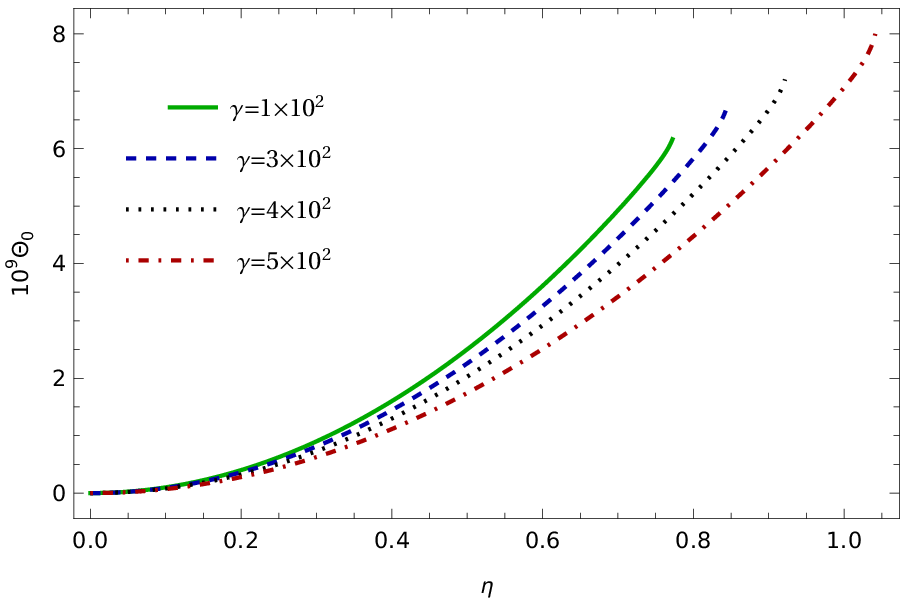}
		\includegraphics[width=8.5cm]{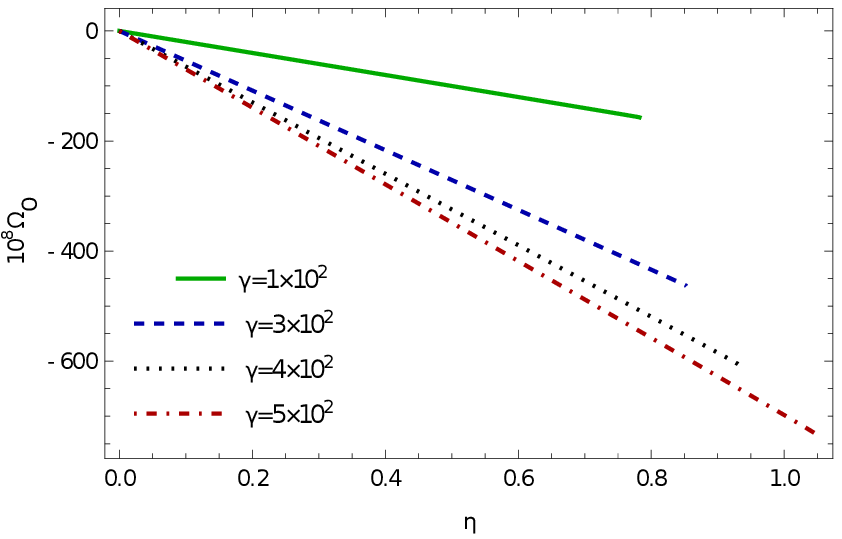}
\newline
\newline
			\includegraphics[width=8.5cm]{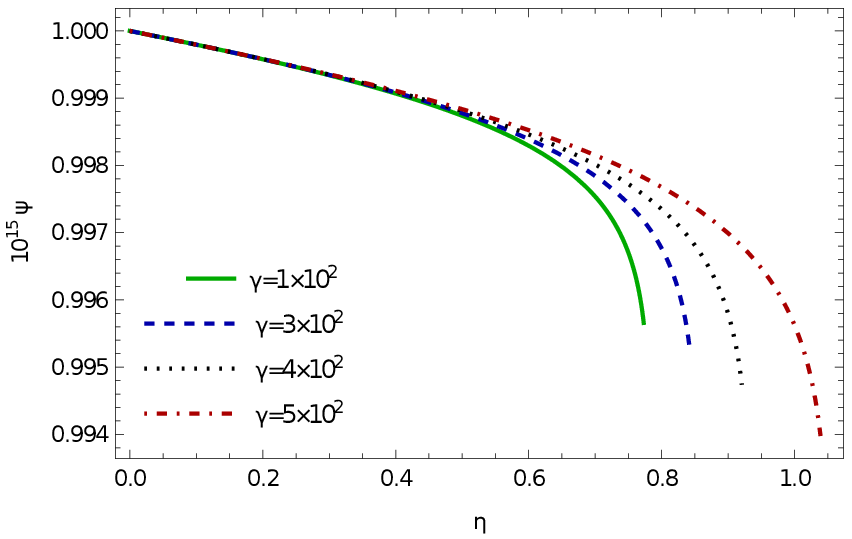}
				\includegraphics[width=8.5cm]{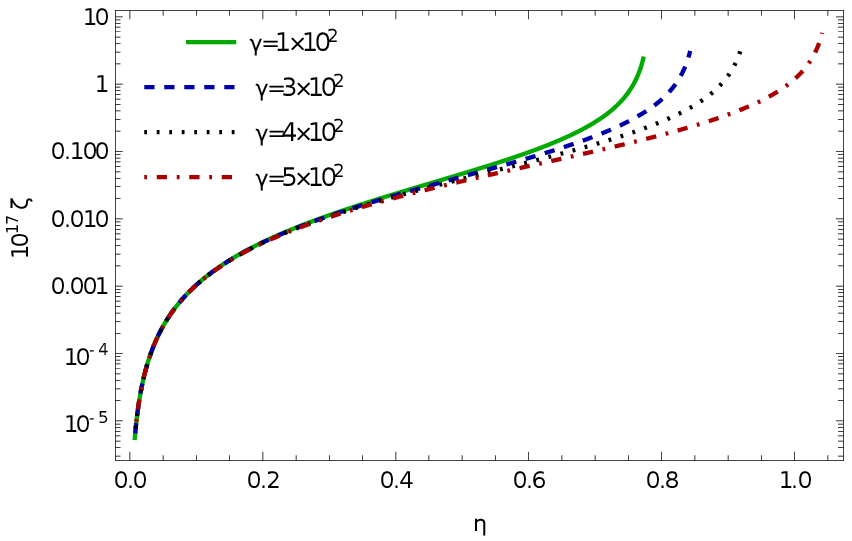}
	\caption{Variation of the metric tensor component  $e^\nu$ (upper left panel), of $\eta m$ (upper right panel),  of $\Theta_0$ (middle left panel), of $\Omega_0$ (middle right panel), of $\psi$ (lower left panel) and of $\zeta$ (lower right panel) as a function of $\eta$ for Weyl geometric black holes in the presence of both temporal and radial components of the Weyl vector for $\psi (0)=1\times 10^{-15}$, $\zeta (0)= 1\times 10^{-28}$, $\Omega_0(0)=1\times 10^{-11}$, $\Theta_0(0)=1\times 10^{-18}$, and for different values of the coupling constant $\gamma$, presented in the legends of the Figures.
}
\label{d19a}
\end{figure*}

\paragraph{Varying the coupling constant $\gamma$.}
We consider first classes of numerical black hole solutions obtained by varying the coupling constant $\gamma$ only, while keeping the initial conditions at infinity of the physical and geometrical quantities fixed.  The variations of the metric tensor components and of the effective mass are represented for different values of $\gamma$, in Fig.~\ref{d19a}. The formation of an event horizon is indicated by the presence of the singularities in the metric tensor components.
 For the adopted values of the coupling constant there is a strong dependence of the position of $\eta_{hor}$ on $\gamma$. On the other hand, the effective mass $m$ of the Weyl black hole is a monotonically increasing function of $\eta$, also showing a stronger dependence on the Weyl couplings.

\begin{table*}[htbp]
	\centering
	\begin{tabular}{|>{\centering}p{50pt}|>{\centering}p{50pt}|>{\centering}p{50pt}|>{\centering}p{50pt}|>{\centering\arraybackslash}p{50pt}|}
		\hline
		$\gamma$ & $100$ & $300$ & $400$ & $500$\\
		\hline
		$\eta_{\textit{hor}}$ & $0.782$ & $0.853$ & $0.932$ & $1.051$\\
		\hline
	\end{tabular}
	\caption{Location of the event horizon of the Weyl geometric black holes in the presence of both temporal and radial components of the Weyl vector, for $\psi (0)=1\times 10^{-15}$, $\zeta (0)= 1\times 10^{-28}$, $\Omega_0(0)=1\times 10^{-11}$, and $\Theta_0(0)=1\times 10^{-18}$, respectively, and for different values of the coupling constant $\gamma$.}
	\label{d21a}
\end{table*}

The positions of the event horizon of the Weyl black holes for different  values of $\gamma$ and fixed initial conditions are presented in Table~\ref{d21a}. The modification of the coupling constant on a large range of values has only a relatively weak effect on the position of the event horizon, as compared to the Schwarzschild case. However, the increase of $\gamma$ leads to a decrease of the value of the physical radius of the event horizon, from $r_{hor}\approx 1.28$ ($\gamma =100$), to $r_{hor}\approx 0.95$, corresponding to $\gamma =500$.

\begin{figure}[htbp]
	\centering
	\includegraphics[width=8.5cm]{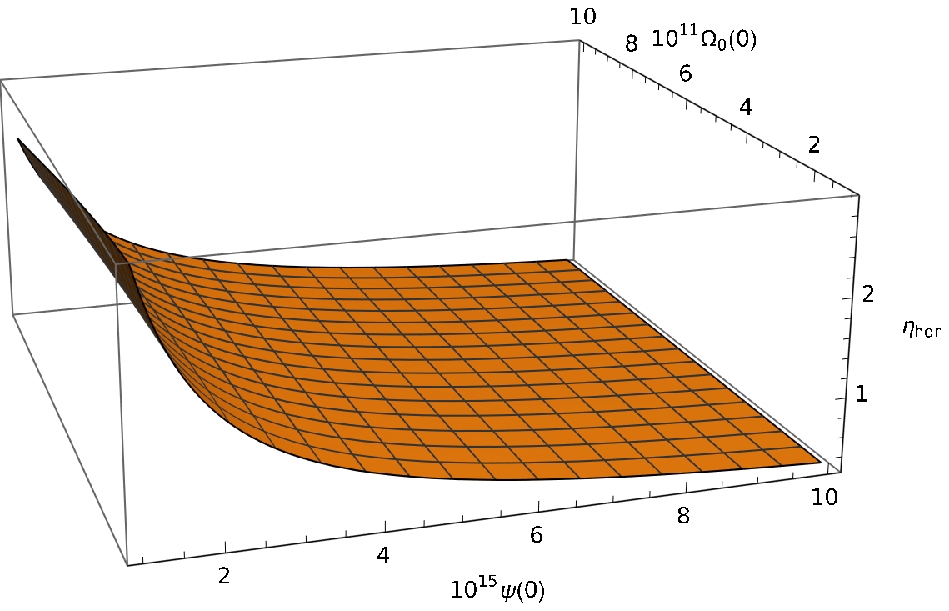}
	\caption{Variation of the position of the event horizon of the Weyl geometric black hole in the presence of both temporal and radial components of the Weyl vector as a function of the initial values of the scalar field $\psi_0(0)$ and of the temporal component of the Weyl vector $\Omega_0(0)$. To obtain the Figure we have assumed $\gamma=10^2$, $\Theta_{0}(0)=10^{-20}$, and $\zeta(0)=10^{-28}$, respectively.} \label{3D1}
\end{figure}

\paragraph{Varying the initial values of the scalar field $\psi (0)$ and of the temporal component of the Weyl vector $\Omega _0(0))$.} We consider now numerical black hole solutions obtained by varying the initial values of the scalar field, and of the temporal component of the Weyl vector, with all the other quantities fixed. The variation of the position of the event horizon is presented in Fig.~\ref{3D1}.

The positions of the event horizon of the Weyl black holes for different values of $\psi(0)$, and fixed initial conditions and values of the coupling constant $\gamma$ are presented in Table~\ref{tabcasecpsi}. As one can see from the Table, very small variations of the initial values of the scalar field can induce very significant changes in the position of the event horizon. Interestingly enough, increasing the value of the scalar field at infinity leads to a decrease of the inverse of the radius of the horizon of the black hole, leading to smaller values of $\eta_{hor}$, as compared to the Schwarzschild case. The physical radius of the event horizon increases with increasing $\psi(0)$, from $r_{hor}\approx 0.33$, corresponding to $\psi(0)=10^{_15}$, to $r_{hor}\approx 1.33$ for $\psi(0)=4\times 10^{-15}$, respectively.

\begin{table*}[htbp]
	\centering
	\begin{tabular}{|>{\centering}p{50pt}|>{\centering}p{50pt}|>{\centering}p{50pt}|>{\centering}p{50pt}|>{\centering\arraybackslash}p{50pt}|}
		\hline
		$\psi(0)$ & $1\times 10^{-15}$ & $2\times 10^{-15}$ & $3\times 10^{-15}$ & $4\times 10^{-15}$\\
		\hline
		$\eta_{\textit{hor}}$ & $2.943$ & $1.493$ & $0.998$ & $0.749$\\
		\hline
	\end{tabular}
	\caption{Location of the event horizon of the Weyl geometric black holes for $\gamma=100$, $\zeta (0)= 1\times 10^{-28}$, $\Omega_0(0)=1\times 10^{-11}$, and $\Theta_0(0)=1\times 10^{-20}$, respectively, and for different values of $\psi(0)$.}
	\label{tabcasecpsi}
\end{table*}

The positions of the event horizon of the Weyl black holes for different values of $\Omega_0(0)$, and fixed initial conditions are presented, for $\gamma =100$,  in Table~\ref{tabcasecomega}. Increasing the values of $\Omega_0(0)$ leads again to a decrease of the inverse of the radius of the Weyl geometric black hole, and to an increase of the physical radius of the event horizon $r_{hor}=1/\eta_{hor}$. While for $\Omega_0(0)=10^{-11}$, the event horizon of the black hole is located at $\eta_{hor}\approx 3$, $r_{hor}=1/\eta_{hor}\approx 0.33$,  for $\Omega _0(0)=6\times 10^{-10}$, the position of the event horizon of the black hole has a value of  $\eta _{hor}\approx 0.8$, leading to a physical radius of the event horizon of the Weyl black hole larger than the value of the horizon of the Schwarzschild black hole, $r_{hor}=1/\eta _{hor}\approx 1.25$.

\begin{table*}[htbp]
	\centering
	\begin{tabular}{|>{\centering}p{50pt}|>{\centering}p{50pt}|>{\centering}p{50pt}|>{\centering}p{50pt}|>{\centering\arraybackslash}p{50pt}|}
		\hline
		$\Omega_0(0)$ & $1\times 10^{-11}$ & $2\times 10^{-10}$ & $4\times 10^{-10}$ & $6\times 10^{-10}$\\
		\hline
		$\eta_{\textit{hor}}$ & $2.943$ & $1.729$ & $1.084$ & $0.783$\\
		\hline
	\end{tabular}
	\caption{Location of the event horizon of the Weyl geometric black holes in the presence of both temporal and radial components of the Weyl vector for $\gamma=100$, $\zeta (0)= 1\times 10^{-28}$, $\psi(0)=1\times 10^{-15}$, and $\Theta_0(0)=1\times 10^{-20}$, respectively, and for different values of $\Omega_0(0)$.}
	\label{tabcasecomega}
\end{table*}

\begin{figure}[htbp]
	\centering
	\includegraphics[width=8.5cm]{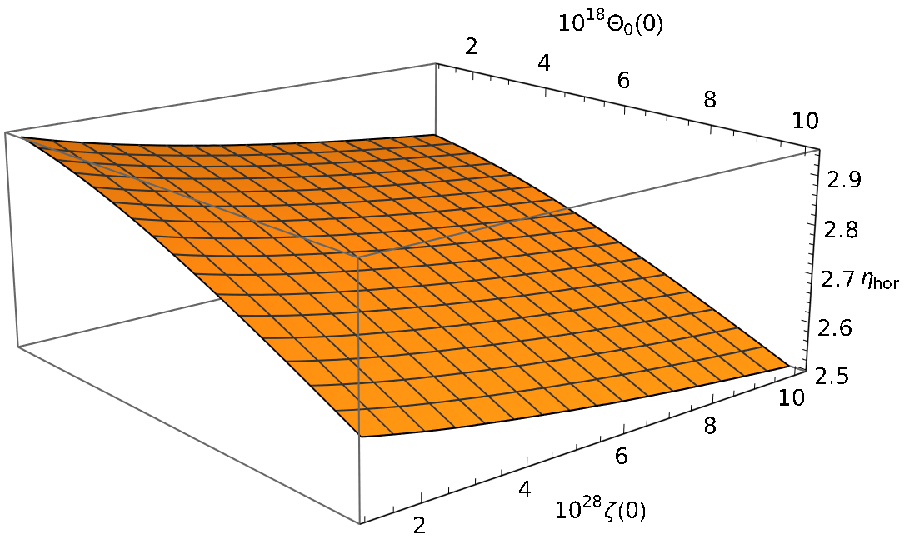}
	\caption{Variation of the position of the event horizon of the Weyl geometric black hole in the presence of both temporal and radial components of the Weyl vector as a function of the initial values of the derivatives of the scalar field $\psi_0(0)$ and of the temporal component of the Weyl vector $\Theta_0(0)$. To obtain the Figure we have assumed $\gamma=10^2$, $\psi(0)=10^{-15}$, and $\Omega_0(0)=10^{-11}$, respectively. } \label{3D2}
\end{figure}

 \paragraph{Varying the initial values of the derivatives of the scalar field  $\zeta (0)$, and of the temporal component of the Weyl vector, $\Theta _0 (0)$.}
 We consider now models obtained by varying the initial values of the derivatives of the scalar field, $\zeta(0)$, and of the derivative of the temporal component of the Weyl vector $\Theta _0(0)$ only, while keeping the initial conditions at infinity of the other physical and geometrical quantities fixed.  The variations of the positions of the position of the event horizon are represented in Fig.~\ref{3D2}.   Our results indicate the formation of an event horizon, due to the presence of the singularities in the metric tensor components.

The positions of the event horizon of the Weyl black holes for different $\zeta(0)$ values and fixed initial conditions are presented in Table~\ref{tabcaseczeta}.

\begin{table*}[htbp]
	\centering
	\begin{tabular}{|>{\centering}p{50pt}|>{\centering}p{50pt}|>{\centering}p{50pt}|>{\centering}p{50pt}|>{\centering\arraybackslash}p{50pt}|}
		\hline
		$\zeta(0)$ & $1\times 10^{-28}$ & $1\times 10^{-27}$ & $3\times 10^{-27}$ & $6\times 10^{-27}$\\
		\hline
		$\eta_{\textit{hor}}$ & $2.943$ & $1.668$ & $1.023$ & $0.727$\\
		\hline
	\end{tabular}
	\caption{Location of the event horizon of the Weyl geometric black holes in the presence of both temporal and radial components of the Weyl vector, for $\gamma=100$, $\Omega_0(0)= 1\times 10^{-11}$, $\psi(0)=1\times 10^{-15}$, $\Theta_0(0)=1\times 10^{-20}$, and for different values of $\zeta(0)$.}
	\label{tabcaseczeta}
\end{table*}

With the increase of $\zeta(0))$, the position of the event horizon increases, in the physical radial distance variable $r$, from $r_{hor}\approx 0.34$, to $r_{hor}\approx 1.375$, indicating the possibility of creation of both smaller and larger black holes as compared to their general relativistic counterparts.

\section{Thermodynamics of the Weyl geometric black holes}\label{sect4}

In our analysis of the vacuum field equations of the linear representation of the quadratic Weyl geometric gravity we have assumed that the metric tensor components $e^{\nu}$ and $e^{\lambda}$, and consequently, the effective mass function $m$ all depend only on the radial coordinate $r$. Therefore, the geometry of the spacetime is static, and, moreover, a timelike Killing vector $t^{\mu}$ always does exist \cite{Wald,Kunst}.

\paragraph{Surface gravity of the Weyl black holes.}

For a static black hole that has a Killing horizon the surface gravity $\tilde{\kappa}$ is generally defined according to \cite{Wald,Kunst}
\be
t^{\mu}\nabla _{\mu}t^{\nu}=t^{\nu}\tilde{\kappa}.
\ee
By adopting  a static, spherically symmetric black hole geometry given by
\be
ds^2=-\tilde{\sigma} ^2 (r)f(r)c^2dt^2+\frac{dr^2}{f(r)}+r^2d\Omega ^2,
\ee
where $\tilde{\sigma}$ and $f$ are functions of the radial coordinate only, and after suitable normalizing the Killing vector $t^{\mu}$ as $t^{\mu}=\left(1/\tilde{\sigma}_{\infty},0,0,0\right)$, the surface gravity of the black hole is given by \cite{Kunst}
\be
\tilde{\kappa}=\left(\frac{\tilde{\sigma} _{hor}}{\tilde{\sigma} _{\infty}}\right)\frac{c^4}{4GM_{hor}}\left.\left[1-\frac{2GM'(r)}{c^2}\right]\right|_{hor},
\ee
where the subscript {hor} specifies that the calculation of all physical quantities must be done  on the outer
apparent horizon of the black hole. If $\tilde{\sigma} \equiv 1$, and $M={\rm constant}$, then the expression of the surface gravity of a Schwarzschild black hole, $\tilde{\kappa}=c^4/4GM_{\textit{hor}}$ \cite{Wald}, is reobtained.

\subsubsection{Hawking temperature of the quadratic Weyl geometric black holes.}

\begin{figure*}[!htb]
	\centering
	\includegraphics[width=8.5cm]{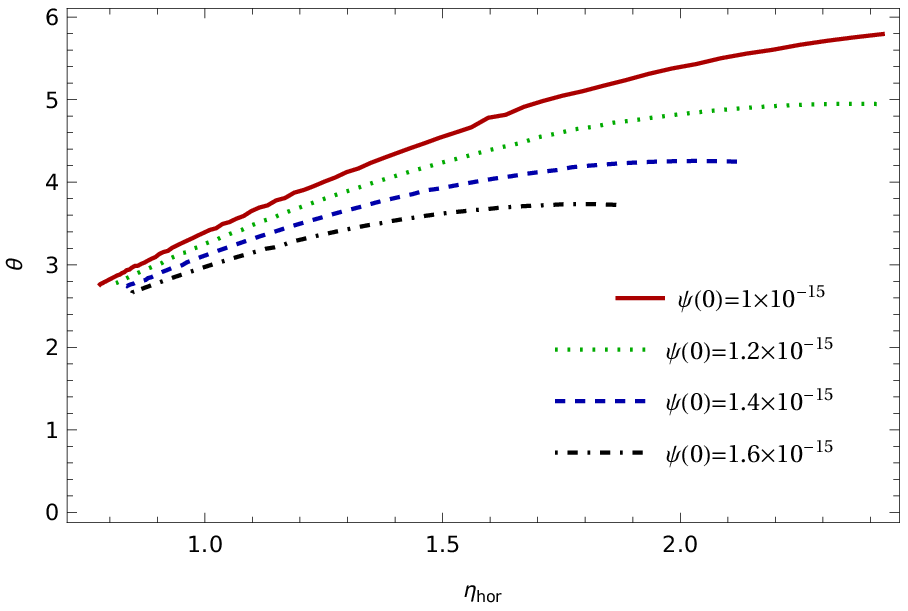}
	\includegraphics[width=8.5cm]{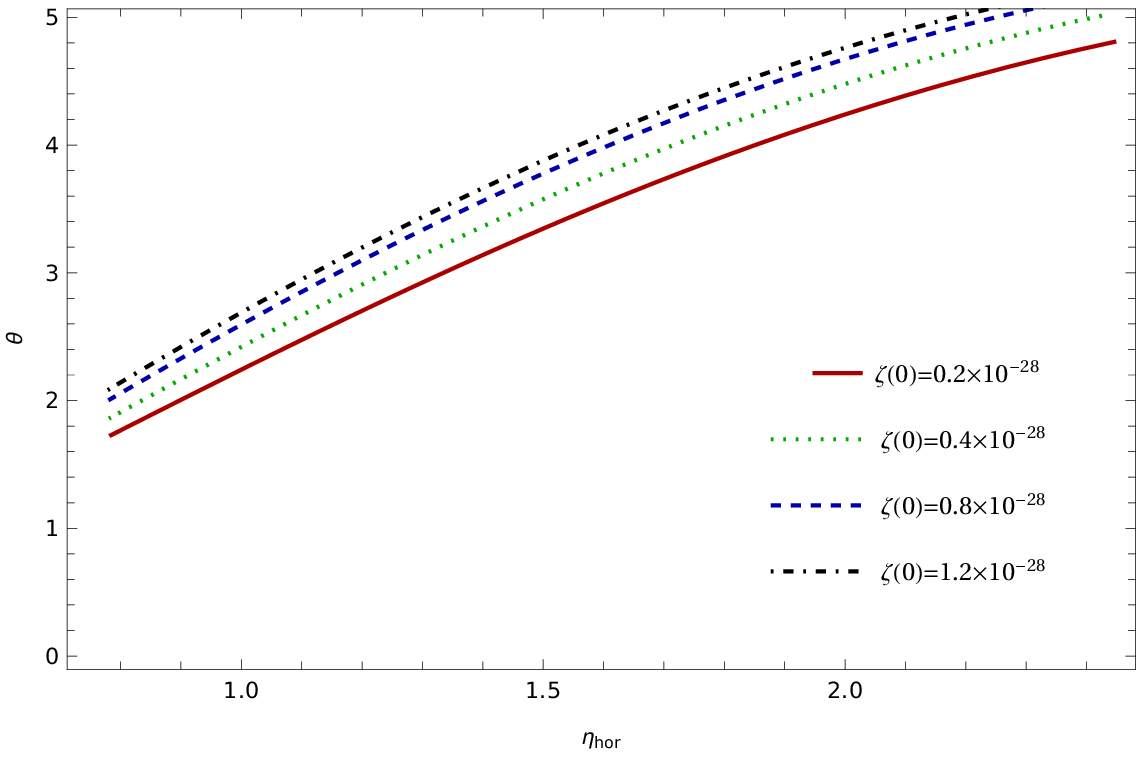}
	\caption{Variation of the dimensionless black hole horizon temperature for a Weyl geometric gravity black hole in the presence of both temporal and radial components of the Weyl vector for $\theta$ for $\Theta_{0}(0)=10^{-18}$, $\Omega_{0}(0)=10^{-11}$, and $\zeta(0)=10^{-28}$, respectively, and different values of $\psi(0)$ (left panel),  and for $\psi(0)=10^{-15}$, and different values of $\zeta(0)$ (right panel). In both panels, the values of $\gamma$ are varied continuously.}
	\label{fi1}
\end{figure*}

The Hawking temperature $T_{BH}$ of the black hole is defined according to  \cite{Wald,Kunst}
\be
T_{BH}=\frac{\hbar}{2\pi ck_B} \tilde{\kappa},
\ee
where by $k_B$ we have denoted Boltzmann's constant. In the system of dimensionless variables defined in Eq.~(\ref{c22}),  the temperature of the black hole is obtained as
\be
T_{BH}=T_H\frac{1}{m\left(\eta _{\textit{hor}}\right)}\left.\left(1+\eta ^2\frac{dm}{d\eta}\right)\right|_{\eta =\eta _{hor}},
\ee
where we have introduced the notation
\be
T_H=\frac{\hbar c^3}{8\pi Gk_BnM_{\odot}},
\ee
corresponding to the Hawking temperature of the standard general relativistic Schwarzschild black hole. The variation of the dimensionless horizon temperature
\be
\theta =\frac{T_{BH}}{T_H},
\ee
of the black holes in quadratic Weyl geometric gravity is represented, for selected values of the model parameters, and for the general case, with the Weyl vector possessing both temporal and radial components, in Fig.~\ref{fi1}.

As one can see from Fig.~\ref{fi1}, , the temperature of the horizon of the Weyl black holes is generally higher than that of their general relativistic counterparts. This is also related to the wider range of event horizon positions, as compared to the Schwarzschild black hole case. The Hawking temperature is a monotonically increasing function of the position of the event horizon, and it has a strong dependence on the initial values of the scalar field, and of its derivatives.,

\subsubsection{Specific heat of the Weyl black holes. } 
\begin{figure*}[!htb]
	\centering
	\includegraphics[width=8.5cm]{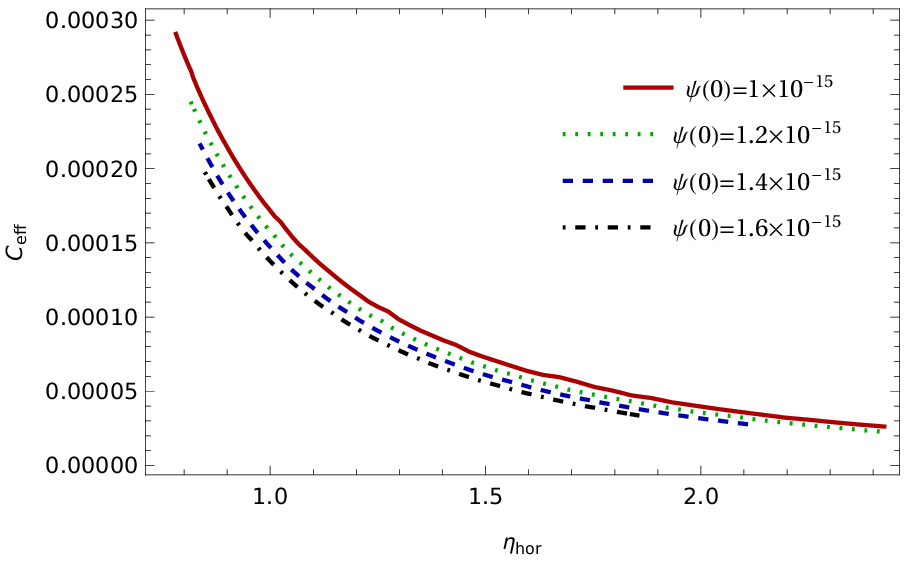}
	\includegraphics[width=8.5cm]{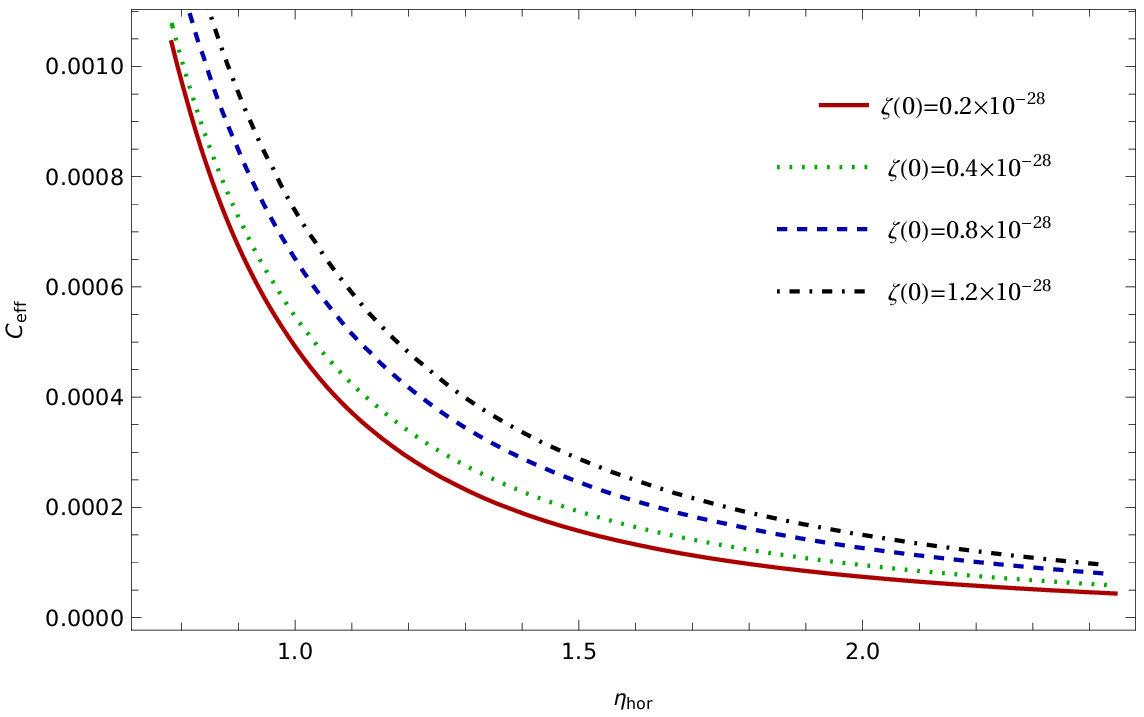}
	\caption{Variation of the dimensionless specific heat $C_{eff}=C_{BH}/C_H$ of the Weyl geometric black holes in the presence of both temporal and radial components of the Weyl vector for $\Theta_{0}(0)=10^{-18}$, $\Omega_{0}(0)=10^{-11}$, and $\zeta(0)=10^{-28}$, respectively, and  for different values of $\psi(0)$ (left panel) and for $\psi(0)=10^{-15}$, and  different values of $\zeta(0)$ (right panel). In both panels, the values of $\gamma$ are varied continuously.}
	\label{fi2}
\end{figure*}

Another important physical quantity characterizing the thermodynamic properties of the black holes, is their specific heat $C_{BH}$, which can be obtained from the definition  \cite{Wald,Kunst},
\bea
C_{BH}&=&\frac{dM}{dT_{BH}}=\left.\frac{dM}{dr}\frac{dr}{dT_{BH}}\right|_{r=r_{hor}}
\nonumber\\
&=&\frac{nM_{\odot}}{T_H} \left.\frac{dm\left(\eta \right)}{d\eta}\frac{d\eta}{d\theta }\right|_{\eta =\eta_{hor}}.
\eea
where we have denoted $C_H=nM_{\odot}/T_H$. The variations of the dimensionless specific heat $C_{eff}=C_{BH}/C_H$ of the Weyl black holes as a function of the dimensionless horizon radius in quadratic Weyl geometric gravity are represented, for some selected values of the model parameters, in Fig.~\ref{fi2}.

Similarly to the Hawking temperature, the numerical values of $C_{eff}$  do depend on the initial conditions at infinity of the scalar field, and of its derivative. The specific heat of the Weyl geometric black holes is a rapidly decreasing function of $\eta_{hor}$.

\subsubsection{The entropy of the Weyl geometric black holes.} 
\begin{figure*}[!htb]
	\centering
	\includegraphics[width=8.5cm]{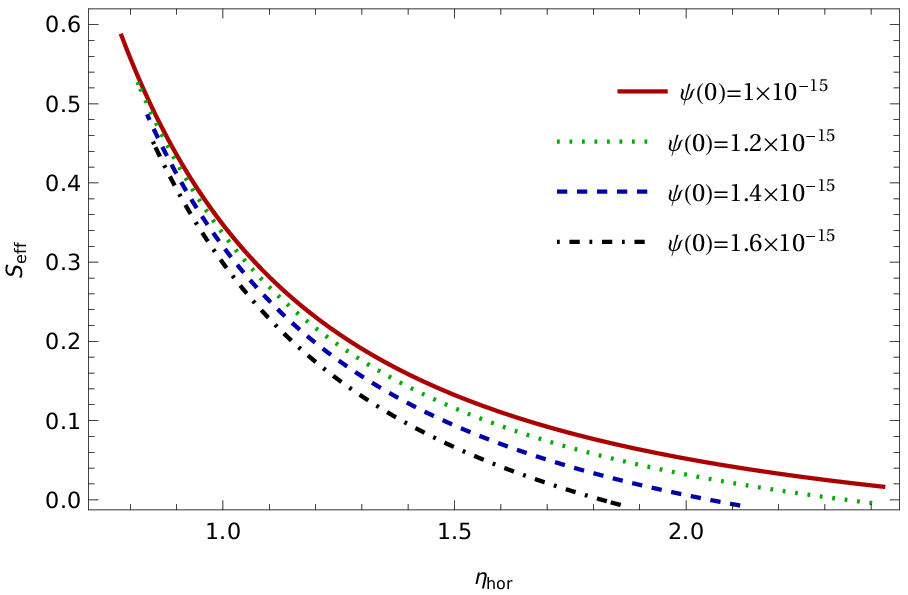}
	\includegraphics[width=8.5cm]{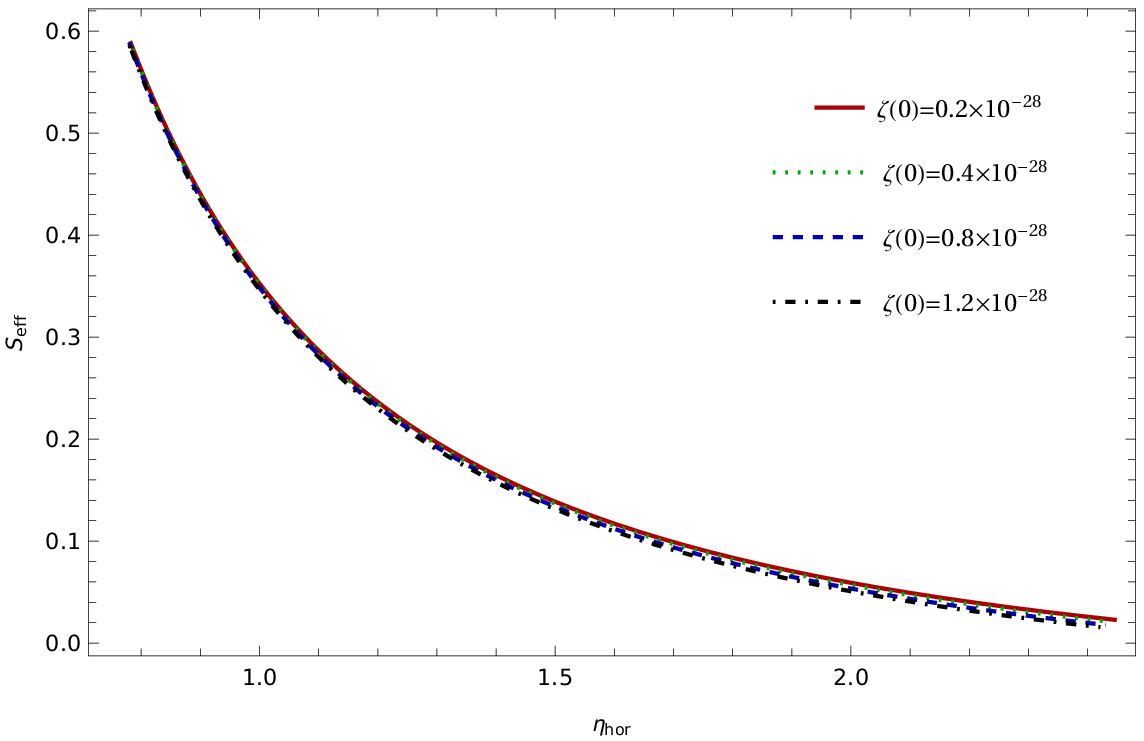}
	\caption{Variation of the dimensionless entropy $S_{eff}=S_{BH}/C_H$ of the Weyl geometric black holes in the presence of both temporal and radial components of the Weyl vector as a function of the position of the event horizon for  $\Theta_{0}(0)=10^{-18}$, $\Omega_{0}(0)=10^{-11}$ and $\zeta(0)=10^{-28}$, respectively, and for different values of $\psi(0)$ (left panel), and for $\psi(0)=10^{-15}$, and different values of $\zeta(0)$ (right panel). In both panels, the values of $\gamma$ are varied continuously.}
	\label{fi3}
\end{figure*}

The entropy $S_{BH}$ of the Weyl geometric black hole is obtained generally as  \cite{Wald,Kunst}
\bea
S_{BH}&=&\int_{\infty}^{r_{\textit{hor}}}{\frac{dM}{T_{BH}}}=\int_{\infty}^{r_{\textit{hor}}}{\frac{1}{T_{BH}}\frac{dM}{dr}dr}.
\eea
In the set of the dimensionless variables considered in the present study we have
\bea
S_{BH}\left(\eta_{hor}\right)=C_H\int_0^{\eta_{hor}}{\frac{1}{\theta \left(\eta \right)}\frac{dm\left(\eta\right)}{d\eta}d\eta}.
\eea
In the following we denote $S_{eff}=S_{BH}\left(\eta_{hor}\right)/C_H$.
The variation  as a function of the dimensionless horizon radius  $\eta_{hor}$ of the entropy $S_{eff}$ of the black holes in the Weyl geometric gravity theory with only radial component in Weyl vector is represented, as a function of the dimensionless horizon radius, for different values of $\psi (0)$ and $\zeta (0)$,  in Fig.~\ref{fi3}.

The black hole entropy decreases with the increase of the radius of the event horizon. While there is a significant dependence on the values at infinity of the scalar field, the dependence of the entropy on the derivatives of the field is weak, and no significant change in its values occurs due to the modifications of $\zeta (0)$.

\subsubsection{The luminosity of the Weyl black holes.} The formation and the evaporation of a spherically symmetric black hole in conformal gravity was investigated in \cite{C3}, by considering the collapse of a spherically symmetric thin shell of radiation, leading to the formation of a singularity-free non-rotating black hole. The black hole has the same Hawking temperature as a Schwarzschild black hole with the same mass, and it completely evaporates either in a finite or in an infinite time, depending on the ensemble. The evaporation process of a spherical neutral AdS black hole in four-dimensional conformal gravity, where the equations of states are branched, was investigated in \cite{C4}.

The luminosity of the Weyl geometric black holes, due to the Hawking evaporation processes, can be obtained according to the relation  \cite{Wald,Kunst},
\be
L_{BH}=-\frac{dM}{dt}=-\sigma A_{BH}T_{BH}^4,
\ee
where $\sigma $ is a parameter that depends on the considered  model, while $A_{BH}=4\pi r_{hor}^2$ is the area of the event horizon. For the black hole evaporation time $\tau $ we thus have
\bea
\hspace{-0.8cm}\tau &=&\int_{t_{\it{in}}}^{t_{\it{fin}}}{dt}=-\frac{1}{4\pi \sigma}\int_{t_{\it{in}}}^{t_{\it{fin}}}{\frac{dM}{r_{\textit{hor}}^2T_{BH}^4}},
\eea
where $t_{{\it in}}$ and $t_{{\it fin}}$ represent the initial and the final times considered for the evaporation process.
Equivalently, we can obtain the evaporation times of the Weyl geometric black hole in the form
\bea
\tau _{BH}\left(\eta_{hor}\right)= \tau _H\int_0^{\eta_{hor}}{\frac{1}{\eta ^2\theta ^4\left(\eta\right)}\frac{dm\left(\eta\right)}{d\eta }d\eta},
\eea
 where we have denoted
\be
\tau _H=\frac{c^4}{8\pi G^2\sigma nM_{\odot}T_{BH}^4}.
\ee

\begin{figure*}[!htb]
	\centering
	\includegraphics[width=8.5cm]{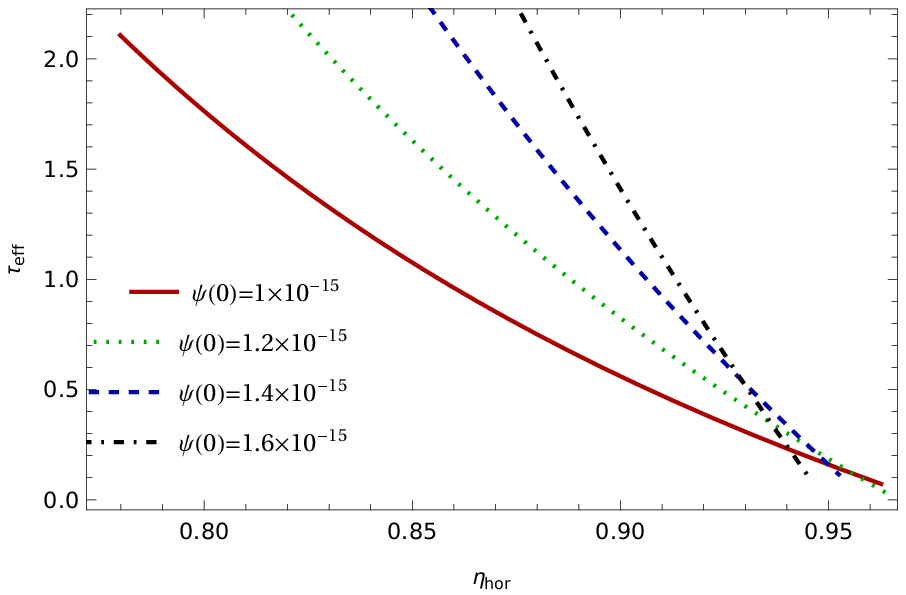}
	\includegraphics[width=8.5cm]{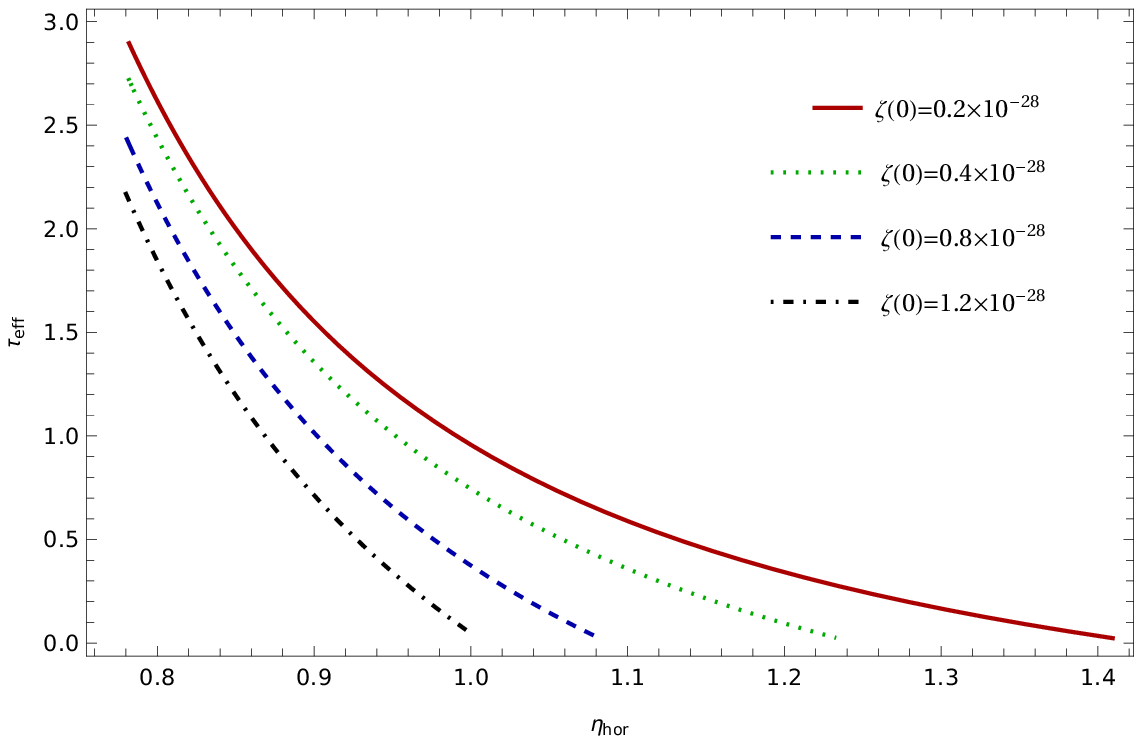}
	\caption{Variation of the evaporation time $\tau _{eff}=\tau _{BH}/\tau_H$ of the Weyl geometric black holes in the presence of both temporal and radial components of the Weyl vector as a function of the position of the event horizon for $\Theta_{0}(0)=10^{-18}$, $\Omega_{0}(0)=10^{-11}$ and $\zeta(0)=10^{-28}$, and for different values of $\psi(0)$ (left panel), and for $\psi(0)=10^{-15}$, and different values of  $\zeta(0)$ (right panel). In both panels, the values of $\gamma$ are varied continuously.}
	\label{fi4}
\end{figure*}

The variations of the dimensionless Hawking evaporation time $\tau _{eff}=\tau_{BH}/\tau_H$ of black holes in the Weyl geometric gravity as a function of the position of the event horizon are  represented in Fig.~\ref{fi4}. In this Figure we have varied the values of $\gamma$ continuously, and kept the values of the other parameters fixed.

The Hawking evaporation times of the Weyl geometric black holes show a strong dependence on the initial conditions at infinity of both the scalar field, and of its derivative. They are rapidly decreasing functions of the radius of the event horizon, and they become negligibly small for enough massive black holes.

\section{Discussions and final remarks}\label{sect5}

The idea that conformal symmetry is an exact, but spontaneously broken symmetry of nature \cite{H1,H2}, has many attractive features. If Einstein's field equations are reformulated by strictly imposing conformal symmetry, then the conformal component of the metric field can be treated as a dilaton field with only renormalizable interactions. Hence, this condition imposes some strong constraints on the gravitational theories, and they are equivalent to demanding regularity of the action as the
dilaton field variable tends to zero. Moreover, such a procedure can turn a black hole into a regular, topologically trivial soliton with no singularities, horizons or firewalls \cite{H2}.

In the present work, we have considered the possible existence of black hole type structures in the framework of the simplest conformally invariant gravitational theory, constructed ab initio in a Weyl geometry. The simplest such gravitational action contains the quadratic Weyl scalar and the electromagnetic type scalar, defined in a way similar to the Faraday tensor in standard electromagnetic theory, with the Weyl vector playing the role of the electromagnetic four-potential. This theory intrinsically contains a scalar degree of freedom, and, once reformulated in the standard Riemannian geometry, it can be linearized in the Ricci scalar. Hence, geometric conformally invariant Weyl gravity is equivalent in the Riemannian space to a scalar-vector-tensor theory, in which, besides the metric tensor, the gravitational properties are determined by a scalar and a tensor field, with both having purely geometric origins.

To investigate the physical properties of the scalar-vector-tensor Weyl geometric theory, in the present work we have considered one of the simplest possible physical and geometrical situations, namely, the case of the vacuum static and spherically symmetric systems. Even by adopting this simple theoretical model, the gravitational field equations are extremely complicated. Hence,  in order to obtain solutions of the field equations one must extensively use numerical methods. In order to do so in a computationally efficient way, one must first rewrite the static spherically symmetric Weyl geometric gravitational field equations in vacuum in a dimensionless form suitable for numerical integration. This can be achieved by introducing as an independent variable the inverse of the radial coordinate. Moreover, we have reformulated the gravitational field equations as a first order dynamical system. In this representation the numerical integration procedure is significantly simplified. In analogy with the Schwarzschild black hole solution of standard general relativity we have also represented the metric tensor component $e^{\lambda}$ in terms of an effective mass. Hence, in order to obtain a numerical solution of the field equations of the Weyl geometric gravitational theory we need to give the initial conditions at infinity of the metric tensor components, of the scalar and of the vector fields, and of their derivatives, respectively. As for the metric tensor components, we have assumed the condition of the asymptotically flat geometry, while the numerical values at infinity of the components of the scalar and vector fields, and of their derivatives, have been chosen in an arbitrary way. We consider that the presence of a singularity in the field equations, or, more precisely, of a singular point in the behavior of the metric tensor components, indicates the formation of an event horizon and, consequently, indicates the presence of a black hole type object. The total mass of the black hole can be determined from the effective mass appearing in the radial metric tensor component, and it can be interpreted as containing the standard (baryonic) mass of the black hole to which  the contributions from the scalar and  Weyl type vector fields are added.

Generally, in spherical symmetry, only two components of the Weyl vector do survive, which are the temporal and the radial components, respectively. Based on this result, we have considered the solutions of the gravitational field equations of the quadratic Weyl geometric gravity in three cases, corresponding to the Weyl vector having only a radial component only, a temporal component only, and in the presence of both components.
As a first general conclusion of our study, we have detected the formation of an event horizon in all three cases. Consequently, these results {\it indicate the formation of black holes within the quadratic geometric Weyl gravity theory}.  The {\it position of the event horizon of the black holes depends significantly on the numerical values of the scalar and vector fields, and of their derivatives at infinity} (representing the initial conditions for our dynamical systems). These results show {\it the existence of an interesting  relation between the asymptotic values of the scalar and Weyl fields, and the black hole properties}. For example, in the case of the presence of the temporal component of the Weyl vector only, for particular values of the vector field, and of its derivative at infinity, the physical position $r_{hor}$ of the event horizon of the black hole can be located at distances of the order of 0.33 - 1.27 of the standard Schwarzschild radius, a result that indicates the possibility of formation of a large variety of Weyl type black  holes, including the presence of  more compact black holes, having the same mass, but a smaller radius, than the standard Schwarzschild black hole of general relativity. On the other hand, black holes with the same mass, but with radii larger than the Schwarzschild radius, can also exist.

As a general rule, the position of the event horizon of the Weyl geometric black holes is also dependent on the coupling parameter $\gamma$ of the model. Thus,  {\it there is a multi-parametric dependence of the Weyl geometric black hole properties on the geometric couplings, and on the asymptotic conditions at infinity of the metric and scalar and vector fields}. We would also like to point out that in our numerical investigations we could not detect any case of the formation of a naked singularity, with the singularity located at the center $r=0$. In all the considered numerical cases the black holes are hidden beyond an event horizon. But, of course, the possible existence in the present conformally invariant quadratic Weyl geometric model of naked singularities, or topologically trivial soliton type structures cannot be excluded a priori. The numerical detection of these types of objects requires a significant extension of the range of values of the parameter space, and more detailed numerical investigations of the dynamical systems describing static spherically symmetric quadratic Weyl geometric gravity structures in vacuum.

In the present work we have presented only the numerical results obtained by considering the asymptotically flat conditions for the metric tensor at infinity. Numerical solutions corresponding to the de Sitter solutions can also be easily obtained. However, if the de Sitter condition is assumed to be of cosmological type, the deviations from the asymptotically flat case are negligible. But if one assumes a significant difference from flatness at infinity, several distinct types of black holes can be obtained. In particular, solutions corresponding to very big "super-massive black holes" can also be constructed, by imposing appropriate initial conditions on the mass function $m$. One such example of  super-massive black hole solutions of the Weyl geometric field equations in the presence  of the radial component of the Weyl vector are presented in Table~\ref{tabcaseapsidesitter}, for $m(0)=0.4$. In the coordinate $\eta$, these black holes have an extremely small value of the position of the event horizon, of the order of $10^{-5}$, much smaller than the position of the inverse of the event horizon of the Schwarzschild black holes. However, the physical radius of these particular types of black holes is very large, ranging from $r_{hor}\approx 10^4$ ($\eta _{hor}=9.96\times 10^{-5}$), to $r_{hor}\approx 3.69\times 10^4$ ($\eta_{hor}=2.71\times 10^{-5}$).  Such Weyl type structures may be considered as possible alternatives to the super massive black holes of standard general relativity.

\begin{table*}[htbp]
		\centering
		\begin{tabular}{|c||c|c|c|c|}
			\hline
			~~~~$\psi (0)$~~~~ & $1\times 10^{-11}$ & $4\times 10^{-11}$ & $7\times 10^{-11}$ & $9\times 10^{-11}$\\
			\hline
			$\eta_{\textit{hor}}$ & $~~2.71\times 10^{-5}~~$ & $~~3.71\times 10^{-5}~~$ & $~~5.96\times 10^{-5}~~$ & $~~9.96\times 10^{-5}~~$ \\
			\hline
		\end{tabular}
		\caption{Variation of the position of the event horizon $\eta_{\textit{hor}}$ of the Weyl geometric black hole with radial component of the Weyl vector only for $\zeta (0)=1\times 10^{-30}$, $m(0)=0.4$ and different initial values of $\psi$.}\label{tabcaseapsidesitter}
	\end{table*}

 Even that most of our investigations have been performed by using a numerical approach, an exact solution of the field equations has also been obtained by using analytical methods. In the case of a Weyl vector having only a radial component, if the metric functions satisfy the condition $\nu +\lambda =0$, an exact solution of the Weyl geometric gravity field equations can be obtained, with the metric given by a function of the form $e^{\nu}=e^{-\lambda}=A-B/r-Cr+Dr^2$, with $A$,  $B$, $C$, and $D$ constants that do not depend on the initial conditions at infinity. Thus, the metric tensor components contain, except the Schwarzschild term $B/r$, two new terms proportional to $r$ and $r^2$, respectively. We could not find similar simple analytic solutions for the cases of the presence of the temporal component of the Weyl only, or for the general case corresponding to the two components Weyl vector. The analytical solutions are extremely useful in the study of the physical properties of the  black holes, and in particular for the study of the dynamics and motion of matter particles around them. They can also be applied for the investigation of the electromagnetic properties of thin accretion disks that may be present around black holes. The obtained exact solution could also help in discriminating quadratic Weyl geometric black holes from standard general relativistic black holes, and for obtaining observational constraints on the coupling constants, and on the Weyl vector components.

We have also investigated in some detail the thermodynamic properties of the numerical black hole solutions of the quadratic Weyl geometric gravity. One of the extremely interesting physical property of the black holes is their Hawking temperature, an essential parameter that has important theoretical implications. As compared to the Hawking temperature of the standard general relativistic Schwarzschild black holes, the horizon temperature of the quadratic Weyl geometric gravity black holes has a strong dependence on the initial conditions at infinity of the scalar and of the Weyl vector fields.  As one can observe from Fig.~\ref{fi1}, the decrease in the physical horizon radius $r_{hor}$ (corresponding to an increase of the coordinate $\eta$), leads to an increase  in the  black hole temperature, a property that is specific to Weyl geometric black holes. In the case of the specific initial conditions for the scalar and Weyl vector fields considered in Fig.~\ref{fi1}, the increase in the temperature is several times higher, as compared to the Schwarzschild general relativistic case. A similar behavior does appear for the specific heat, entropy, and evaporation time, respectively, of quadratic Weyl geometric black holes, with all these physical quantities strongly dependent on the initial conditions of the scalar and of the Weyl vector field at infinity. The specific heat and the entropy of the Weyl black holes decrease with decreasing $r_{hor}$ (increasing $\eta _{hor}$). The black hole  evaporation times may be very different for Weyl type black holes, as compared to the Schwarzschild black holes, decreasing with increasing $r_{hor}$. However, we would like to point out that {\it our results on the thermodynamics of Weyl gravity black holes are obtained  for a limited range of initial conditions at infinity for the scalar field and for the Weyl vector, and therefore they may be considered as having a qualitative interpretation only}. But, even within this limited level of numerical investigation, they provide {\it an indicator of the complex physical behavior of the Weyl black holes, and of the interesting and novel properties related to them}.

The no-hair theorem is an important result in black hole physics \cite{Bek, Bek1, Adler, Bek2}. For short, this theorem asserts that asymptotically flat black holes cannot have external nontrivial  scalar fields possessing a non-negative field potential $V (\phi)$. It would be interesting to consider the no-hair theorem for Weyl geometric black holes. The preliminary, and mostly numerical results obtained in the present work seem to point towards the conclusion that the no-hair theorem in its standard formulation may not be valid for quadratic Weyl black holes. All the  black hole solutions we have considered have an asymptotically flat geometry, and scalar and vector fields do exist around them. However, the question if these results follow from the particular choice of the model parameters (coupling constant and initial conditions at infinity), or they are essential properties of the theory deserves further studies, and investigations.

Quadratic Weyl gravity black holes have more variability as associated to their basic properties, leading to a more complicated external dynamics, as compared with the Schwarzschild black holes of general relativity. These richer properties do follow from the presence of the scalar and vector degrees of freedom, resulting in very complicated and  strongly nonlinear field equations. The effects associated  with the scalar field and Weyl vector degrees of freedom  could also lead to some specific astrophysical imprints  and signatures, whose observational detection could open some new perspectives on the possibilities of testing Weyl geometry, and its associated gravitational theory on cosmic scales. The possible observational/astrophysical implications  of the existence of quadratic Weyl geometric black holes  will be considered
in a future investigation.

\section*{Acknowledgments}

We would like to thank the anonymous referee for comments and suggestions that helped us to significantly improve our work. TH is supported by a grant of the Romanian Ministry of Education and Research, CNCS-UEFISCDI, project number PN-III-P4-ID-PCE-2020-2255 (PNCDI III).

\end{document}